\documentclass[trackchanges,twocolumn]{aastex701}

\usepackage{amsmath}
\usepackage{ragged2e}
\usepackage{booktabs}
\usepackage{subfig}
\usepackage{caption}
\usepackage[english]{babel}
\PassOptionsToPackage{disable}{longtable} 
\usepackage{adjustbox} 
\usepackage{graphicx}  
\usepackage{tabularx}

\begin{document}

\title{Oblique Shocks at Supernova Remnants in Massive Star Clusters: A Model for the Cosmic-Ray Knee Observed by LHAASO}

\author[0000-0003-2508-6157]{Luana N. Padilha}
\affiliation{Programa de Pós-graduação em Física \& Departamento de Física, Universidade Estadual de Londrina (UEL), 86057-970 Londrina, PR, Brazil}
\affiliation{INAF-Osservatorio Astrofisico di Arcetri, Largo Enrico Fermi 5, Firenze, Italy}
\email[show]{luana.natalie.padilha@uel.br}

\author[0000-0002-6463-2272]{R. C. Dos Anjos}
\affiliation{Programa de Pós-graduação em Física \& Departamento de Física, Universidade Estadual de Londrina (UEL), 86057-970 Londrina, PR, Brazil}
\affiliation{Departamento de Engenharia e Ciências Exatas, Universidade Federal do Paraná (UFPR), Pioneiro, 2153, 85950-000 Palotina, PR, Brazil}
\affiliation{Núcleo de Astrofísica e Cosmologia, Universidade Federal do Esp\'irito Santo (UFES), 29075--910, Vit\'oria, ES, Brazil}
\affiliation{Programa de Pós-graduação em Física e Astronomia, Universidade Tecnológica do Paraná (UTFPR), 80230-901, Curitiba, PR, Brazil}
\affiliation{Programa de Pós-graduação em Física Aplicada, Universidade Federal da Integração Latino-Americana (UNILA), 85867-670, Foz do Igua\c{c}u, PR, Brazil}
\email[show]{ritacassia@ufpr.br}

\begin{abstract}
This work establishes oblique shocks in Massive Star Clusters (MSC) as a primary mechanism for accelerating cosmic rays (CR) up to the knee of the energy spectrum. We develop a model that incorporates the combined contribution of supernova and collective wind shocks, emphasizing the critical role of the shock obliquity angle in determining the maximum particle energy. We illustrate, within our model that oblique shocks can significantly enhance acceleration efficiency, allowing particles to reach multi-PeV energies in a rigidity-dependent manner. Our preferred model, which incorporates oblique shocks, reproduces the all-particle spectrum and composition observed by The Large High Altitude Air Shower Observatory (LHAASO), interpreting the knee as arising from a sequence of rigidity-dependent cutoffs. The model also predicts subdominant but detectable gamma-ray and neutrino emissions. This study provides an attempt at building a unified framework connecting MSC particle acceleration to the observed features of the cosmic-ray knee.
\end{abstract}

\keywords{\uat{Cosmic rays}{329}; \uat{Particle acceleration}{1164}; \uat{Gamma-ray sources}{633}; \uat{Shock}{2086}; \uat{Termination shock}{1690}; \uat{Supernova remnants}{1667}; \uat{Young massive clusters}{2049}}

\section{Introduction} 
\label{sec:intro}

In recent years, numerous measurements have refined our knowledge of the primary spectrum and chemical composition of cosmic rays (CR), spanning direct observations with stratospheric balloons and space missions and indirect reconstructions with ground arrays \citep{2002PhRvD..66h3004K, 2009PrPNP..63..293B}. Together, these datasets allow a coherent reconstruction of the CR spectrum across broad energy ranges. The all-particle spectrum follows an approximate power law with characteristic structures, notably the knee and the ankle. The knee, at $(2-5)\times 10^{15}$~eV, marks a steepening of the spectrum \citep{1996ApJ...461..408A, 2012APh....35..660K} and encodes key information on sources, transport, and the maximum energies achieved by Galactic accelerators \citep{2002PhRvD..66h3004K, 2019NatAs...3..561A}.

A variety of acceleration mechanisms have been proposed to explain the origin of cosmic rays within our Galaxy, particularly around the knee region of the spectrum \citep{2019A&A...623A..86A, 2022hxga.book...52V, 2013Sci...339..807A, 2006A&A...449..223A, 2023MNRAS.519..136V}. From the foundational concept introduced by Fermi in 1949, to contemporary models involving diffusive shock acceleration at supernova remnants (SNRs), stochastic acceleration by turbulence, magnetic reconnection, and acceleration in magnetospheric gaps, numerous scenarios contribute to particle energization \citep{1949PhRv...75.1169F, 2023EPJWC.28304001G, 2020LRCA....6....1M}. SNRs remain the leading Galactic candidates responsible for accelerating cosmic rays up to PeV energies, consistent with rigidity-dependent spectral steepening observed near the knee \citep{2021A&A...650A..62C, 2023PPCF...65a4002B}. Detailed models of particle acceleration in SNRs have been developed over the past decades, incorporating treatments of shock evolution, magnetic field amplification via cosmic-ray-driven streaming instabilities, and particle escape \citep{2010ApJ...718...31P, 2013MNRAS.431..415B, 2019IJMPD..2830022G, 2020APh...123j2492C}. The detection of neutral pion decay gamma rays from middle-aged SNRs strongly supports a hadronic component in cosmic ray production \citep{2013Sci...339..807A}. However, the complexity of SNR environments and source distributions introduces challenges in fully reproducing all aspects of the observed CR spectrum.

Complementary ideas combine supernova shocks with the collective winds of massive, young stars, where repeated shocks propagate in a turbulent, magnetized medium inside superbubbles and MSC \citep{2001AstL...27..625B, 2004A&A...424..747P, 2010A&A...510A.101F, 2010ARA&A..48..431P, 2018NPPP..297..183B}. In this framework, MSCs provide the physical conditions, such as enhanced magnetic field strength, elevated turbulence levels, and efficient particle confinement, that modify and regulate the acceleration process at SNR shocks evolving within the cluster environment. This distinction is crucial: while isolated SNRs rely primarily on cosmic-ray-driven streaming instabilities for magnetic field amplification \citep{2004MNRAS.353..550B, 2013MNRAS.431..415B}, cluster environments offer an alternative amplification mechanism through the confinement and conversion of stellar wind kinetic energy, as demonstrated by three-dimensional magnetohydrodynamic (MHD) simulations \citep{2022MNRAS.517.2818B}.
Building on this framework, oblique shocks, where the magnetic field intersects the shock at a finite angle, can modify spectral features and raise maximum energies in a rigidity-dependent fashion, providing a natural route to the knee~\citep{2002PhRvD..66h3004K}. Recent particle-in-cell simulations have confirmed that shock obliquity affects both spectral indices and maximum energies. However, the spectral index dependence on obliquity is complex: depending on the angle and plasma parameters, oblique configurations can produce spectra harder than $p^{-4}$~\citep{2011MNRAS.418.1208B} or even curved spectra~\citep{2025MNRAS.544L.160S}, though they may also yield steeper indices under certain conditions~\citep{2015ApJ...798L..28C}.

The Large High Altitude Air Shower Observatory (LHAASO), located at an altitude of 4410~m, is a hybrid detector array encompassing $\sim 1.3$~km$^2$ dedicated to detecting CR and gamma rays \citep{2019arXiv190502773C}. Its Square Kilometer Array (KM2A) component is designed for cosmic-ray composition studies around the knee region. KM2A combines a surface array of 5216 electromagnetic detectors (EDs, 15--30~m spacing) with an underground array of 1188 muon detectors (MDs, 36~m$^2$ each), situated at a depth of 2.5~m of soil. This design provides high-quality measurements of the lateral distributions of electrons and muons from air showers in the energy range from $\sim 30$~TeV to $\sim 100$~PeV \citep{2019ChPhC..43g5001Y, 2020ChPhC..44f5002J, 2022PhRvD.106l3028Z, 2024PhRvL.132m1002C, 2024PhRvD.110d3030T}.

LHAASO has recently reported a high-precision proton and helium spectrum at the knee and delivered TeV - PeV spectra of diffuse Galactic gamma rays. The combined dataset offers a direct test of the relation between local hadronic spectra and large-scale gamma-ray emission and motivates a careful, rigidity-aware interpretation of the knee in terms of energy and composition groups. Critically, the measured mean logarithmic mass of $\langle \ln A \rangle \approx 1.3$ at the knee provides a powerful constraint on the composition. This value is remarkably close to that of pure helium ($\langle \ln A \rangle \approx 1.39$) and is far heavier than pure proton. This indicates that the flux is dominated by light elements, with the combined proton and helium contribution accounting for 65--95\% of the total particle flux at the knee, proving conclusively that this feature marks the cutoff of these light, high-rigidity components \citep{2022PhRvD.106l3028Z, 2024PhRvL.132m1002C, 2024PhRvD.110d3030T}.

This work investigates the role of oblique shocks in massive star clusters (MSCs), where particle acceleration occurs at supernova shocks that evolve within the cluster environment, to explain the observed cosmic-ray spectrum and composition around the knee. Previous approaches to this problem \citep{2023MNRAS.519..136V, 2025MNRAS.luana} relied 
on strong assumptions that may not be physically realistic, most notably the requirement for unusually high magnetic field values to reach the energies observed at the knee. These assumptions were necessary within a purely parallel-shock framework, in which the shock normal is aligned with the upstream magnetic field, but they introduce significant model dependence. Oblique shocks, where the shock normal is inclined with respect to the magnetic field, provide a physically motivated way to relax these constraints: the effective magnetic field component relevant for particle confinement is enhanced geometrically, allowing maximum energies consistent with the knee to be reached without invoking extreme field amplification.

Our approach is phenomenological and designed to explore the statistical contribution of MSC populations to the knee region. We consider two classes of clusters: dispersed massive systems, and young compact massive systems, which differ primarily in the 
environments experienced by SNR shocks. We further derive conservative estimates for the high-energy gamma-ray and neutrino fluxes from the collective contribution of such 
clusters. Section~\ref{analysis_catalog} introduces the observed cluster sample and the synthetic population used to supplement it. Section~\ref{sec:spectrum} presents the particle spectra and composition analysis anchored to LHAASO results, as well as the 
rigidity-dependent shock model \citep{1987ApJ...313..842J, 2006A&A...454..687M}. Section~\ref{sec:gama_neutrino} provides upper limits for gamma-ray and neutrino emission, and Section~\ref{sec:Conclusions} summarizes the main findings.

\section{Classification of Massive Star Clusters: observational and synthetic hybrid Catalog}
\label{analysis_catalog}

Star clusters serve as powerful astrophysical laboratories for probing the structure and history of our Galaxy. They span a broad range of ages, from young clusters just a few million years old, which are essential for understanding ongoing star formation and early stellar evolution, to ancient globular and open clusters billions of years in age, whose origins trace back to the early stages of the Galactic disk. By examining these systems across cosmic time, we acquire a comprehensive understanding of the dynamical and chemical evolution of the Milky Way, thereby elucidating the processes that govern the formation and aging of its stellar populations~\citep{1995ARA&A..33..381F, 2010ARA&A..48..431P,2013A&A...558A..53K, 2018A&A...618A..93C, 2020A&A...640A...1C, 2020A&A...633A..99C}.

Galactic catalogs have been instrumental in analyzing and understanding the physical processes occurring within star clusters. Among the most widely used are the catalog by \citet{2020A&A...640A...1C}, which is based on GAIA DR2 data, and the compilation by \citet{2013A&A...558A..53K}, which incorporates data from the COCD (Catalogue of Open Cluster Data), DAML02 (version 3.1 from 2010), as well as several catalogs focused on infrared-detected clusters \citep{2003A&A...404..223B, 2003A&A...397..177B, 2003A&A...400..533D, 2007MNRAS.374..399F, 2010MNRAS.409.1281F, 2009MNRAS.400..518M, 2011AcA....61..231B}. The latter also integrates astrometric and photometric data from the PPMXL and 2MASS surveys.

Although the catalog by \citet{2020A&A...640A...1C} is more recent, the one by \citet{2013A&A...558A..53K} is particularly notable for providing a broad range of derived parameters and covering a substantial number of clusters, including 3006 objects confirmed as real. This makes it especially well suited for the present study. Nevertheless, the \citet{2013A&A...558A..53K} catalog shares a common limitation with many other catalogs: its observational coverage is incomplete. It is nearly complete only within approximately 2 kpc from the Sun, and the authors note that older clusters with weak emission may still be missing \cite{2013A&A...558A..53K}. This is an issue we aim to address in the subsequent sections. In Section \ref{class.}, we describe our classification approach and, using empirical estimates, extend the catalog beyond a reliable radius with a synthetic population, as detailed in Section \ref{pop.syn}.

\subsection{Cluster selection and classification criteria}
\label{class.}

\begin{figure*}[ht!]
    \centering
    \includegraphics[width=0.9\linewidth]{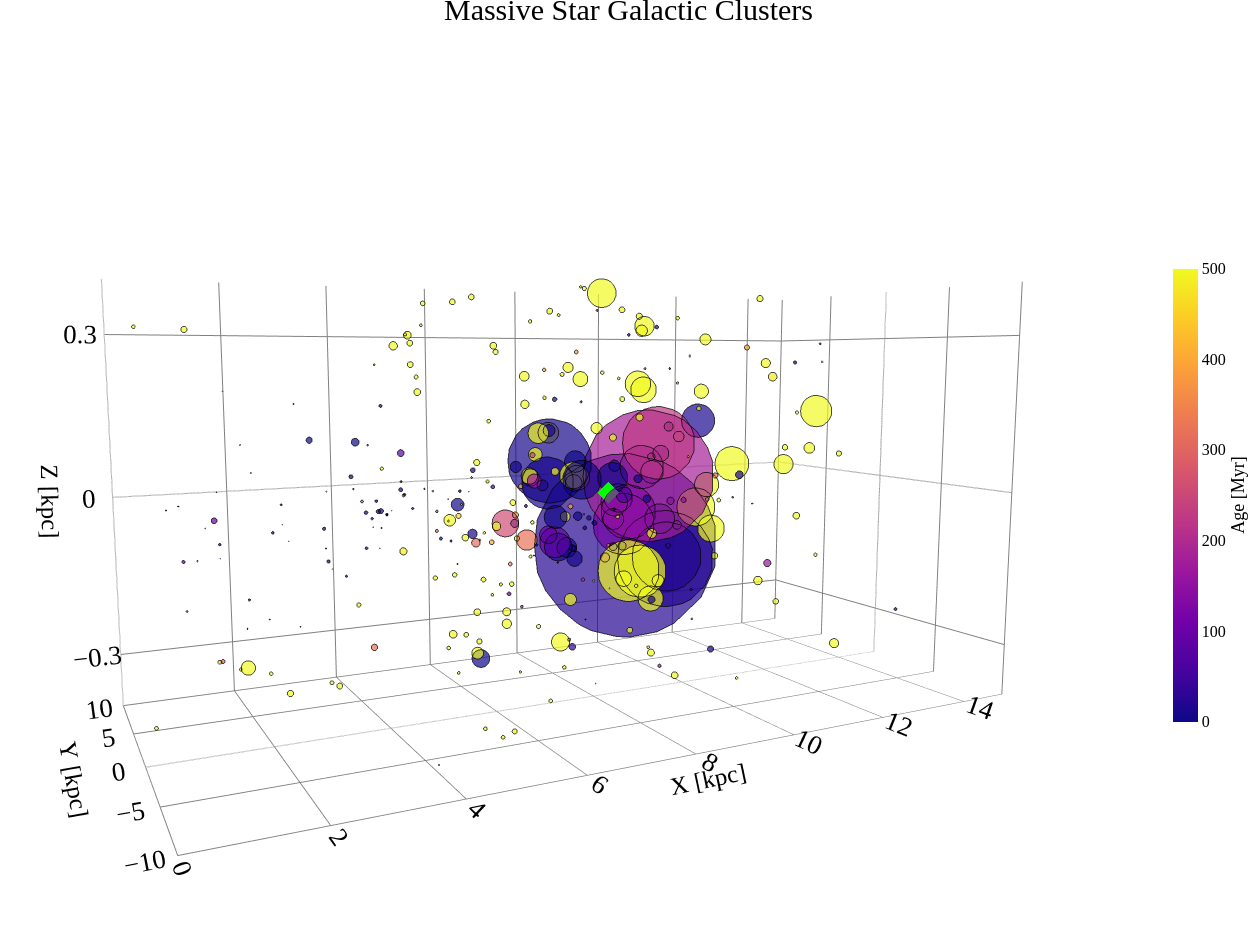}
    \caption{\small \justifying Three-dimensional distribution of the selected massive star clusters from the \citet{2013A&A...558A..53K} catalog in Sun-centered Cartesian coordinates (X, Y, Z in kpc). Marker colour encodes the cluster age, and larger markers indicate higher estimated cluster mass \(M_{\mathrm{cl}}\) (selection requires \(M_{\mathrm{cl}}>10^{3}\,M_{\odot}\)); the concentration near \(Z\simeq 0\) traces the Galactic mid-plane.}
    \label{fig:msc}
\end{figure*}

This work focuses on MSC as efficient particle accelerators powered by their populations of young, massive stars. Candidate MSC were identified through a mass proxy derived from the tidal radius \(r_{\mathrm{t}}\) following \cite{1957ApJ...125..451V} and \cite{2013ApJ...764..124W}:

\begin{equation}
r_{\mathrm{t}} \;=\; R_{\mathrm{gc}}\left(\frac{M_{\mathrm{cl}}}{M_{\mathrm{gal}}}\right)^{1/3},
\label{eq:mass}
\end{equation}
where \(R_{\mathrm{gc}}\) is the Galactocentric distance, \(M_{\mathrm{cl}}\) is the cluster mass, and \(M_{\mathrm{gal}}\) is the total Galactic mass \citep{2025MNRAS.luana}. We adopt \(M_{\mathrm{gal}}=10^{11}\,M_{\odot}\), appropriate for systems within \(100\,\mathrm{kpc}\) \citep{1998MNRAS.294..429D}. Inverting Eq.~\eqref{eq:mass} yields \(M_{\mathrm{cl}} = M_{\mathrm{gal}}\,(r_{\mathrm{t}}/R_{\mathrm{gc}})^{3}\), which we apply to the catalog of \citet{2013A&A...558A..53K}. We select clusters with \(M_{\mathrm{cl}}>10^{3}\,M_{\odot}\), as shown in Fig.~\ref{fig:msc} \footnote{An interactive 3D visualization of the catalog map is accessible at \href{https://anonymous-astro2025.github.io/3D_maps/MSC_age}{MSC age}. Complementary interactive views emphasizing (i) age and (ii) King radius, with the powerful and soft classes highlighted, are accessible at \href{https://anonymous-astro2025.github.io/3D_maps/powsoft_age}{age-emphasis} and \href{https://anonymous-astro2025.github.io/3D_maps/powsoft_kingradius}{King-radius}, respectively.}. This first cut yields \(598\) MSC, approximately \(20\%\) of the full catalog. 

To sustain strong collective winds, MSC must contain massive stars; stars with \(M\geq 8\,M_{\odot}\) have lifetimes shorter than \(40\,\mathrm{Myr}\)~\citep{2008ApJ...675..614P}. We therefore impose an age cut at 40 Myr, which reduces the sample to \(98\) clusters (\(\simeq 16\%\) of the MSC set). Cluster compactness enhances acceleration efficiency by strengthening wind–wind interactions and turbulence. We quantify compactness via the King radius \(r_{0}\), the radius where the stellar surface density drops to roughly one-third of its central value \citep{1962AJ.....67..471K}. Following \cite{2023MNRAS.519..136V} and \cite{2025MNRAS.luana}, we adopt \(r_{0}\lesssim 5\,\mathrm{pc}\) as the compactness threshold.

From the compact subsample we further select the youngest systems, with ages \(\leq 20\,\mathrm{Myr}\), consistent with \cite{2023MNRAS.519..136V},\cite{2025A&A...695A.175M},\cite{2024A&A...686A.118C},\cite{2024ApJ...972L..22P} and \cite{2025MNRAS.luana}. Based on these criteria we define two categories: \emph{Powerful clusters:} young (\(\leq 20\,\mathrm{Myr}\)) and compact (\(r_{0}\lesssim 5\,\mathrm{pc}\)) MSC with intense collective winds and conditions conducive to forthcoming supernovae; no supernova has yet occurred in these systems; and \emph{Soft clusters:} older systems with ages \(20\)–\(40\,\mathrm{Myr}\), which may be compact or extended.

Our model includes acceleration at \emph{wind termination shocks} formed where the cluster wind meets the interstellar medium (ISM), an efficient site for energizing particles \citep{2023MNRAS.519..136V,2025MNRAS.luana}. 
When supernova activity is present, the combination of wind and supernova shocks provides optimal conditions for cosmic-ray acceleration \citep{2023MNRAS.519..136V,2025MNRAS.luana}. In contrast, soft clusters are primarily powered by supernova shocks because their winds have weakened with age.

The \citet{2013A&A...558A..53K} catalog is fully reliable only out to \(\sim 2\,\mathrm{kpc}\). To assess local contributions to the CR population, we additionally impose a spatial cut of \(3\,\mathrm{kpc}\), retaining the nearest clusters to maximize completeness and reduce systematics. Within this volume, our classification yields \(15\%\) soft clusters and \(11\%\) powerful clusters. This distinction enables us to assign distinct acceleration channels and to evaluate their respective contributions to cosmic rays (Sec.~\ref{ssec:inj_prop}) as well as to associated gamma-ray and neutrino emission (Sec.~\ref{sec:gama_neutrino}).

\subsection{Synthetic population modeling and Catalog completion}
\label{pop.syn}

Observational constraints, in particular sensitivity limits and interstellar extinction, hinder the detection of distant or intrinsically faint stellar systems. As a consequence, available compilations of Galactic clusters are incomplete, which complicates any attempt to derive their true spatial distribution~\citep{2025A&A...695A.175M}.

Independent estimates of the cluster formation rate in the Milky Way suggest rates between \(0.2\) and \(0.5\,\mathrm{Myr^{-1}\,kpc^{-2}}\), equivalent to about \(200\,M_{\odot}\,\mathrm{Myr^{-1}\,kpc^{-2}}\) \citep{2006A&A...445..545P,1991MNRAS.249...76B}. Over at least \(250\,\mathrm{Myr}\), this implies that roughly \(2.3\times10^{4}\) to \(3.7\times10^{4}\) clusters should have formed \citep{2010ARA&A..48..431P}. By contrast, the most comprehensive modern catalogs \citep{2020A&A...640A...1C,2013A&A...558A..53K} contain only about ten to fifteen percent of that expectation.

To mitigate this bias, we extend the \cite{2013A&A...558A..53K} sample with a synthetic population. We concentrate on the region beyond a radius of \(3\,\mathrm{kpc}\) from the Sun. Although the catalog is regarded as complete within \(2\,\mathrm{kpc}\), our science targets are young systems with ages up to \(40\,\mathrm{Myr}\), which reduces the impact of potential incompleteness among faint old clusters.

\begin{figure}[ht!]
    \centering
    \includegraphics[width=0.9\linewidth]{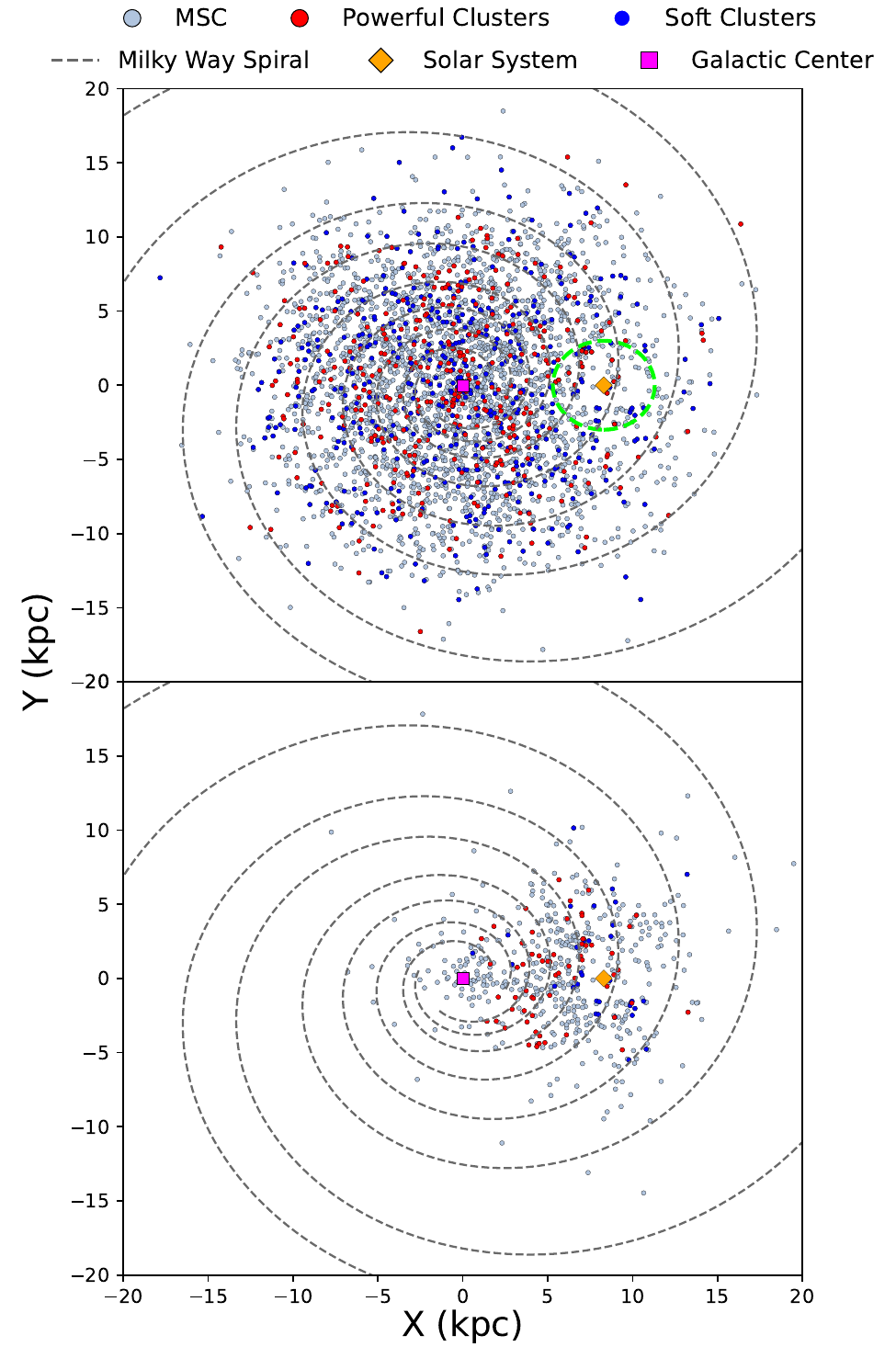}
    \caption{\small \justifying
    Spatial distribution of massive star clusters in the Galactic plane. Upper panel: combined sample that includes the simulated population and the observed objects, projected onto the \(X\)\(Y\) plane. The green circle marks the sources inside a radius of \(3.0\,\mathrm{kpc}\) from the Sun as listed by \citet{2013A&A...558A..53K}. Lower panel: the observational catalog alone with the same projection. Grey curves indicate the spiral pattern adopted for visual reference.}
    \label{fig:mapXY}
\end{figure}

We first characterize the radial distribution of observed objects with respect to the Galactic center using a Kernel Density Estimation (KDE)\citep{rosenblatt1956remarks}, which yields a continuous probability density curve. We then fit a Gaussian model to this KDE, obtaining peak amplitude \(a \simeq 0.108\), mean position \(\mu \simeq 6.924\,\mathrm{kpc}\), and dispersion \(\sigma \simeq 3.775\,\mathrm{kpc}\). The fit is performed by least squares with physically motivated initial guesses to ensure convergence. This procedure validates a simple Gaussian description of the radial density and provides a continuous representation along the galactocentric radius. 

Next, we generate a synthetic population of massive clusters distributed in the Galactic plane. Galactocentric radii are drawn from a Gaussian truncated between \(0\) and \(20\,\mathrm{kpc}\), using the parameters inferred from the KDE analysis. Longitudes are sampled uniformly, and the vertical coordinate \(z\) is chosen at random within a symmetric interval of \(\pm 0.5\,\mathrm{kpc}\), thereby confining the sources to the thin disk. We retain only objects with \(|z|\le 0.5\,\mathrm{kpc}\) and Galactic latitude \(|b|<5^{\circ}\) as seen from the Sun, which correspond to those with the highest detection probability.

The total number of simulated massive clusters is set to \(3200\). This follows from an estimated Galactic population of about \(4\times10^{5}\) massive stars, of which approximately \(80\%\) are assumed to reside in clusters with a mean of one hundred massive stars per cluster, yielding an expected number of massive clusters close to \(3200\) \citep{2023MNRAS.519..136V,2025MNRAS.luana}. To avoid double counting, we subtract objects already present in the observational sample within \(3\,\mathrm{kpc}\) of the Sun.

We then assign evolutionary classes to the synthetic objects according to the fractions derived in Section~\ref{class.}: about \(11\%\) powerful clusters and about \(15\%\) soft clusters. These proportions are applied randomly to the simulated set so that the combined distribution reproduces the nearby statistics. The outcome of merging the synthetic population with the observed catalog is shown in Figure~\ref{fig:mapXY}. The combined catalog, integrating both observed and synthetic sources, provides a realistic representation of the expected Galactic distribution of MSC and serves as the foundation for subsequent emission calculations. In particular, it is fundamental to deriving constraints on gamma-ray and neutrino fluxes discussed in Section~\ref{sec:gama_neutrino}.

\section{Modeling of Particle Acceleration in MSC}
\label{sec:spectrum}

\subsection{Maximum Energy from Oblique Shocks}
\label{ssec:obliquo}

The efficiency of particle acceleration is governed not only by the properties of the source and its ambient medium but also by the geometry of the shock. In particular, the angle between the shock normal and the local magnetic field, known as the obliquity, controls cross–field transport and the microphysics of injection. Changes in obliquity modify the parallel and perpendicular diffusion coefficients and activate different pre–acceleration channels, which in turn regulate both the acceleration rate and the maximum rigidity that particles can reach \citep{1987ApJ...313..842J, 2002PhRvD..66h3004K, 2003APh....19..649M, 2006A&A...454..687M, 2011MNRAS.418.1208B, 2022RvMPP...6...29A}.

In this work, particle acceleration is assumed to take place at supernova shocks evolving inside massive star clusters. MSC provide a natural laboratory where shocks of many obliquities coexist. Cluster winds, supernova blast waves, and their interactions drive a turbulent, magnetized environment in which repeated compressions and shear flows favor efficient acceleration \citep{2019NatAs...3..561A, 2023MNRAS.519..136V, 2024hegr.confE..16G, 2025MNRAS.luana}. In what follows, we examine the role of oblique shocks in these systems and explain why they are especially relevant to cluster conditions. We then apply a rigidity dependent acceleration model, assuming that particle acceleration occurs at supernova shocks evolving inside massive star clusters, to estimate the maximum particle energies reachable in these environments.

\subsubsection{Oblique shocks in particle acceleration}
\label{oblique}

Studies such as~\citet{2011MNRAS.418.1208B} identify shock obliquity as a controlling parameter for acceleration outcomes, because the angle between the shock normal and the magnetic field fixes the degree of cross-field transport and governs particle return to the upstream. Particle-in-cell simulations confirm that shock obliquity can affect both spectral indices and maximum energies. Depending on the angle and plasma parameters, oblique configurations can produce spectra harder than $p^{-4}$~\citep{2011MNRAS.418.1208B} or even curved spectra, as demonstrated using the publicly available code \textsc{Sapphire++}~\citep{2025JCoPh.52313690S, 2025MNRAS.544L.160S}. However, the actual spectral shape and maximum energy depend sensitively on the injection mechanism, magnetic field turbulence, and shock parameters.

Nevertheless, in regimes dominated by anisotropic transport, strongly oblique shocks can achieve faster energy gains under suitable conditions. Building on systematic studies of nonrelativistic shocks with obliquities from \(0^{\circ}\) to \(90^{\circ}\), \citet{2006A&A...454..687M} emphasized subluminal configurations and showed that the acceleration rate is controlled by the relative strength of the diffusion coefficients parallel and perpendicular to the mean magnetic field, \(k_{\parallel}\) and \(k_{\perp}\). In the regime \(k_{\parallel}\gg k_{\perp}\), large obliquity promotes efficient confinement near the discontinuity and enables repeated shock encounters, leading to rapid energy growth. This behavior is consistent with the classical rigidity-dependent arguments of \citet{1987ApJ...313..842J}.

Such conditions are naturally expected in MSC, where feedback from massive stars drives a turbulent, magnetized medium. Ionizing radiation and powerful stellar winds inject energy and momentum into the gas: photoionization heats and overpressurizes H\,\textsc{ii} regions, radiation pressure and wind ram pressure stir the surrounding material, and turbulence and magnetic amplification follow, as observed for the Pismis~24 region in NGC~6357~\citep{2015A&A...573A..95M}. Young clusters in the inner Milky Way show similar behavior: gravitationally unstable gas forms condensations that evolve under intense feedback, carving ionized cavities and distorting the magnetic field, which promotes magnetic turbulence~\citep{2015EAS....75...43L,2024JApA...45...17S}.
 
In this physical context, several models of clusters estimate maximum energies in superbubbles by adopting a parametric form of diffusive shock acceleration (DSA) that depends on confinement scales and the finite lifetime of the accelerating shock. Examples include the applications of \cite{2023MNRAS.519..136V}, \cite{2025MNRAS.luana} and \cite{2022MNRAS.512.1275V}, which follow the classical prescriptions of \cite{1983A&A...125..249L} and \cite{2005JPhG...31R..95H}. These approaches capture global energetics in environments driven by supernova shocks but do not incorporate the explicit dependence on shock obliquity. While more detailed treatments of supernova remnant evolution exist in the literature \citep{2010ApJ...718...31P, 2013MNRAS.431..415B, 2019IJMPD..2830022G, 2020APh...123j2492C, 2021A&A...650A..62C}, our phenomenological approach is designed to explore the statistical contribution of MSC populations to the cosmic-ray spectrum around the knee, rather than to model individual remnant evolution in detail.

The high magnetic field strengths required to accelerate particles to PeV energies present a significant challenge. In our model, we adopt a mean magnetic field strength of $\sim 100\,\mu$G in the non-oblique shock scenario \citep{2023MNRAS.519..136V}. However, for the oblique-shock models we assume a lower mean field of $\sim 50\,\mu$G motivated by recent three-dimensional MHD simulations of young massive star clusters. In particular, simulations by \citet{2024MNRAS.527.3749B} show that magnetic field strengths above $100\,\mu$G occur only locally in the cluster core, with a low filling factor upstream of the shock. Moreover, \citet{2025A&A...698A...6H} find that although magnetic field strengths can reach milligauss levels in the immediate vicinity of magnetic stars, the volume-filling magnetic field in the subsonic cluster core is much lower, with mean values in the range $\sim 30-200\,\mu$G and median values of $\sim 8-35\,\mu$G. In the surrounding subsonic superbubble, even lower mean field strengths of $\sim 5-25\,\mu$G are reported. These results indicate that fields of a few tens of microgauss provide a realistic description of the upstream medium encountered by shocks propagating through the collective cluster wind. Unlike isolated supernova remnants, where cosmic-ray-driven streaming instabilities \citep{2004MNRAS.353..550B, 2013MNRAS.431..415B} are the primary mechanism for magnetic field amplification, we invoke a different process in cluster environments. The large kinetic energy flux from stellar winds of O and Wolf-Rayet stars, totaling \(\sim10^{38}\)--\(10^{39}\,\text{erg\,s}^{-1}\) in the most massive young clusters, is deposited and efficiently trapped within the compact cluster volume \citep{2022MNRAS.515.2256V, 2023MNRAS.519..136V, 2025MNRAS.luana}. This energy is converted into magnetic energy through the Axford--Cranfill mechanism \citep{1977ICRC....2..273A, 1978ApJ...226..650C}, in which compressions and shear flows in the turbulent, shocked wind material amplify seed magnetic fields.

Three-dimensional MHD simulations of MSC cores support this scenario \citep{2022MNRAS.517.2818B}. These simulations demonstrate that kinetic energy injected by stellar winds is efficiently converted into magnetic energy through turbulent dynamo action and shock compression, generating magnetic fields organized in filamentary structures with peak values reaching several hundred \(\mu\)G. Although these simulations do not fully resolve the smallest turbulent scales, they confirm that the confined energy budget and the intense shock activity within clusters can sustain field strengths sufficient for PeV acceleration without invoking cosmic-ray-driven instabilities. We emphasize that this mechanism differs fundamentally from that operating in isolated SNRs: it relies on the collective wind power and confinement provided by the cluster environment, rather than on the resonant growth of waves driven by escaping cosmic rays. This distinction is crucial, as streaming instabilities are inefficient at oblique and perpendicular shocks \citep{2004MNRAS.353..550B, 2013MNRAS.431..415B}, whereas the wind-driven turbulent amplification operates independently of shock geometry.

Motivated by this set of evidence, our analysis focuses on oblique shocks in MSC, where the complex magnetic topology and vigorous turbulence naturally yield a broad distribution of shock angles. To estimate the maximum particle energies achievable under the combined effects of anisotropic diffusion and constraints from finite lifetimes, we adopt the prescriptions of \cite{2006A&A...454..687M} and \cite{1987ApJ...313..842J}, and we apply these models to the cluster samples introduced in the previous sections.

\subsubsection{Estimation of maximum energies}
\label{emaxoblique}

To estimate the maximum particle energy achieved at oblique shocks, we follow \citet{2006A&A...454..687M}, who models the time momentum growth up to knee energies by balancing systematic acceleration with adiabatic losses:
\begin{equation}
    p(t) \;=\; p_{0}\,\left(\frac{t}{t_{0}}\right)^{1.25}.
    \label{bierman}
\end{equation}
Equation~\eqref{bierman} encapsulates the effect of the acceleration rate and expansion losses during the initial phase of energization \citep{2006A&A...454..687M}. Because precise start and stop times for efficient acceleration are difficult to specify in realistic flows, it is convenient to relate the evolution time to the shock dynamical scale, \(t \simeq r_{\mathrm{sh}}/u_{\mathrm{sh}}\), where \(r_{\mathrm{sh}}\) is the shock radius and \(u_{\mathrm{sh}}\) is the shock speed \citep{2006A&A...454..687M}. The reference time \(t_{0}\) can be identified with the acceleration time \(t_{\mathrm{acc}}\), which represents the lower bound required for multiple shock crossings in diffusive acceleration \citep{1987ApJ...313..842J}.

\begin{figure}[ht!] 
\centering 
\includegraphics[width=0.95\linewidth]{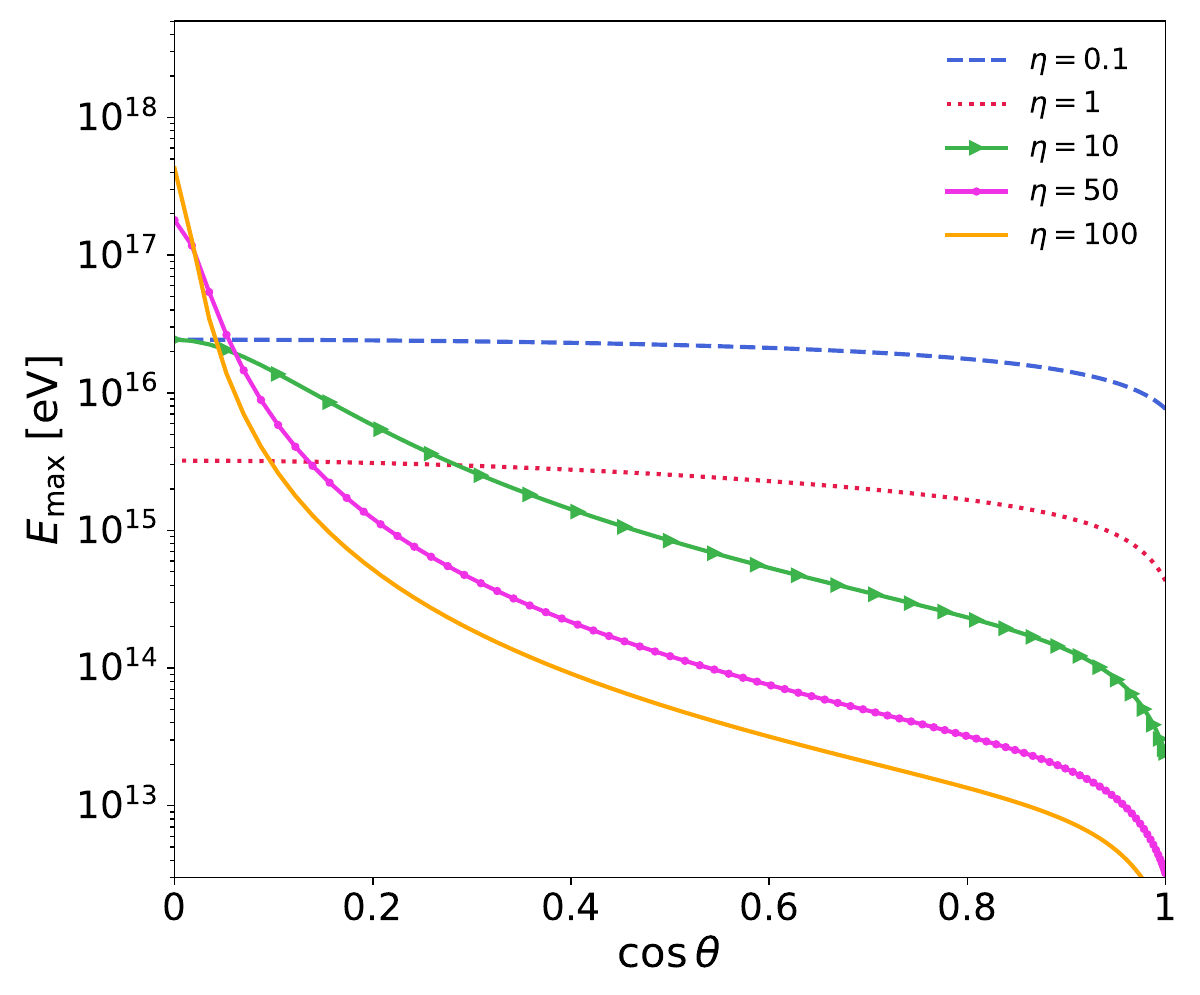} 
\caption{\small \justifying
Maximum energy \(E_{\max}\) as a function of the obliquity, shown as \(\cos\theta\). Curves are computed from Eq.~\eqref{emax} using the diffusion coefficients in Eq.~\eqref{eq:k1k2} at fixed shock and environmental parameters (see Sec.~\ref{emaxoblique}).
As \(\eta\) increases, diffusion becomes strongly anisotropic and \(E_{\max}\) rises toward more perpendicular shocks, reflecting faster acceleration due to reduced effective cross–field transport; in the limit \(\eta\!\ll\!1\) the angle dependence weakens and the curves flatten.}
\label{fig:n}
\end{figure}

The acceleration time derived from the transport equation for energetic particles, including spatial diffusion and advection on both sides of the discontinuity, is
\begin{equation}
t_{\mathrm{acc}} \;=\; \frac{3}{v_{1}-v_{2}}
\int_{p_{0}}^{p_{1}}
\left(\frac{k_{1}}{v_{1}}+\frac{k_{2}}{v_{2}}\right)\,\frac{dp}{p},
\label{t_acc}
\end{equation}
where \(k_{1}\) and \(k_{2}\) are the effective diffusion coefficients upstream and downstream, \(v_{1}\) and \(v_{2}\) are the corresponding flow speeds in the shock frame, and \(r\equiv v_{1}/v_{2}\) is the compression ratio.
Assuming that \(k_{1}\), \(k_{2}\), \(v_{1}\), and \(v_{2}\) vary slowly with momentum over the interval of interest, Eq.~\eqref{t_acc} reduces to
\[
t_{\mathrm{acc}} \;\simeq\; \frac{3}{\,v_{1}-v_{2}\,}\left(\frac{k_{1}}{v_{1}}+\frac{k_{2}}{v_{2}}\right).
\]
For an oblique shock, the normal diffusion coefficients combine the parallel and perpendicular components with the upstream obliquity \(\theta\) (angle between the mean magnetic field and the shock normal). Following \citet{1987ApJ...313..842J} and \citet{2006A&A...454..687M}, the diffusion coefficients are given by
\begin{equation*}
k_{1} = \frac{1}{3}\,\eta\,r_{g}\,v \left[\cos^{2}\theta+\frac{\sin^{2}\theta}{1+\eta^{2}}\right],
\end{equation*}
\begin{equation}
k_{2} = \frac{1}{3}\,\eta\,r_{g}\,v \left[\cos^{2}\theta+r^{2}\sin^{2}\theta\right]^{-3/2} \left[\cos^{2}\theta+\frac{r^{2}\sin^{2}\theta}{1+\eta^{2}}\right],
\label{eq:k1k2}
\end{equation}
where \(v\) is the particle speed, \(r\) is the shock compression ratio (taken as \(r=4\) in our calculations), and \(\eta \equiv \lambda_{\parallel}/r_{g}\) is the standard anisotropy parameter. Here, \(r_g = pc/(ZeB)\) is the gyroradius, \(B\) is the upstream magnetic field strength, and \(Z\) is the particle charge number. Within the formalism of \citet{1987ApJ...313..842J}, the parameter \(\eta\) is defined by the relation \(\lambda_{\parallel}=\eta r_g\). By adopting \(\eta=1\), we assume that the parallel mean free path reaches its minimum physically allowed value, \(\lambda_{\parallel}\sim r_g\), often associated with the Bohm limit. However, this choice constrains only the efficiency of scattering along the magnetic field and does not imply the absence of a mean field or a fully isotropic magnetic configuration. Even for \(\eta=1\), particle transport remains defined with respect to a local mean magnetic field, while perpendicular diffusion is still reduced by a factor \(\sim 1/(1+\eta^2)\). Consequently, the concept of shock obliquity remains well defined, as originally discussed by \citet{1987ApJ...313..842J}. In our simulations, we adopt \(\eta\simeq 1\), corresponding to an intermediate diffusion regime, in which Bohm diffusion should be interpreted as an effective lower limit for the diffusion coefficient, rather than as the assumption of a fully isotropic turbulent field with \(\Delta B/B\gg 1\) at all scales. We note that Eq.~(\ref{eq:k1k2}) is based on older models and does not include the full 3D structure of stellar wind magnetic fields. In reality, particles may escape along the polar regions of Parker spirals \citep{1958ApJ...128..664P}, which could reduce the maximum energy compared to the extreme values shown in Fig.~\ref{fig:n}. However, our choice of conservative diffusion efficiency and realistic shock obliquities ensures that the maximum energies considered remain physically plausible and consistent with the Hillas criterion.

Inserting \(t \simeq r_{\mathrm{sh}}/u_{\mathrm{sh}}\) and \(t_{0}=t_{\mathrm{acc}}\) into Eq.~\eqref{bierman} and converting \(p\) to energy via \(E\simeq pc\) for the high-energy regime leads to
\begin{equation}
E_{\max}^{\mathrm{soft/pow}}
\;=\;
p_{\mathrm{inj}}\,c\,
\left[
\frac{r_{\mathrm{sh}}\,u_{\mathrm{sh}}\left(1-\frac{1}{r}\right)}
{3\left(k_{1}+r\,k_{2}\right)}
\right]^{1.25},
\label{emax}
\end{equation}
where \(p_{\mathrm{inj}}\) is the injection momentum. For definiteness in our estimates, we adopt \(p_{\mathrm{inj}}\simeq 3\,m_{p}\,u_{\mathrm{sh}}\) \citep{2006A&A...454..687M}, which captures the standard condition that injected particles exceed a few times the downstream thermal momentum.

Figure~\ref{fig:n} shows Eq.~\eqref{emax} as a function of \(\cos\theta\) for representative values of \(\eta\). As \(\eta\) increases, diffusion becomes strongly anisotropic and the dependence on \(\theta\) steepens: nearly parallel configurations (\(\cos\theta\to 1\)) yield larger \(k_{1,2}\) and therefore longer \(t_{\mathrm{acc}}\), while more perpendicular geometries favor smaller effective diffusion across the shock and hence faster acceleration. In the opposite limit \(\eta\ll 1\), diffusion is nearly isotropic, the \(\theta\) dependence weakens, and the curves flatten. The intermediate case \(\eta\simeq 1\) marks the transition between these regimes and is a practical choice for cluster conditions in which diffusion is restricted but not fully aligned with the magnetic field. In this context, Bohm-like diffusion should be understood as an effective lower limit on the diffusion coefficient, while a local mean magnetic field and a well-defined shock obliquity are preserved.

The maximum energy also scales with the injection momentum and thus with rigidity \(R=pc/(Ze)\). For fixed shock and field parameters, nuclei with larger \(Z\) can reach higher energies than protons. For example, with \(u_{\mathrm{sh}}\sim 10^{-2}c\) in young compact clusters, the model admits proton maxima well below the corresponding heavy nuclei limits, while iron can approach the upper range implied by our acceleration and confinement constraints. Overall, Eq.~\eqref{emax} reproduces the expected behaviour of diffusive acceleration with finite lifetimes and escape: energy gain saturates when acceleration timescales compete with losses or with the finite duration and size of the system, and the resulting species–dependent limits follow directly from the rigidity scaling built into \(k_{1}\) and \(k_{2}\).

\subsection{Injection and propagation framework}
\label{ssec:inj_prop}

To represent the population of nuclei accelerated at cluster shocks, we follow the general framework of \citet{2023MNRAS.519..136V} and \citet{2025MNRAS.luana}, , and write the differential spectrum of a nucleus with charge \(Z\) as
\begin{equation}
\begin{cases}
\phi_{\rm soft} \;=\; A_{\rm soft}\,f_{Z}\,\dfrac{E_{\rm SN}}{(p_{0}c)^{\,2-\alpha}}\; E_{p}^{-\alpha}\;\exp\!\bigl[-E_{p}/E_{\rm max}^{\rm soft}(u)\bigr], \\[1.0em]
\phi_{\rm pow} \;=\; A_{\rm pow}\,n_{\rm sn}\,f_{Z}\,\dfrac{E_{\rm SN}}{2}\; E_{p}^{-\beta}\;\exp\!\bigl[-E_{p}/E_{\rm max}^{\rm pow}(u)\bigr],
\end{cases}
\label{eq:spectrum}
\end{equation}
where \(p_{0}\) is the reference injection momentum, \(E_{\rm SN}\) is the supernova energy (\(10^{51}\,\mathrm{erg}\)), \(\alpha\) and \(\beta\) are spectral indices, and \(A_{\rm soft}\), \(A_{\rm pow}\), \(A_{\rm wind}\) are the normalizations for each component. The  \(f_{Z}\) encodes the average acceleration efficiency for each nucleus. The proton efficiency is set as \(f_{1} \sim 0.1\), while the efficiencies for heavier nuclei are adjusted afterwards to reproduce the local cosmic-ray composition at \(1\,\mathrm{TeV}\), based on the flux ratios tabulated in \citet{2003APh....19..193H}.

\begin{table}[ht]
\centering
\renewcommand{\arraystretch}{1.3}
\setlength{\tabcolsep}{20pt}\small
\begin{tabular}{ll}
\hline\hline
Parameter  & Value \\
\hline
$E_{\max}^{\rm wind}$ & \(1\) PeV \\
$f_{\star}$           & \(0.8\) \\
$H$                   & \(3.0\) kpc \\
$V$                   & \(400\) kpc$^{3}$ \\
$N_{\rm gal}$         & \(4\times 10^{5}\) \\
$D_{0}$               & \(9 \times 10^{28}\,\mathrm{cm^{2}\,s^{-1}}\) \\
$B_{\rm soft}$        & \(5\,\mu\mathrm{G}\) \\
$B_{\rm pow}$         & \(50\,\mu\mathrm{G}\) \\
$n_{\rm sn}$          & \(0.50\) \\
$\eta$                & \(1\) \\
$p_{0}$               & \(10\,\mathrm{MeV}\,c^{-1}\) \\
$\beta$               & 2.00 \\
$f_{pow}$               & 0.11 \\
$f_{soft}$               & 0.15 \\
\hline\hline
\end{tabular}
\caption{\small \justifying Common physical parameters adopted in all acceleration models. The table lists fixed quantities used throughout the analysis, including the maximum wind energy ($E_{\mathrm{max}}^{\mathrm{wind}}$), 
stellar clustering fraction ($f_{\star}$), Galactic halo’s scale height ($H$) and volume ($V$), total number of massive stars in the Galaxy ($N_{\mathrm{gal}}$), reference diffusion coefficient ($D_{0}$), 
characteristic magnetic fields for soft and powerful clusters ($B_{\mathrm{soft}}$, $B_{\mathrm{pow}}$), 
probability that a supernova occurs within the collective wind region ($n_{\mathrm{sn}}$), 
diffusion anisotropy parameter ($\eta$), and reference injection momentum ($p_{0}$).}
\label{tab:parameterscomuns}
\end{table}

The factor \(n_{\rm sn}\) expresses the probability that a supernova launches a fast shock within the collective wind region of a cluster. Not every explosion satisfies this geometric and timing requirement. A practical estimate follows from the expected number of explosions in a cluster, \(N_{\star}\,f_{\rm sn}\), where \(N_{\star}\) is the stellar count and \(f_{\rm sn}\) is the fraction of stars that end as core collapse events. Since direct determination of \(f_{\rm sn}\) requires detailed knowledge of the high mass content, we use the initial mass function to obtain a theoretical estimate. Adopting a Chabrier initial mass function for low masses, with a slope following the Salpeter coefficient \citep{2003PASP..115..763C,1955ApJ...121..161S}, restricting the progenitors to \(8\) to \(20\,M_{\odot}\) \citep{2009MNRAS.395.1409S}, by integrating over this mass range and normalizing by the total number of stars between \(1–150M_{\odot}\), we get \(f_{\rm sn}\approx 0.043\) \citep{2025MNRAS.luana}. This result is consistent with the supernova rates reported by \citet{2021MNRAS.506.4131W}. For massive young clusters, this implies at least two explosions per system over the relevant time window. If about half of these events occur within the collective wind region, a working value \(n_{\rm sn}\simeq 0.5\) is justified.

The normalizations \(A_{\rm soft}\), \(A_{\rm pow}\), and \(A_{\rm wind}\) are set by energy balance. For the soft component, we impose a total injected energy \(f_{Z}E_{\rm SN}\). For the powerful component, we use one half of this budget, \(f_{Z}E_{\rm SN}/2\), which reflects partial transfer to energetic particles in crowded shock environments. Finally, for the wind termination contribution, we adopt a global normalization based on \(f_{Z}\,f_{\rm pow}\,f_{\star}\,N_{\rm Gal}\,P_{w}\), where \(N_{\rm Gal}\) is the Galactic number of massive stars, \(P_{w}\) is the mean mechanical wind power in young clusters, and \(f_{\star}\) is the probability that a massive star resides in a dense cluster environment \citep{2023MNRAS.519..136V,2025MNRAS.luana}. In each case, the constant is obtained by enforcing
\[
\int_{E_{\min}}^{E_{\max}} E\,\phi(E)\,dE \;=\; E_{\rm tot},
\]
with \(E_{\rm tot}\) equal to the appropriate budget.
The combined flux of Galactic cosmic rays from clusters is
\begin{equation}
\begin{cases}
\Phi_{\rm soft} \;=\; f_{\rm soft}\,f_{\star}\,\displaystyle\int \phi_{\rm soft}\,\upsilon(u)\,du, \\[1.0em]
\Phi_{\rm pow} \;=\; f_{\rm pow}\,f_{\star}\,\displaystyle\int \phi_{\rm pow}\,\upsilon(u)\,du, \\[1.0em]
\Phi_{\rm wind} \;=\; A_{\rm wind}\,f_{Z}\; E_{p}^{-2}\;\exp\!\bigl[-E_{p}/E_{\rm max}^{\rm wind}\bigr],
\end{cases}
\label{eq:fluxes}
\end{equation}
where \(f_{\rm soft}\) and \(f_{\rm pow}\) are the population fractions defined in Section~\ref{class.}. The function \(\upsilon(u)\) represents the distribution of shock speeds. We use the functional form \(\upsilon(u) \propto 1/u\) from \citet{2023MNRAS.519..136V} and \citet{2025MNRAS.luana}, which models the low probability of fast shocks (\(\sim\)2 events per millennium) and is tuned to reproduce the Auger spectrum around 1 EeV \citep{2021EPJC...81..966A}.

Propagation in the Galactic disk yields the observed spectrum
\begin{equation}
J(E,Z) \;=\; \left(\frac{c}{4\pi\,D_{\rm gal}}\right)\left(\frac{H^{2}}{V}\right)_{\rm disk}\,\bigl[\Phi_{\rm soft}+\Phi_{\rm pow}+\Phi_{\rm wind}\bigr],
\label{eq:totalflux}
\end{equation}
where H and V are the Galactic halo’s scale height and volume,
respectively. The diffusion coefficient is written as $D_{\rm{gal}} = D_0 \; (\mathcal{R}_p/\mathcal{R}_0)^s$, where $s$ is the scaling index, $D_0$ is the reference diffusion coefficient, $\mathcal{R}_p = E_p/Z$ is the rigidity of the particles, and $\mathcal{R}_0 \sim 1$ GV is the reference rigidity corresponding to $D_0 \sim 9 \times 10^{28} \rm{cm^2/s}$ \citep{2007ARNPS..57..285S,2011ApJ...729..106T,2016ApJ...824...16J,2024A&A...692A..20R}. 

We consider four models that differ in the relative roles of the soft and powerful cluster components while sharing a wind contribution with fixed maximum energy $E_{\max}^{\rm wind} \simeq 1.0\,Z~\mathrm{PeV}$ \citep{2021MNRAS.504.6096M, 2024arXiv240316650M, 2016A&A...591A..71M, 2024NatAs...8..530P}. This limiting value is set by particle confinement at the wind termination shock, where acceleration can proceed as long as the diffusion length of the particles remains smaller than the characteristic size of the shock. In the Bohm diffusion limit, this condition can be expressed in terms of the Larmor radius, since the two quantities are directly related in that regime. An equivalent constraint arises from adiabatic losses upstream in the stellar wind, where particles diffusing inward subsequently lose energy when advected outward. Strong magnetic turbulence and the high speed of stellar winds enhance scattering and confinement, allowing particles to remain efficiently trapped and accelerated up to PeV energies before diffusive escape from the shock region becomes dominant \citep{2021MNRAS.504.6096M,2024arXiv240316650M,2016A&A...591A..71M,2024NatAs...8..530P}. Model A assumes strictly parallel shocks in all components. Model B includes oblique shocks in both the soft and powerful components, as developed in Section~\ref{emaxoblique}.
Model C follows Model B but with a larger initial power at the wind termination shock.
Model D applies oblique shocks only in the powerful component and keeps parallel shocks in the soft component, thereby combining features of Models A and B.

\begin{figure*}[!htb]
   \centering
   \subfloat[Model A]{\includegraphics[width=0.53\textwidth]{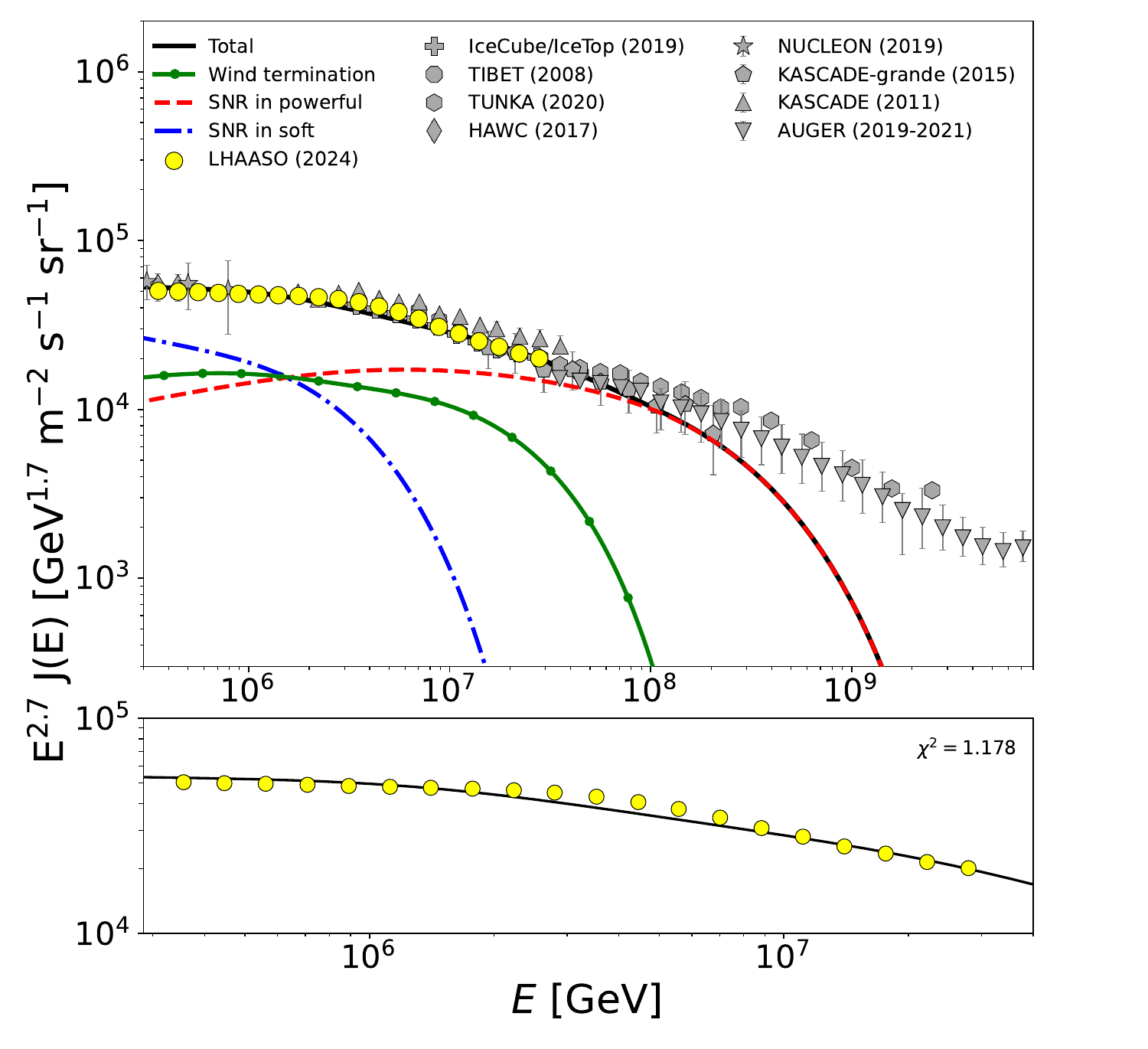}}
   \subfloat[Model B]{\includegraphics[width=0.50\textwidth]{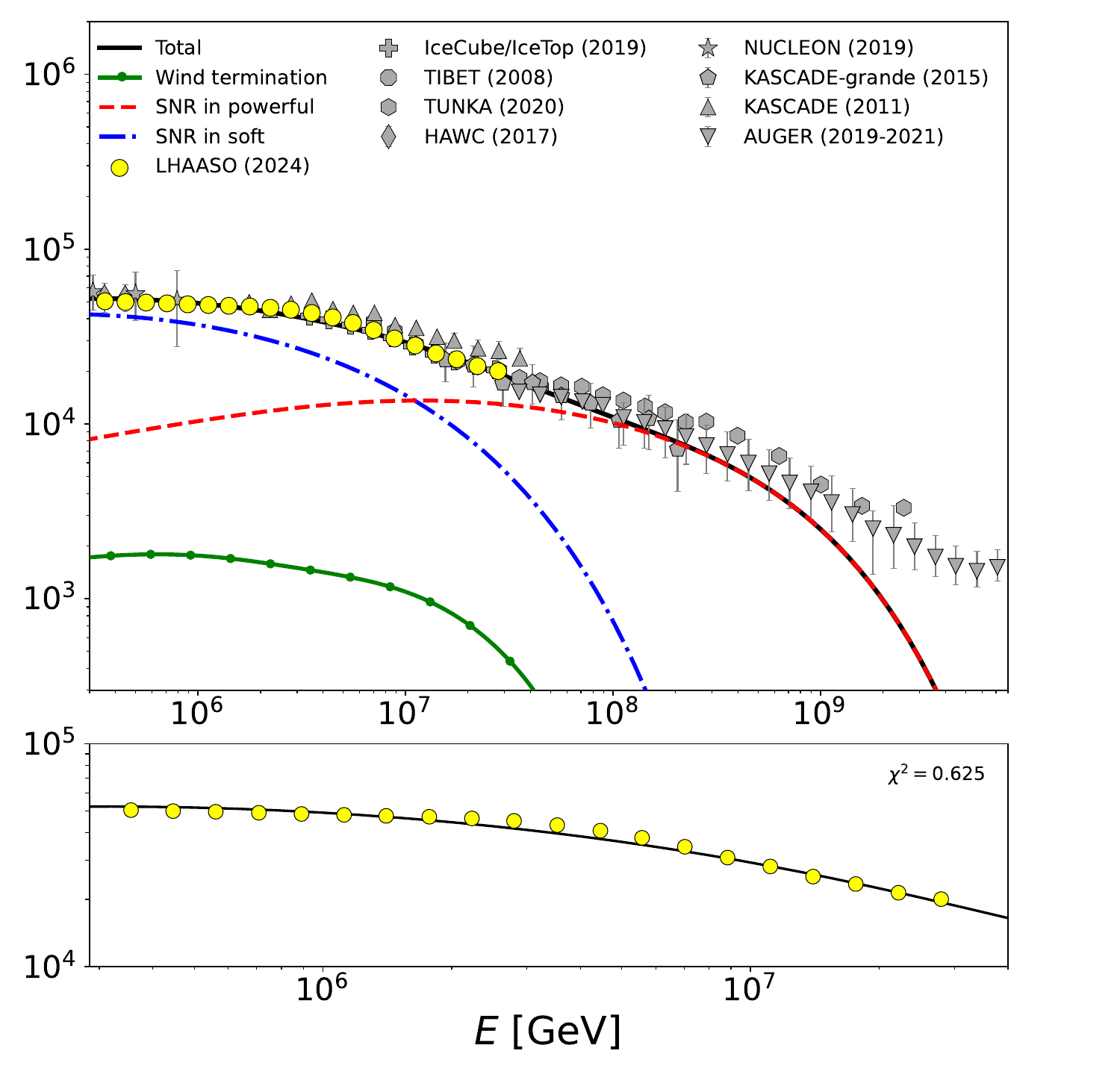}}\hfill
   \subfloat[Model C]{\includegraphics[width=0.53\textwidth]{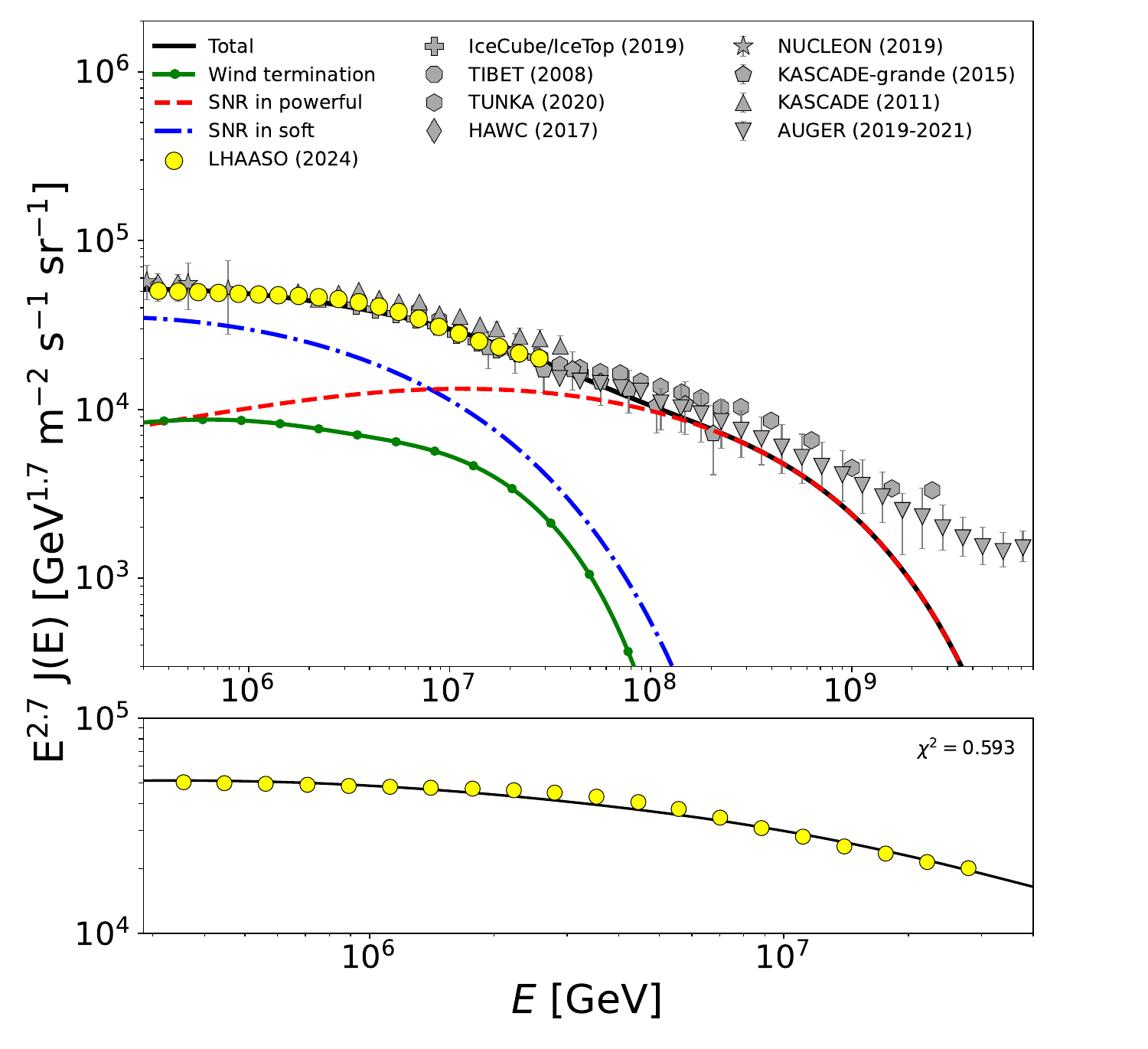}}
   \subfloat[Model D]{\includegraphics[width=0.50\textwidth]{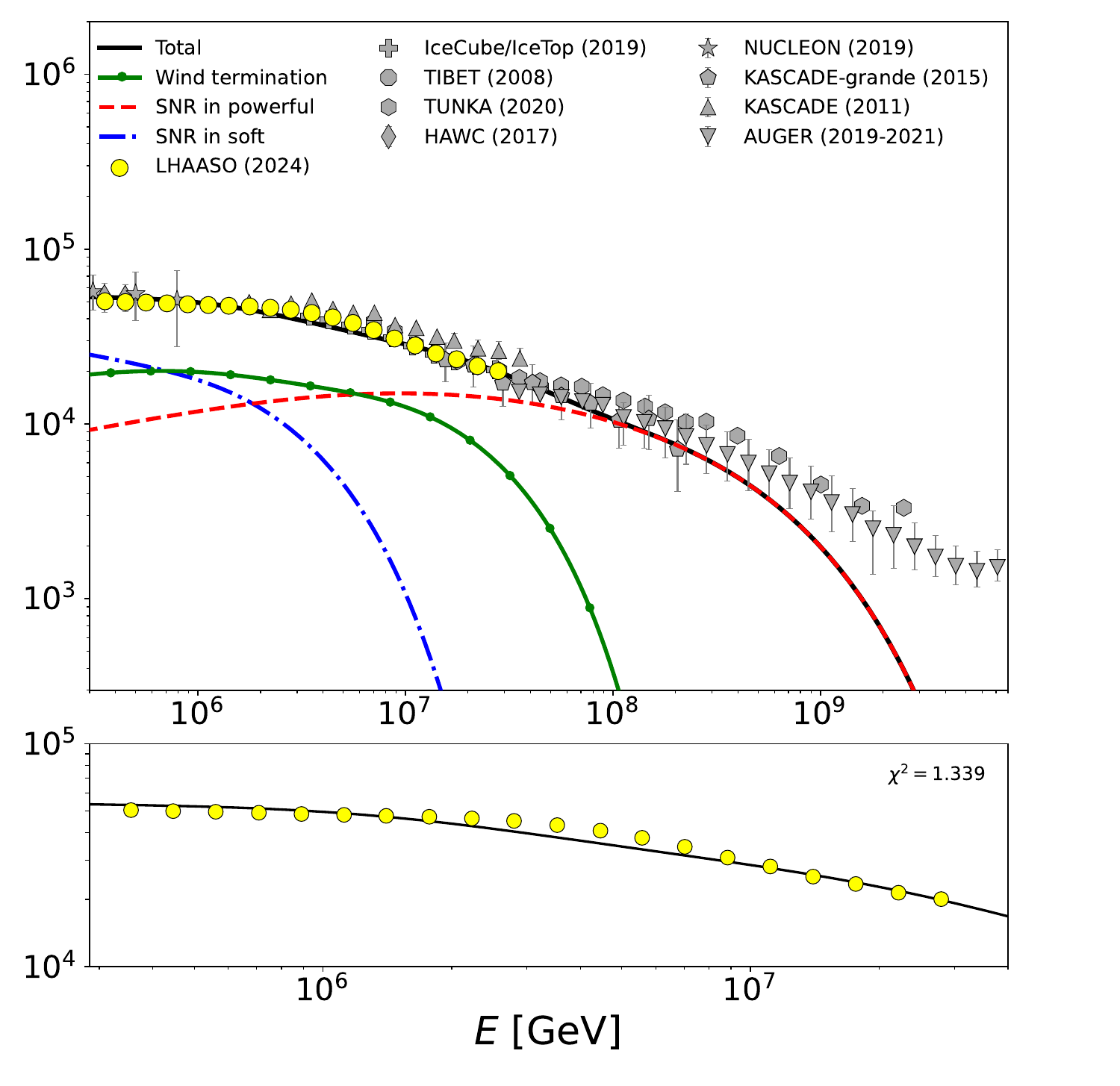}}\hfill
   \caption{\small \justifying
   All particle cosmic ray spectra for Models A, B, C, and D. Each panel displays the predicted flux \(J(E)\) for nuclei up to \(Z=40\) (solid curve) and the corresponding measurements from LHAASO \citep{2024PhRvL.132m1002C}, the Pierre Auger Observatory \citep{2019ICRC...36..450V,2021EPJC...81..966A}, KASCADE Grande \citep{2015ICRC...34..359B}, IceCube/IceTop \citep{2019PhRvD.100h2002A}, Tibet AS\(\gamma\) \citep{2019PhRvD.100h2002A}, Tunka \citep{2020APh...11702406B}, HAWC \citep{2017PhRvD..96l2001A}, NUCLEON \citep{2017PhRvD..96l2001A}, and KASCADE \citep{The_CR_spectrum}. Model definitions and parameters are given in Secs.~\ref{ssec:inj_prop}–\ref{emaxoblique} and Tables~\ref{tab:parameterscomuns}-\ref{tab:angle}.}
   \label{fig:spectrum}
\end{figure*}

\begin{table}[ht]
\centering
\renewcommand{\arraystretch}{1.2}
\setlength{\tabcolsep}{10pt}
\begin{tabular}{l cc cc c c c c }
\hline\hline
 Model & $\alpha$ & $s$ & $r_{\rm sh}$ & $P_{w}\times 10^{35}$ \\
 &     &          &     pc            & erg s$^{-1}$ \\
\hline
A & 2.0650 & 0.4502 & \textit{n.a.} & 6.6448 \\
B & 2.0333 & 0.4787 & 0.7612 & 1.0150 \\
C & 2.0467 & 0.4805 & 0.7564 & 5.0186 \\
D & 2.0562 & 0.4672 & 0.5953 & 9.9422 \\
\hline\hline
\end{tabular}
\caption{\small \justifying The table lists the spectral index ($\alpha$), the Galactic diffusion coefficient scaling ($s$), the shock radius ($r_{\rm sh}$), and the stellar wind power ($P_{w}$) for each model. These quantities characterize the efficiency of particle acceleration.}
\label{tab:parameters}
\end{table}

In Model A, less energetic star clusters without collective winds can accelerate particles up to a maximum energy of \(E_{\max}^{\rm soft}\simeq 0.1\,Z\,B\,{E_SN}^{0.33}\,\rho^{-0.33}\,u_{5}^{0.33}\,\mathrm{PeV}\)\footnote{Obtained from the Sedov–Taylor solutions \citet{2009MNRAS.396.2065C} applied
to the maximum energy from \citet{1983A&A...125..249L}.} \citep{1983A&A...125..249L,2005JPhG...31R..95H}. In this expression, \(B\) is the magnetic field, \(E_{SN}\) is the typical energy released by a core-collapse supernova, \(\rho\) is the ambient density, \(u_{5}\) denotes the shock speed in units of \(5000~\mathrm{km\,s^{-1}}\), and \(Z\) is the atomic number of the particle. Using typical environmental values for clusters in the Galaxy (\(B\sim 2 \mu G\) and \(\rho\sim 0.01 \rm cm^2\)), and assuming that small variations of these parameters have a negligible effect, the maximum energy can be approximated as \(E_{\max}^{\rm soft}\simeq 0.1\,Z\,u_{5}^{0.33}\,\mathrm{PeV}\). In the most powerful clusters, young, compact systems with strong collective winds, the maximum energy can reach  \(E_{\mathrm{pow}}^{\mathrm{max}} \sim 4\,\mathrm{Z}\,\upsilon_5 \; f_c (R_c,N_{*},n_c,\eta_T)\) PeV~\citep{2023MNRAS.519..136V,2025MNRAS.luana}. The function \(f_c\) combines several cluster properties, such as the cluster radius (\(R_c\)), , the number of massive stars (\(N_{*}\)), the fraction of stellar mechanical power injected, and the level of MHD turbulence (\(\eta_T\)). Using typical values for the parameters in \(f_c\) and assuming \(\eta_T \sim 10\%\), the expression can be approximated as \(E_{\mathrm{pow}}^{\mathrm{max}} \sim 4\,\mathrm{Z}\,\upsilon_5 \). Particle acceleration follows Bohm diffusion, sustained by turbulence and magnetic amplification.

In Model B we compute \(E_{\max}^{\rm soft}\) and \(E_{\max}^{\rm pow}\) from Eq.~\eqref{emax} for obliquity angles from \(0^{\circ}\) to \(90^{\circ}\).
Model C adopts the same procedure as Model B but increases the initial power at the wind termination region.
Model D keeps the soft component at the parallel limit of Model A and applies Eq.~\eqref{emax} only to the powerful component.

\begin{table}[ht]
\centering
\renewcommand{\arraystretch}{1.3}
\setlength{\tabcolsep}{5pt}
\begin{tabular}{l ccc}
\hline\hline
\textbf{Model} &  & \textbf{Angle range (\(\theta\))} & \\
  &\(0^\circ\)&\(\in [10^\circ, 50^\circ]\)&\(\in [60^\circ, 90^\circ]\)\\
\hline
\multicolumn{4}{c}{ \qquad \qquad \textbf{Soft cluster}} \\
B and C   & 20\% & 80\% & 0\% \\
\hline
\multicolumn{4}{c}{ \qquad \qquad \textbf{Powerful cluster}} \\
B, C and D & 4\% & 26\%  & 70\% \\
\hline\hline
\end{tabular}

\caption{\small \justifying Fractional contribution of each shock obliquity angle range to the soft and powerful components considered in the acceleration models.}
\label{tab:angle}
\end{table}

We aim to reproduce the recent LHAASO measurements \citep{2024PhRvL.132m1002C}. Consequently, we treat \(P_{w}\), \(s\), \(r_{\rm sh}\), and \(\alpha\) as free parameters, together with the fractional weights assigned to each shock angle interval, were defined based on the physical properties of each environment and later used as free parameters in the fit. For soft clusters we impose a preference for angles up to \(50^{\circ}\) over larger angles, motivated by the expectation of lower turbulence where collective winds are weaker and highly oblique shocks are less frequent.
For powerful clusters, where winds are stronger and turbulence is higher, we favor larger angles and constrain the fitted fractions accordingly.
The recovered angular fractions are summarized in Table~\ref{tab:angle}.
Table~\ref{tab:parameters} shows the parameters adjusted for each model, and Table~\ref{tab:parameterscomuns} lists the parameters that are common to all cases considered.

We assume an average magnetic field of \(100\,\mu\)G for model A, following \citet{2023MNRAS.519..136V}, while for the other models we use \(50\,\mu\)G for powerful clusters, as discussed earlier. This choice is based on the confinement of kinetic energy injected by stellar winds from O and Wolf-Rayet (WR) stars within the cluster, which favors its conversion into magnetic energy \citep{2022MNRAS.515.2256V}. This scenario is supported by 3D MHD simulations of the inner core of these systems, which show efficient magnetic field amplification through the Axford-Cranfill effect, reaching values on the order of several hundred $\mu G$ in filamentary structures \citep{2022MNRAS.517.2818B}.

In Models B, C, and D, the shock radius $r_{\rm sh}$ is treated as a dynamic fit parameter in the range $0.5-1.0$ pc. This ensures that the SNR shock has expanded beyond the progenitor wind cavity (typically $\sim 0.2$ pc)~\citep{2022MNRAS.517.2818B, 2022MNRAS.512.1275V} and is interacting with the subsonic, turbulent intracluster medium dominated by the collective cluster wind. This regime is physically distinct from acceleration within the individual progenitor wind bubble. The upper limit of $1.0$ pc is motivated by the typical sizes of young massive clusters in our catalog. Beyond this scale, shock deceleration becomes significant and the upstream magnetic field weakens substantially, which would reduce the maximum particle energy. Therefore, our adopted range captures the phase where PeV acceleration is most efficient.

Figure~\ref{fig:spectrum} shows all particle spectra for Models A through D, with the individual contributions from soft clusters, powerful clusters, and the wind termination region (nuclei up to \(Z=40\)). In every case, the knee emerges from a rigidity dependent cutoff of the light groups, together with a transfer of dominance from the soft to the powerful cluster component, while the wind term primarily sets the normalization below the knee and remains subdominant at multi PeV energies. Model A, which assumes only parallel shocks, reproduces the broad trend through the knee but relies on a relatively strong wind input to match the sub PeV flux. Its high energy tail is limited by the ceiling on \(E_{\max}\) for parallel geometry, so the powerful cluster term carries most of the flux at and above the knee, as described in \citet{2023MNRAS.519..136V} and \citet{2025MNRAS.luana}. Model B, which includes oblique shocks in both cluster classes, gives the closest overall agreement with the LHAASO data \citep{2024PhRvL.132m1002C} from \(10^{6}\) to \(10^{8}\,\mathrm{GeV}\). Allowing a distribution of angles increases \(E_{\max}\) for a sizable fraction of shocks and naturally softens the transition through the knee; the preferred solution uses a compact acceleration scale, a modest wind power, and a transport index near the canonical value. Model C follows the same geometry as Model B but starts from a larger wind power; this raises the flux below the knee and slightly hardens the approach to the knee, yet it does not improve the global match, indicating that extra wind luminosity cannot replace the gains produced by obliquity. Model D applies obliquity only in the powerful component and keeps the soft component at the parallel limit; its performance is intermediate, with the knee still governed by the powerful clusters and with reduced flexibility below the knee that the fit compensates by increasing the wind power and the characteristic shock radius.

The parameter distributions shown in the corner plot in Appendix~A (Figure~\ref{fig:corners}) represent a local exploration of the parameter space around the best-fit solution, weighted by the variation in $\chi^2$. The fit was performed primarily in the energy range $10^6$--$10^7$ GeV (1--10 PeV), corresponding to the knee region where our Galactic cluster model is expected to dominate. We used the full LHAASO dataset \citep{2024PhRvL.132m1002C}, together with a limited number of additional data points from KASCADE-Grande \citep{2015ICRC...34..359B} and the Pierre Auger Observatory \citep{2019ICRC...36..450V,2021EPJC...81..966A} extending up to $\sim 10^8$ GeV (100 PeV). These higher-energy points help constrain the high-energy cutoff behavior and guide the spectral shape in the transition region to the extragalactic component, but they contribute little to the $\chi^2$ calculation. We explicitly avoided fitting the ultra-high-energy regime beyond $\sim 100$ PeV, where extragalactic contributions become dominant and our purely Galactic model is not expected to apply. In Figure~\ref{fig:spectrum}, the lower panel of each model shows a zoom into the LHAASO energy range ($10^6$--$10^7$ GeV). The $\chi^2$ value displayed in the upper-right corner corresponds to the best fit obtained using only the LHAASO data in this range, allowing a direct assessment of the fit quality in the knee region where our model is most applicable. This approach ensures that the statistical evaluation reflects the model's performance in reproducing the knee structure itself, rather than forcing agreement at ultra-high energies where the Galactic-to-extragalactic transition occurs. These distributions are consistent across model variants and suggest a common physical picture. 

The source index is tightly concentrated around \(\alpha\simeq 2.0\), and the transport index clusters near \(s\simeq 0.45\) to \(0.46\). The spectral slope of powerful clusters, $\beta = 2.0$, is fixed in all models and arises naturally from DSA at strong shocks in turbulent magnetized environments, producing hard spectra consistent with standard DSA predictions (e.g., \cite{2018A&A...611A..77Y},\cite{2021NatAs...5..465A},\cite{2022A&A...666A.124A}). When the shock radius is part of the model, the fits prefer \(r_{\rm sh}\) of order \(0.5-0.8\,\mathrm{pc}\). The favored wind power depends on the shock geometry prescription: it is lowest when oblique shocks operate in both cluster classes (Model B) and highest when only the powerful clusters include obliquity (Model D). The joint contours show trends that are physically reasonable: a mild positive correlation between \(P_{w}\) and \(r_{\rm sh}\), and a weak anti-correlation between \(P_{w}\) and \(s\), reflecting compensation between source power, acceleration scale, and transport. In the configuration that omits \(r_{\rm sh}\) (Model A), the proximity factor for supernovae remains centered near one half with limited spread, in line with the prior construction of the cluster population. 

Although Models A, B, and C provide similarly good statistical fits to the LHAASO data, with comparable $\chi^2$ values, the underlying physical scenarios differ in important ways, as shown in Fig.~\ref{fig:spectrum}. In Models B and C, the maximum proton energies reach about $\sim 6$ PeV in powerful clusters and $\sim 0.2$ PeV in soft clusters. These values, which are higher than in Model A, arise naturally from the shock properties and from the dependence of the Larmor radius on the ambient magnetic field (Section~\ref{oblique}). Unlike simpler approaches that assume relations such as $E_{\rm max} \propto Z$, we adopt more conservative and physically motivated parameters. Including shock obliquity also changes the physical conditions favored by the model. In Models B and C, the required magnetic field strengths are lower (mean field of $50\,\mu$G) than in Model A ($100\,\mu$G), while larger shock radii ($r_{\rm sh} = 0.5-1.0$ pc) are considered to ensure that the shock has expanded beyond the progenitor star's wind bubble into the region where the collective stellar wind interacts with the supernova shock. This combination is still consistent with reaching high maximum energies without requiring extremely strong magnetic fields, pointing to solutions that rely less on extreme parameter values. The role of the collective stellar wind also varies between the models. In Model B, it makes a relevant contribution, especially at lower energies. In Model C, its contribution is smaller, and the spectrum becomes dominated by the other components. Even so, the comparison between Models B and C shows that, despite different wind contributions, the total spectrum changes only slightly (Fig.~\ref{fig:spectrum}). This indicates low sensitivity to moderate variations in stellar wind power, consistent with the wide range of stellar types in massive clusters. Therefore, clusters with typical wind powers in the range of $\sim (1-6) \times 10^{35}$ erg s$^{-1}$ tend to produce similar contributions to the total spectrum.

In addition, Model D serves as a physical consistency test for the acceleration scenario. By describing the soft clusters with a purely parallel shock regime, without obliquity, the model introduces a fundamental difference in the acceleration mechanism between soft and powerful environments. However, this hybrid approach results in the worst statistical fit among the models considered (Fig.~\ref{fig:corners}). This suggests that the physics of particle acceleration may be more uniform across different types of clusters than previously assumed in Model D, and that including shock obliquity, even in less energetic environments, may be important for reproducing the observed spectrum shape.

Overall, the results indicate that the models based on the scenario proposed in this work, especially Models B and C, provide the best descriptions of the observed spectrum. Although they show very similar statistical performance in the LHAASO energy range, Model B shows slightly better agreement when the broader energy range (\(\sim 10^{5}-10^{8}\) GeV) is considered.

\subsection{Composition analysis}
\label{ssec:composition}

The knee of the all–particle spectrum has been probed by several observatories, including KASCADE \citep{2005APh....24....1A}, ARGO–YBJ \citep{2016NPPP..279....7M}, CASA–MIA \citep{1999APh....10..291G}, LHAASO \citep{2024PhRvL.132m1002C}, and IceCube/IceTop \citep{2019PhRvD.100h2002A,2020PhRvD.102l2001A}. The physical origin and elemental makeup remain under discussion. A long–standing view links the knee to source acceleration and Galactic transport with breaks that scale with particle rigidity \citep{2001AdSpR..27..803E,1961NCim...22..800P}. Alternative interpretations invoke a mass–dependent cutoff, in which the limiting energy for each element grows in proportion to nuclear mass \citep{2004APh....21..241H}.

\begin{figure*}[!htb]
   \centering
   \subfloat[Model B]{\includegraphics[width=0.475\textwidth]{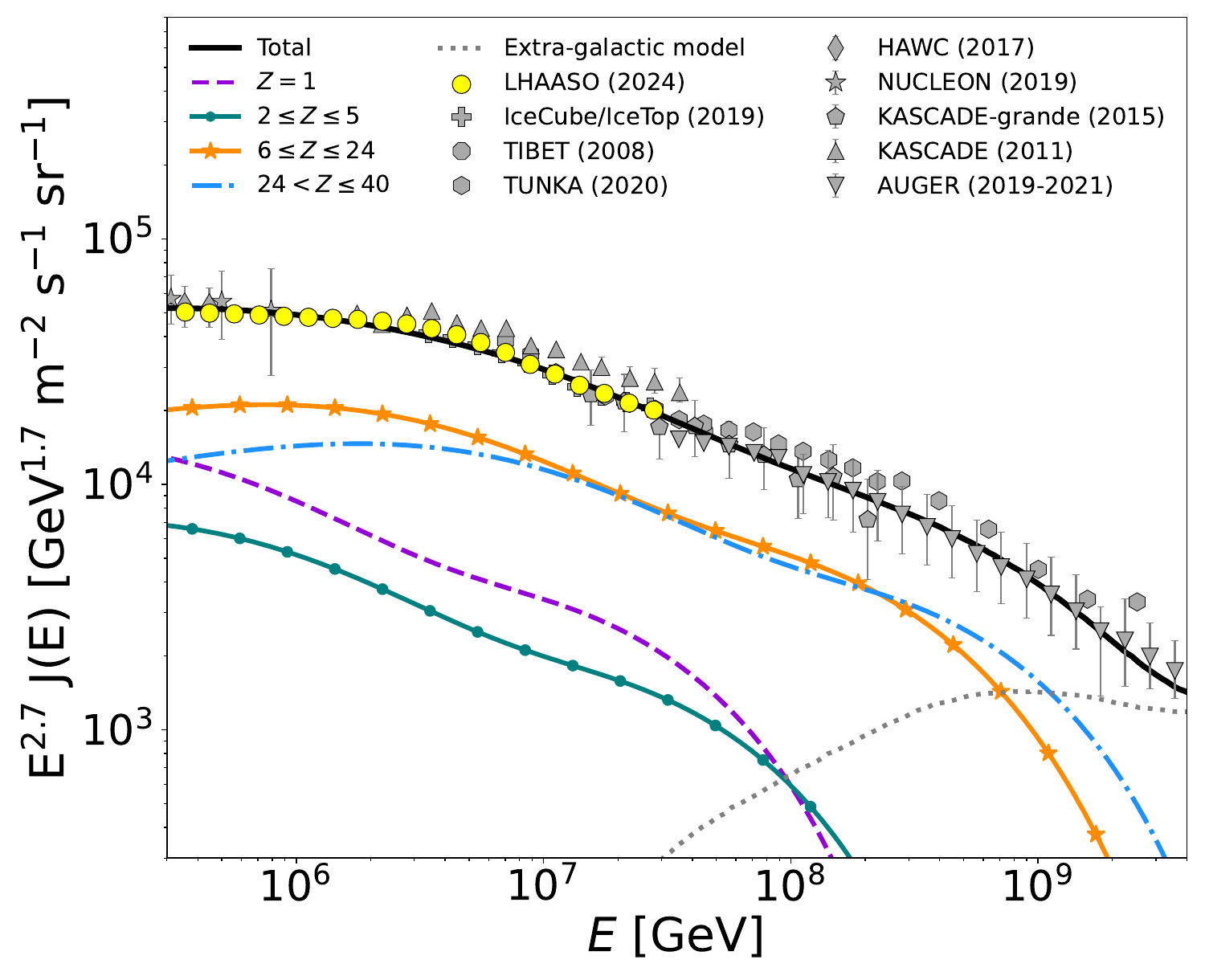}}\hfill
    \subfloat[Model C]{\includegraphics[width=0.458\textwidth]{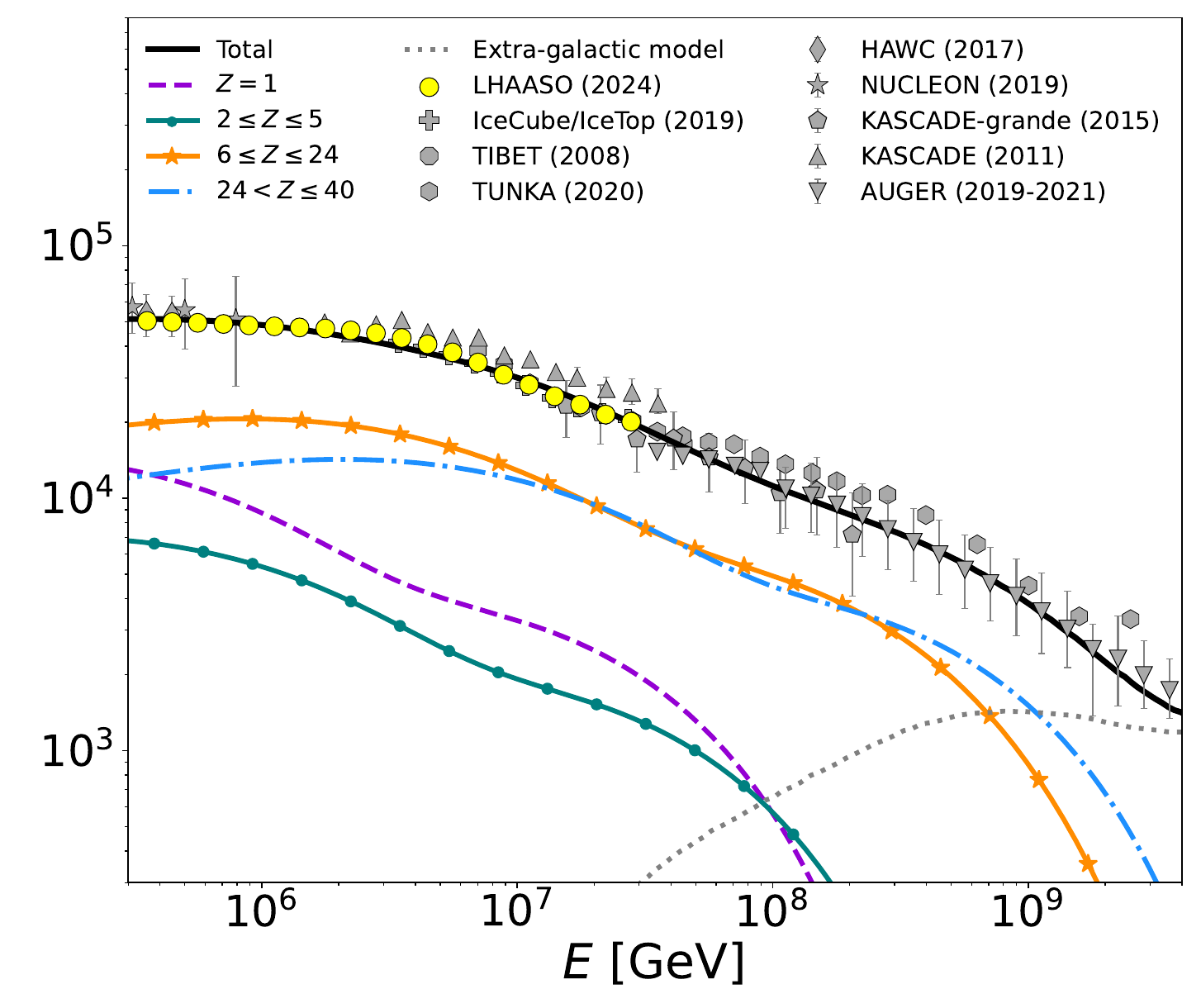}} \hfill
    \caption{\small \justifying
    Spectral composition for Models B and C. Each panel shows the energy flux for the total all–particle together with the contributions from four charge groups: protons \((Z=1)\), light nuclei \((2\le Z\le 5)\), intermediate nuclei \((6\le Z\le 24)\), and heavy nuclei \((24<Z\le 40)\). Solid curves give the model prediction for each group and their sum. Extragalactic component (gray dotted line) from \citet{2016A&A...595A..33T} (EG-RSB93): pure protons, \(E^{-2}\) injection spectrum, exponential cutoff at \(10^{11}\;\mathrm{GeV}\). Data are from LHAASO \citep{2024PhRvL.132m1002C}, the Pierre Auger Observatory \citep{2019ICRC...36..450V,2021EPJC...81..966A}, KASCADE Grande \citep{2015ICRC...34..359B}, IceCube/IceTop \citep{2019PhRvD.100h2002A}, Tibet AS\(\gamma\) \citep{2019PhRvD.100h2002A}, Tunka \citep{2020APh...11702406B}, HAWC \citep{2017PhRvD..96l2001A}, NUCLEON \citep{2017PhRvD..96l2001A}, and KASCADE \citep{The_CR_spectrum}. Model definitions and parameters are given in Secs.~\ref{ssec:inj_prop}–\ref{emaxoblique} and Tables~\ref{tab:parameterscomuns}-\ref{tab:angle}.}
    \label{fig:composition}
\end{figure*}

\begin{figure}[ht!]
\raggedright
\includegraphics[width=1.02\linewidth]{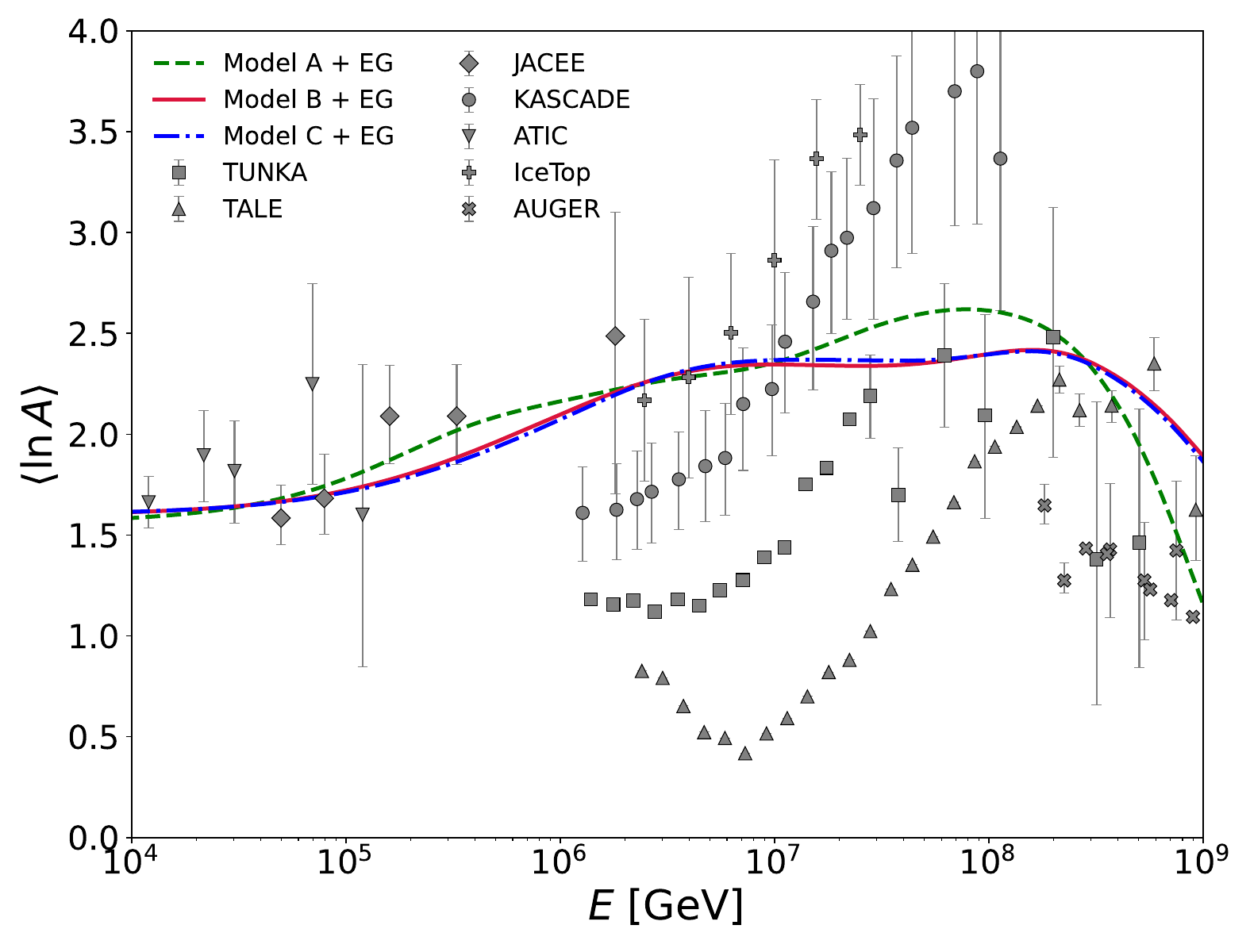}
\caption{\small \justifying Cosmic-ray composition in terms of the mean logarithmic atomic mass ⟨ln A⟩, derived from Models A, B and C and compared with data from AUGER \citep{2019ICRC...36..482Y,2023EPJWC.28302010P}, TALE \citep{2021ApJ...909..178A}, TUNKA \citep{2022icrc.confE.731B,2018PhRvD..97l2004B}, IceTop \citep{2013APh....42...15I}, KASCADE \citep{2013ICRC...33.1538S}, JACEE \citep{takahashi1998elemental} and ATIC \citep{2009BRASP..73..564P}.}
\label{fig:lnA}
\end{figure}

Recent LHAASO results sharpen this picture \citep{2024PhRvL.132m1002C}. \citet{2024arXiv241113793H} introduced an analysis based on the total mass–log energy spectrum and found that protons dominate the formation of the knee, with a characteristic cutoff near \(3.2\;\mathrm{PeV}\). The same analysis reveals an excess of iron around \(9.7\;\mathrm{PeV}\). Taken together, these features disfavor a mass–scaled cutoff and instead support a rigidity–ordered sequence of elemental knees. With the spectral indices constrained by LHAASO, the fit further indicates a proton fraction of at least \(52.7\%\pm1.2\%\) at the knee. In what follows, we adopt this rigidity–dependent interpretation to anchor the composition in our injection and propagation framework.

\begin{table}[ht]
\centering
\renewcommand{\arraystretch}{1.3}
\setlength{\tabcolsep}{9pt}
\begin{tabular}{l cc}
\hline
\hline
    & \textbf{Model}  \\
  &  \textbf{B} & \textbf{C}  \\
\hline
$Z = 1$ \% & 34.4\% & 34.3\% \\
$2 \leq Z \leq 5$  & 14.8\% & 14.8\% \\
$6 \leq Z \leq 24$ & 32.1\% & 32.2\% \\
$24 < Z \leq 40$  & 18.7\% & 18.8\%  \\
\hline
\hline
\end{tabular}
\caption{\small \justifying
Fractional contribution (\%) of each charge group to the total all-particle cosmic-ray flux for models. }
\label{tab:composition}
\end{table}

Figure~\ref{fig:composition} presents the spectra for Models B and C together with the composition fractions reported in Table~\ref{tab:composition}. The table shows that the global elemental budget over the modeled range is stable across all variants, with protons contributing about \(34\%\), light nuclei with \(2\le Z\le 5\) near \(14\%\), the intermediate group with \(6\le Z\le 24\) around \(32\%\), and the heavy group with \(24<Z\le 40\) close to \(18\%\). These values summarize the full energy interval; at a few PeV, the proton share increases, while at higher energies, the intermediate and heavy groups become more prominent through rigidity ordering.

The panel~\ref{fig:composition} for Model B, where oblique shocks operate in both soft and powerful clusters, the increase in maximum energy for a significant fraction of shocks produces a smoother transition across the knee. In addition, the extended contribution of heavy nuclei at high energies can produce a mild ``shoulder'' in the heavy component near \(\sim 10\;\mathrm{PeV}\), qualitatively consistent with a trend toward a heavier composition in this region. In Model C, the stronger wind slightly raises the normalization below \(1\,\mathrm{PeV}\), mainly for the light groups, making the transition through the knee smoother. The relative weights of the charge groups change little, and the sequence of cutoffs that follow rigidity is preserved. 

\begin{table*}[ht]
\centering
\renewcommand{\arraystretch}{1.3}
\setlength{\tabcolsep}{9pt}
\begin{tabular}{l cccc}
\hline
\hline
&  & \textbf{MODEL B} & \\ 
\hline
\textbf{Energy range [PeV]}  & $Z = 1$ & $2 \leq Z \leq 5$ & $6 \leq Z \leq 24$  & $24 < Z \leq 40$ \\
\hline
\;\;\;\;\;\;\;\;0.359 - 27.0 & 20.4\% & 11.6\% & 41.0\% & 27.0\% \\
\;\;\;\;\;\;\;\;0.359 - 1.0  & 21.4\% & 12.0\% & 40.5\% & 26.1\% \\
\;\;\;\;\;\;\;\;1.0 - 10.0 & 15.0\% & 9.3\% & 43.7\% & 32.0\% \\
\;\;\;\;\;\;\;\;10.0 - 27.0 & 11.6\% & 6.9\% & 41.7\% & 39.8\% \\
\hline
&  & \textbf{MODEL C} & \\ 
\hline
\;\;\;\;\;\;\;\;0.359 - 27.0 & 21.1\% & 11.9\% & 40.6\% & 26.4\% \\
\;\;\;\;\;\;\;\;0.359 - 1.0  & 22.2\% & 12.4\% & 40.0\% & 25.5\% \\
\;\;\;\;\;\;\;\;1.0 - 10.0 & 15.1\% & 9.7\% & 43.8\% & 31.3\% \\
\;\;\;\;\;\;\;\;10.0 - 27.0 & 11.0\% & 6.5\% & 42.0\% & 40.5\% \\
\hline
\hline
\end{tabular}
\caption{\small \justifying The composition is shown across the LHAASO \citep{2024PhRvL.132m1002C} energy range and within its sub-intervals.}
\label{tab:composition_lhaaso}
\end{table*}

The results in Table \ref{tab:composition_lhaaso} show a clear trend towards a heavier composition with increasing energy in both Models B and C. This behavior arises naturally from rigidity-dependent maximum energies. The proton fraction decreases from about \(21–22\%\) below \(1\;\mathrm{PeV}\) to around \(11–12\%\) above \(10\;\mathrm{PeV}\), indicating that protons soften first near the knee. The light intermediate group (\(2 \le  Z \le  5\)) shows something similar, dropping from about \(12\%\) to \(\sim 6–7\%\) across the same range. The intermediate-mass group (\(6 \le Z \le 24\)) remains the dominant contributor over most of the range, staying close to \(40–44\%\). This stability suggests it plays a key role in shaping the smooth spectral transition around the knee. In contrast, the heavy group (\(24 < Z \le 40\)) increases significantly with energy, rising from about \(25–27\%\) below \(1\;\mathrm{PeV}\) to nearly \(40\%\) above \(10\; \mathrm{PeV}\). Differences between Models B and C are small, at the few percent level, indicating that the overall compositional evolution is robust and not strongly sensitive to specific model details. Overall, the table supports a picture in which the knee arises from successive rigidity-dependent cutoffs, producing a gradual transition from light to heavy composition across the PeV range.

Figure~\ref{fig:composition} supports a knee that follows rigidity. The proton channel breaks near \(3\,\mathrm{PeV}\); helium turns down at slightly higher energy; intermediate nuclei sustain the spectrum above the knee; and the heaviest group adds a shoulder close to \(10\,\mathrm{PeV}\). This pattern agrees with recent LHAASO results, which provide high precision spectra for protons and helium at the knee and TeV to PeV spectra of Galactic gamma rays~\citep{2024PhRvL.132m1002C}. The combined data link local hadronic spectra to large scale gamma ray emission and point to a composition that is ordered by rigidity. In particular, the mean logarithmic mass at the knee is very close to the value for pure helium and is much larger than for pure proton, showing that the flux is dominated by light elements; a proton plus helium fraction of \(65\%\) to \(95\%\) at the knee confirms that this feature marks the cutoff of the light, high rigidity components \citep{2024PhRvL.132m1002C,2024PhRvD.110d3030T,PhysRevD.106.123028}. This behaviour is also the expected outcome of acceleration in non-parallel shocks, where the maximum energy rises with obliquity and elemental knees appear in a rigidity ordered sequence \citep{2002PhRvD..66h3004K}.

Figure~\ref{fig:lnA} shows the cosmic-ray composition in terms of the mean logarithmic atomic mass, $\left\langle \ln A \right\rangle$, for Models A, B, and C, including the extragalactic contribution \citep{2016A&A...595A..33T}. At low energies, all models predict a light composition, in good agreement with experimental data. As the energy increases, $\left\langle \ln A \right\rangle$ rises gradually, reflecting a growing contribution from heavier nuclei, as expected in rigidity-limited Galactic scenarios. At the highest energies, the inclusion of a proton-dominated extragalactic component 
produces a decrease in $\left\langle \ln A \right\rangle$, signaling a transition toward a lighter composition. Although Model A shows slightly better agreement with the data compared to Models B and C, the purpose of this comparison is not to invalidate Model A, but rather to demonstrate that Models B and C are physically viable and reproduce the observed composition consistently. We note that Model A is already theoretically disfavoured by the arguments discussed in Section~\ref{ssec:inj_prop}, so the comparison presented here is intended to highlight the observational viability of the proposed oblique shock alternatives. The results show good agreement with observations up to approximately 100~PeV. Above this energy, however, the predicted composition becomes heavier than indicated by Auger data. This discrepancy likely reflects an overestimate of the maximum energy reached in the oblique-shock framework, which shifts the Galactic-to-extragalactic transition to somewhat higher energies than observed. We treat this as a current limitation of the model and a direction for future refinement.

\section{High-Energy Emission from cluster populations}
\label{sec:gama_neutrino}

Gamma rays in MSC arise from leptonic and hadronic channels. The main processes are neutral pion decay, bremsstrahlung and inverse Compton scattering~\citep{2019A&A...623A..86A,2022hxga.book...52V}. Their relative importance depends on the local conditions. Shock-accelerated ions may interact with the surrounding gas, producing neutral pions that decay into gamma rays. This process can, in principle, lead to a characteristic spectral feature around \(67\,\mathrm{MeV}\), often referred to as the pion bump, which is considered a theoretical signature of hadronic interactions~\citep{1998ApJ...492..219G}. However, the efficiency of this mechanism in stellar cluster environments remains uncertain, since strong stellar winds can reduce the ambient gas density, and in many observed systems, leptonic emission scenarios also provide viable explanations for the gamma-ray emission. In the following, we focus on the hadronic channel to compute the gamma-ray emission from the cluster population and the corresponding neutrino flux.

For the gamma-ray source term we adopt the parametrization of the differential cross section \(d\sigma/dE_{\gamma}\) given by \citet{2014PhRvD..90l3014K}, which is calibrated to accelerator data and high energy interaction models, including SIBYLL \citep{1992PhRvD..46.5013E,2009PhRvD..80i4003A}. For each cluster in the catalog of Section~\ref{analysis_catalog} (including the synthetic extensions beyond \(3\,\mathrm{kpc}\)) we compute the gamma-ray flux at Earth as \citep{2025MNRAS.luana,2025A&A...695A.175M}
\begin{equation}
\Phi_{\gamma}(E_{\gamma}) \;=\; \frac{n_{\rm H}\,c}{4\pi d^{2}}
\int \phi(E_{p})\,\epsilon(E_{p})\,\frac{d\sigma}{dE_{\gamma}}(E_{p},E_{\gamma})\,dE_{p},
\end{equation}
where \(E_{p}\) is the kinetic energy of protons, \(\phi(E_{p})\) is the injected proton spectrum from Eq.~\eqref{eq:spectrum}, \(d\) is the source distance, and \(\epsilon(E_{p}) \sim 1.15\) represents the nuclear enhancement factor, accounting only for the most common nuclei, including hydrogen, helium, carbon, and oxygen. The CR flux (\(\phi(E_{p})\)) used in this calculation corresponds to the injection flux from each cluster, without propagation through the Galactic disk. This approach therefore does not constitute a full calculation of the diffuse gamma-ray flux in the sense of convolving the Galactic cosmic-ray density from MSCs with the mean interstellar gas density. Rather, it estimates the hadronic emission produced within the cluster environments and serves as a consistency check to verify that the model does not overproduce the observed diffuse flux at the source level. The local gas near the clusters can differ from the Galactic average. Therefore, we adopt a density of \(n_{\rm H} = 1\,\rm cm^{-3}\), considering both the typical diffuse ISM and potentially denser regions of the Galaxy \citep{2016SAAS...43...85K}, as well as the environment immediately surrounding the clusters \citep{2025MNRAS.luana,2025A&A...695A.175M,2025A&A...694A.244B,2024MNRAS.533..561B,2022A&A...667A..69S}. 

\begin{figure*}[htb!]
   \centering
   \subfloat[gamma]{\includegraphics[width=0.51\textwidth]{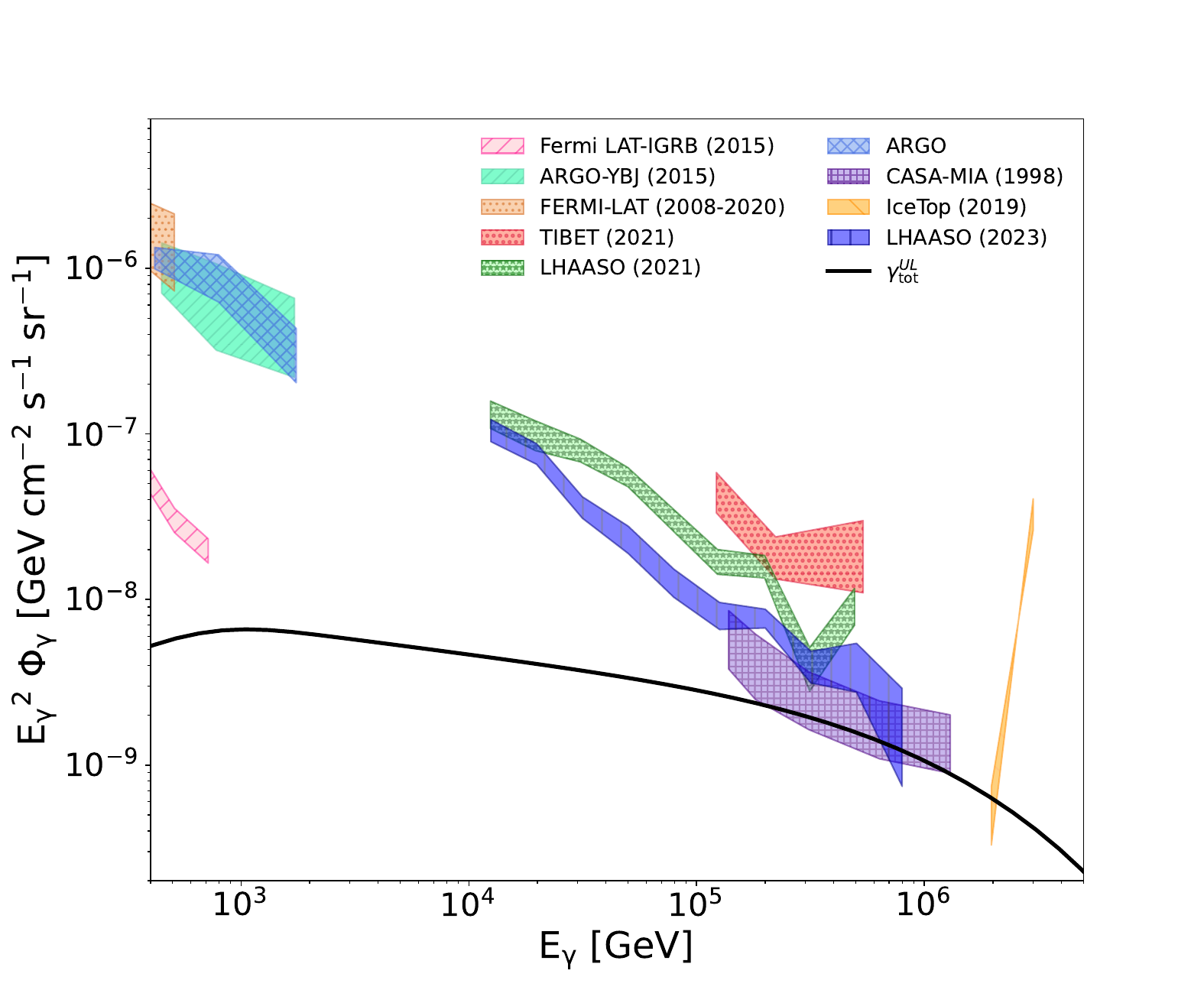}}
   \subfloat[neutrino]{\includegraphics[width=0.51\textwidth]{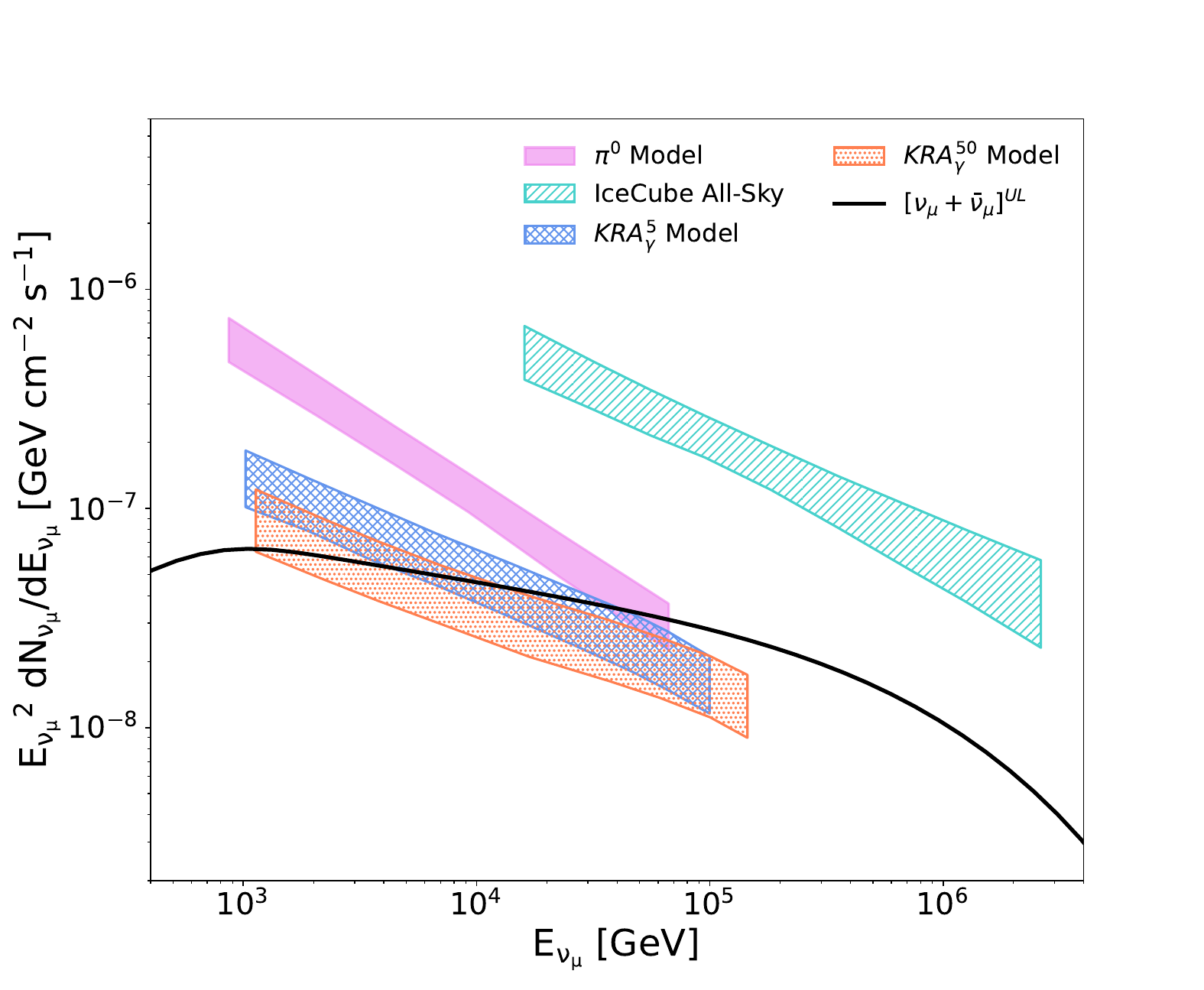}}\hfill
       \caption{\small \justifying (a) Contribution from MSC to gamma-ray emission, compared with observations of the diffuse Galactic gamma-ray emission \citep{2015ApJ...806...20B,2017NPPP..291....9G,2021PhRvL.126n1101A,2009ApJ...692...61A,1998ApJ...493..175B,2017ApJ...849...67A,2023A&A...672A..58D,2015ApJ...799...86A,2023PhRvL.131o1001C}. (b) Muon neutrino flux from the combined contribution of the powerful and soft MSC, compared with predictions of the diffuse Galactic neutrino flux and the IceCube sensitivity to neutrinos from the Galactoc plane \citep{2020arXiv200109520I}. The $KRA_{\gamma}$ models provide neutrino flux predictions based on gamma-ray data, assuming cutoff energies pf 5 PeV and 50 PeV \citep{2023Sci...380.1338I}. The $\pi^0$ model uses low-energy gamma-ray observations to estimate the neutrino flux, assuming that the same physical process produces both particles, with a smooth decay and no strong cutoffs \citep{2023Sci...380.1338I}.}
    \label{fig:multimessenger}
\end{figure*}

The total flux is obtained by summing the contributions from the different components: powerful clusters, soft clusters, and stellar wind termination shocks. The gamma-ray flux is calculated cluster by cluster, following the spatial distribution derived in Section \ref{analysis_catalog}, and then integrated over the entire population. The gamma-ray emission from wind termination shocks is estimated assuming full confinement of cosmic rays within the cluster bubble, which provides an upper limit on the expected flux. Even under this maximum assumption, the wind contribution is negligible compared to that from powerful and soft clusters. In addition, the model does not include an explicit treatment of particle escape from the cluster. The total gamma-ray flux is obtained by evaluating \(\Phi_{\gamma}\) cluster by cluster and summing the individual contributions of all clusters and all physical components over the entire sample of massive star clusters in the Galaxy.

Charged pions produced in the same hadronic interactions decay into neutrinos. The relation between the gamma-ray and neutrino energy fluxes can be written in a compact form that includes gamma-ray absorption along the line of sight,
\begin{equation}
E_{\gamma}\,\Phi_{\gamma}(E_{\gamma}) \;\simeq\; \frac{1}{3}\,e^{-\tau_{\gamma\gamma}(E_{\gamma},R,z,\alpha)}\,
\sum_{f} E_{\nu_{f}}\,\frac{dN_{\nu_{f}}}{dE_{\nu_{f}}}(E_{\nu_{f}}),
\end{equation}
where \(\tau_{\gamma\gamma}=d/\lambda_{\gamma\gamma}\) is the optical depth for pair production on the interstellar radiation fields and the cosmic microwave background, and the sum runs over neutrino flavors \citep{2014PhRvD..90b3010A,2023MNRAS.521.1144S}. The optical depth depends on position in the Galaxy and on energy; we use the tabulated values of \citet{2006ApJ...640L.155M} to model this attenuation. For convenience, we bin sources by galactocentric azimuth \(\alpha=0^{\circ},\,90^{\circ},\,180^{\circ}\) with the intervals \(0^{\circ}\le\alpha<45^{\circ}\), \(45^{\circ}\le\alpha<135^{\circ}\), and \(135^{\circ}\le\alpha\le 180^{\circ}\), and by galactocentric radius in the ranges \(0\)–\(5\), \(5\)–\(10\), \(10\)–\(15\), and \(15\)–\(20\,\mathrm{kpc}\). We compute the gamma-ray flux for each bin, convert it to a neutrino flux with the relation above, and then sum the contributions to obtain the final neutrino spectrum shown in Fig.~\ref{fig:multimessenger}-(b).

Figure~\ref{fig:multimessenger} summarizes the predicted emission from the full population of MSC in gamma rays and neutrinos. Panel ~\ref{fig:multimessenger}-(a) displays the integrated gamma-ray flux from neutral pion decay. A direct, quantitative comparison with data is challenging because the model provides a full-sky flux, while observations are confined to specific regions containing diffuse Galactic emission and are subject to instrumental variations. Therefore, the following analysis should be interpreted qualitatively. The predicted flux exhibits the hadronic spectrum typical of p–p collisions and remains sub-dominant to the measured diffuse emission from GeV to PeV energies. Notably, in the hundred TeV range and beyond, our prediction lies well under the LHAASO measurement~\citep{2023PhRvL.131o1001C}, suggesting MSCs contribute at most a modest fraction to the Galactic diffuse flux at these energies. The model's normalization is primarily governed by the target gas density in the cluster environment, while the cosmic-ray injection power for each cluster class is fixed by the fit to the observed cosmic-ray spectrum. The calculation also includes the effects of pair-production attenuation.

Panel ~\ref{fig:multimessenger}-(b) presents the associated muon neutrino flux obtained from the same hadronic channel. The prediction is shown together with the current IceCube limits for the Galactic plane~\citep{2020arXiv200109520I,2023Sci...380.1338I} and with representative diffuse models inferred from gamma-ray data. The cluster contribution lies below the IceCube sensitivity across the energy range considered, and it is smaller than the diffuse models based on CR interactions in the ISM. The spectral shape follows the parent proton distribution with a gradual softening above tens of TeV as the maximum energy of the sources is approached, and without a sharp cutoff. Since neutrinos do not suffer absorption during propagation, the difference between the gamma-ray and neutrino panels at high energy reflects the gamma-ray attenuation included in panel~\ref{fig:multimessenger}-(a). Figure~\ref{fig:multimessenger} shows that the hadronic output from MSC is consistent with present gamma-ray constraints and remains below current neutrino bounds, while providing a well-defined target for future wide field searches at energies from a few tens of TeV to the PeV scale.

\section{Summary and Conclusions}
\label{sec:Conclusions}

In this work, we have developed a unified model to interpret the origin of cosmic rays around the knee region \((2\text{--}5) \times 10^{15}\,\mathrm{eV}\) assuming that particle acceleration occurs at SNR shocks propagating inside the cores of MSCs, where they interact with the collective stellar wind and subsonic intracluster medium. By combining observational catalogs with a synthetic population to account for incompleteness, we classified MSC into two categories: powerful clusters, which are young, compact, and characterized by strong collective winds, and soft clusters, which are older systems with weakened winds~\citep{2023MNRAS.519..136V,2025MNRAS.luana}. These two classes contribute differently to CR acceleration, reflecting their distinct physical conditions.

We have shown that oblique shocks significantly enhance the maximum particle energies while allowing more realistic magnetic field values. Using the formalism of~\citet{2006A&A...454..687M} and~\citet{1987ApJ...313..842J}, we demonstrate that oblique shocks, which naturally emerge in the turbulent environments of MSC, can accelerate particles to energies compatible with the knee. This framework therefore enables a more realistic description of acceleration in the turbulent, magnetized environments of MSC, while remaining consistent with multi-messenger constraints and yielding improved agreement with LHAASO composition data when extragalactic contributions are included \citep{2024PhRvL.132m1002C}. By treating the shock radius as a dynamic parameter ($r_{\rm sh} = 0.5$--$1.0$ pc), we ensure that SNR shocks have expanded beyond individual progenitor wind cavities ($\sim 0.2$ pc) into the collective intracluster medium. This approach, combined with magnetic field strengths of $\sim 50\,\mu$G motivated by recent MHD simulations~\citep{2024MNRAS.527.3749B,2025A&A...698A...6H}, provides a physically consistent framework that does not rely on extreme parameter values. The maximum energy \(E_{\text{max}}\) increases with shock obliquity and scales with rigidity, allowing heavier nuclei to reach higher energies than protons. When comparing the predicted spectra to observational data, we found that our model successfully reproduces the LHAASO CR spectrum and composition~\citep{PhysRevD.106.123028,2024PhRvL.132m1002C}. Among the four acceleration scenarios considered (Models A--D), Model B, which incorporates oblique shocks in both cluster types, best reproduces the all-particle spectrum measured by LHAASO in the knee region ($10^6$--$10^7$ GeV)~\citep{2019ChPhC..43g5001Y}. In this framework, the knee arises as a rigidity-dependent cutoff, with protons dominating up to \(\sim 3\,\mathrm{PeV}\) while heavier nuclei become increasingly important at higher energies.

The spectral fits, performed primarily in the 1--10 PeV range where our Galactic model is expected to dominate, indicate that the inferred parameter constraints remain stable throughout the explored model space. The source index consistently clusters around \(\alpha \simeq 2.0\), while the transport index lies in the range \(s \simeq 0.45\)--\(0.46\). The wind power required to reproduce the sub-PeV normalization remains modest when obliquity effects are accounted for in both cluster categories (Model B: $\sim 1 \times 10^{35}$ erg s$^{-1}$), but increases when this physical leverage is excluded (Model D: $\sim 10 \times 10^{35}$ erg s$^{-1}$). These behaviors illustrate natural compensations among acceleration scale, transport properties, and source energetics.

The analysis of composition, including the mean logarithmic mass $\langle \ln A \rangle$, further supports a rigidity-ordered knee. When the extragalactic proton component is included, our preferred Models B and C predict a gradual increase in $\langle \ln A \rangle$ up to the knee region, followed by a transition to lighter composition at higher energies ($> 100$ PeV), consistent with measurements from Auger and other experiments. Our results indicate \(\langle \ln A \rangle \approx 1.3\) at the knee, consistent with a light-dominated composition composed mostly of protons and helium \citep{2012APh....35..660K, 2019ChPhC..43g5001Y,2019PhRvD.100h2002A,2024PhRvL.132m1002C}. The models predict a smooth transition between elemental groups, with heavier nuclei contributing significantly around \(10\,\mathrm{PeV}\), in agreement with the latest LHAASO observations~\citep{2024PhRvL.132m1002C}.

We also estimated the associated high-energy photon and neutrino signatures. Our calculation represents the contribution from the massive star cluster population to the Galactic gamma-ray sky. Since this estimate uses the CR injection flux with local cluster gas densities, rather than the propagated CR density convolved with the mean interstellar gas density, the hadronic emission from MSCs through neutral pion decay appears subdominant compared to the observed diffuse flux in the TeV to PeV range.~\citep{2023PhRvL.131o1001C}. The corresponding neutrino flux remains below present IceCube sensitivity limits \citep{2020arXiv200109520I,2023Sci...380.1338I}, rendering MSC a subdominant yet well-defined target for future multi-messenger searches.

Finally, the construction of a hybrid catalog, which supplements the \citet{2013A&A...558A..53K} catalog with a synthetic cluster population based on Galactic structure, has enabled population studies. This methodological improvement provides a more complete spatial and evolutionary representation of MSC. In summary, SNR shocks evolving inside MSC cores, influenced by oblique geometries and cluster environments, offer a physically motivated mechanism to explain the spectral and compositional properties of cosmic rays around the knee. The framework we have developed not only reproduces the latest measurements from LHAASO~\citep{2024PhRvL.132m1002C} but also delivers clear and testable predictions for the next generation of high-energy gamma-ray and neutrino observatories.

Future work can refine the angle distribution with specific magnetization and turbulence models, include time evolution of repeated shocks, consider possible differences between core-collapse and Type Ia supernovae and their impact on particle acceleration, incorporate detailed gas density distributions around clusters, and exploit joint fits to the LHAASO composition data, gamma-ray maps, and next-generation neutrino limits.

\section*{Acknowledgements}

We thank the anonymous referee for a thorough and constructive review that significantly improved the clarity and physical interpretation of this work. L.N.P acknowledges financial support from the Coordenação de Aperfeiçoamento de Pessoal de Nível Superior – Brasil (CAPES) – Finance Code 001. L.N.P. and R.C.A. acknowledge the support of the NAPI “Fenômenos Extremos do Universo” of Fundação de Apoio à Ciência, Tecnologia e Inovação do Paraná. R.C.A. research is supported by CAPES/Alexander von Humboldt Program (88881.800216/2022-01), CNPq (308859/2025-1) and (4000045/2023-0), Araucária Foundation (698/2022) and (721/2022) and FAPESP (2021/01089-1). The authors acknowledge the AWS Cloud Credit/CNPq and the National Laboratory for Scientific Computing (LNCC/MCTI, Brazil) for providing HPC resources of the SDumont supercomputer, which have contributed to the research results reported in this paper. URL: https://sdumont.lncc.br.

\bibliography{sample701}{}

@ARTICLE{1996ApJ...461..408A,
       author = {{Amenomori}, M. and {Cao}, Z. and {Dai}, B.~Z. and {Ding}, L.~K. and {Feng}, Y.~X. and {Feng}, Z.~Y. and {Hibino}, K. and {Hotta}, N. and {Huang}, Q. and {Huo}, A.~X. and {Jia}, H.~Y. and {Jiang}, G.~Z. and {Jiao}, S.~Q. and {Kajino}, F. and {Kasahara}, K. and {Labaciren} and {Liu}, S.~M. and {Mei}, D.~M. and {Meng}, L. and {Meng}, X.~R. and {Mimaciren} and {Mizutani}, K. and {Mu}, J. and {Nanjo}, H. and {Nishizawa}, M. and {Ohnishi}, M. and {Ohta}, I. and {Ouchi}, T. and {Ren}, J.~R. and {Saito}, To. and {Sakata}, M. and {Shi}, Z.~Z. and {Shibata}, M. and {Shiomi}, A. and {Shirai}, T. and {Sugimoto}, H. and {Sun}, X.~X. and {Taira}, K. and {Tan}, Y.~H. and {Tateyama}, N. and {Torii}, S. and {Wang}, H. and {Wen}, C.~Z. and {Yamamoto}, Y. and {Yu}, G.~C. and {Yuan}, P. and {Yuda}, T. and {Zhang}, C.~S. and {Zhang}, H.~M. and {Zhang}, L. and {Zhasang} and {Zhaxiciren} and {Zhou}, W.~D. and {Tibet As Gamma Collaboration}},
        title = "{The Cosmic-Ray Energy Spectrum between 10 14.5 and 10 16.3 eV Covering the ``Knee'' Region}",
      journal = {\apj},
         year = 1996,
        month = apr,
       volume = {461},
        pages = {408},
          doi = {10.1086/177069},
       adsurl = {https://ui.adsabs.harvard.edu/abs/1996ApJ...461..408A}
}

@ARTICLE{2012APh....35..660K,
       author = {{Kampert}, Karl-Heinz and {Unger}, Michael},
        title = "{Measurements of the cosmic ray composition with air shower experiments}",
      journal = {Astroparticle Physics},
         year = 2012,
        month = may,
       volume = {35},
       number = {10},
        pages = {660-678},
          doi = {10.1016/j.astropartphys.2012.02.004},
archivePrefix = {arXiv},
       eprint = {1201.0018},
 primaryClass = {astro-ph.HE},
       adsurl = {https://ui.adsabs.harvard.edu/abs/2012APh....35..660K}
}

@ARTICLE{2002PhRvD..66h3004K,
       author = {{Kobayakawa}, K. and {Honda}, Y.~S. and {Samura}, T.},
        title = "{Acceleration by oblique shocks at supernova remnants and cosmic ray spectra around the knee region}",
      journal = {\prd},
         year = 2002,
        month = oct,
       volume = {66},
       number = {8},
          eid = {083004},
        pages = {083004},
          doi = {10.1103/PhysRevD.66.083004},
archivePrefix = {arXiv},
       eprint = {astro-ph/0008209},
 primaryClass = {astro-ph},
       adsurl = {https://ui.adsabs.harvard.edu/abs/2002PhRvD..66h3004K}
}

@ARTICLE{2019NatAs...3..561A,
       author = {{Aharonian}, Felix and {Yang}, Ruizhi and {de O{\~n}a Wilhelmi}, Emma},
        title = "{Massive stars as major factories of Galactic cosmic rays}",
      journal = {Nature Astronomy},
         year = 2019,
        month = mar,
       volume = {3},
        pages = {561-567},
          doi = {10.1038/s41550-019-0724-0},
archivePrefix = {arXiv},
       eprint = {1804.02331},
 primaryClass = {astro-ph.HE},
       adsurl = {https://ui.adsabs.harvard.edu/abs/2019NatAs...3..561A}
}

@ARTICLE{2019A&A...623A..86A,
       author = {{Ambrogi}, L. and {Zanin}, R. and {Casanova}, S. and {De O{\~n}a Wilhelmi}, E. and {Peron}, G. and {Aharonian}, F.},
        title = "{Spectral and morphological study of the gamma radiation of the middle-aged supernova remnant HB 21}",
      journal = {\aap},
         year = 2019,
        month = mar,
       volume = {623},
          eid = {A86},
        pages = {A86},
          doi = {10.1051/0004-6361/201833985},
archivePrefix = {arXiv},
       eprint = {1902.06064},
 primaryClass = {astro-ph.HE},
       adsurl = {https://ui.adsabs.harvard.edu/abs/2019A&A...623A..86A}
}

@INCOLLECTION{2022hxga.book...52V,
       author = {{Vink}, Jacco and {Bamba}, Aya},
        title = "{Nonthermal Processes and Particle Acceleration in Supernova Remnants}",
    booktitle = {Handbook of X-ray and Gamma-ray Astrophysics},
         year = 2022,
       editor = {{Bambi}, Cosimo and {Sangangelo}, Andrea},
          eid = {52},
        pages = {52},
          doi = {10.1007/978-981-16-4544-0_90-1},
       adsurl = {https://ui.adsabs.harvard.edu/abs/2022hxga.book...52V}
}

@ARTICLE{2013Sci...339..807A,
       author = {{Ackermann}, M. and {Ajello}, M. and {Allafort}, A. and {Baldini}, L. and {Ballet}, J. and {Barbiellini}, G. and {Baring}, M.~G. and {Bastieri}, D. and {Bechtol}, K. and {Bellazzini}, R. and {Blandford}, R.~D. and {Bloom}, E.~D. and {Bonamente}, E. and {Borgland}, A.~W. and {Bottacini}, E. and {Brandt}, T.~J. and {Bregeon}, J. and {Brigida}, M. and {Bruel}, P. and {Buehler}, R. and {Busetto}, G. and {Buson}, S. and {Caliandro}, G.~A. and {Cameron}, R.~A. and {Caraveo}, P.~A. and {Casandjian}, J.~M. and {Cecchi}, C. and {{\c{C}}elik}, {\"O}. and {Charles}, E. and {Chaty}, S. and {Chaves}, R.~C.~G. and {Chekhtman}, A. and {Cheung}, C.~C. and {Chiang}, J. and {Chiaro}, G. and {Cillis}, A.~N. and {Ciprini}, S. and {Claus}, R. and {Cohen-Tanugi}, J. and {Cominsky}, L.~R. and {Conrad}, J. and {Corbel}, S. and {Cutini}, S. and {D'Ammando}, F. and {de Angelis}, A. and {de Palma}, F. and {Dermer}, C.~D. and {do Couto e Silva}, E. and {Drell}, P.~S. and {Drlica-Wagner}, A. and {Falletti}, L. and {Favuzzi}, C. and {Ferrara}, E.~C. and {Franckowiak}, A. and {Fukazawa}, Y. and {Funk}, S. and {Fusco}, P. and {Gargano}, F. and {Germani}, S. and {Giglietto}, N. and {Giommi}, P. and {Giordano}, F. and {Giroletti}, M. and {Glanzman}, T. and {Godfrey}, G. and {Grenier}, I.~A. and {Grondin}, M.-H. and {Grove}, J.~E. and {Guiriec}, S. and {Hadasch}, D. and {Hanabata}, Y. and {Harding}, A.~K. and {Hayashida}, M. and {Hayashi}, K. and {Hays}, E. and {Hewitt}, J.~W. and {Hill}, A.~B. and {Hughes}, R.~E. and {Jackson}, M.~S. and {Jogler}, T. and {J{\'o}hannesson}, G. and {Johnson}, A.~S. and {Kamae}, T. and {Kataoka}, J. and {Katsuta}, J. and {Kn{\"o}dlseder}, J. and {Kuss}, M. and {Lande}, J. and {Larsson}, S. and {Latronico}, L. and {Lemoine-Goumard}, M. and {Longo}, F. and {Loparco}, F. and {Lovellette}, M.~N. and {Lubrano}, P. and {Madejski}, G.~M. and {Massaro}, F. and {Mayer}, M. and {Mazziotta}, M.~N. and {McEnery}, J.~E. and {Mehault}, J. and {Michelson}, P.~F. and {Mignani}, R.~P. and {Mitthumsiri}, W. and {Mizuno}, T. and {Moiseev}, A.~A. and {Monzani}, M.~E. and {Morselli}, A. and {Moskalenko}, I.~V. and {Murgia}, S. and {Nakamori}, T. and {Nemmen}, R. and {Nuss}, E. and {Ohno}, M. and {Ohsugi}, T. and {Omodei}, N. and {Orienti}, M. and {Orlando}, E. and {Ormes}, J.~F. and {Paneque}, D. and {Perkins}, J.~S. and {Pesce-Rollins}, M. and {Piron}, F. and {Pivato}, G. and {Rain{\`o}}, S. and {Rando}, R. and {Razzano}, M. and {Razzaque}, S. and {Reimer}, A. and {Reimer}, O. and {Ritz}, S. and {Romoli}, C. and {S{\'a}nchez-Conde}, M. and {Schulz}, A. and {Sgr{\`o}}, C. and {Simeon}, P.~E. and {Siskind}, E.~J. and {Smith}, D.~A. and {Spandre}, G. and {Spinelli}, P. and {Stecker}, F.~W. and {Strong}, A.~W. and {Suson}, D.~J. and {Tajima}, H. and {Takahashi}, H. and {Takahashi}, T. and {Tanaka}, T. and {Thayer}, J.~G. and {Thayer}, J.~B. and {Thompson}, D.~J. and {Thorsett}, S.~E. and {Tibaldo}, L. and {Tibolla}, O. and {Tinivella}, M. and {Troja}, E. and {Uchiyama}, Y. and {Usher}, T.~L. and {Vandenbroucke}, J. and {Vasileiou}, V. and {Vianello}, G. and {Vitale}, V. and {Waite}, A.~P. and {Werner}, M. and {Winer}, B.~L. and {Wood}, K.~S. and {Wood}, M. and {Yamazaki}, R. and {Yang}, Z. and {Zimmer}, S.},
        title = "{Detection of the Characteristic Pion-Decay Signature in Supernova Remnants}",
      journal = {Science},
         year = 2013,
        month = feb,
       volume = {339},
       number = {6121},
        pages = {807-811},
          doi = {10.1126/science.1231160},
archivePrefix = {arXiv},
       eprint = {1302.3307},
 primaryClass = {astro-ph.HE},
       adsurl = {https://ui.adsabs.harvard.edu/abs/2013Sci...339..807A}
}

@ARTICLE{2006A&A...449..223A,
       author = {{Aharonian}, F. and {Akhperjanian}, A.~G. and {Bazer-Bachi}, A.~R. and {Beilicke}, M. and {Benbow}, W. and {Berge}, D. and {Bernl{\"o}hr}, K. and {Boisson}, C. and {Bolz}, O. and {Borrel}, V. and {Braun}, I. and {Breitling}, F. and {Brown}, A.~M. and {Chadwick}, P.~M. and {Chounet}, L.-M. and {Cornils}, R. and {Costamante}, L. and {Degrange}, B. and {Dickinson}, H.~J. and {Djannati-Ata{\"\i}}, A. and {O'C. Drury}, L. and {Dubus}, G. and {Emmanoulopoulos}, D. and {Espigat}, P. and {Feinstein}, F. and {Fontaine}, G. and {Fuchs}, Y. and {Funk}, S. and {Gallant}, Y.~A. and {Giebels}, B. and {Glicenstein}, J.~F. and {Goret}, P. and {Hadjichristidis}, C. and {Hauser}, D. and {Hauser}, M. and {Heinzelmann}, G. and {Henri}, G. and {Hermann}, G. and {Hinton}, J.~A. and {Hofmann}, W. and {Holleran}, M. and {Horns}, D. and {Jacholkowska}, A. and {de Jager}, O.~C. and {Kh{\'e}lifi}, B. and {Klages}, S. and {Komin}, Nu. and {Konopelko}, A. and {Latham}, I.~J. and {Le Gallou}, R. and {Lemi{\`e}re}, A. and {Lemoine-Goumard}, M. and {Lohse}, T. and {Martin}, J.~M. and {Martineau-Huynh}, O. and {Marcowith}, A. and {Masterson}, C. and {McComb}, T.~J.~L. and {de Naurois}, M. and {Nedbal}, D. and {Nolan}, S.~J. and {Noutsos}, A. and {Orford}, K.~J. and {Osborne}, J.~L. and {Ouchrif}, M. and {Panter}, M. and {Pelletier}, G. and {Pita}, S. and {P{\"u}hlhofer}, G. and {Punch}, M. and {Raubenheimer}, B.~C. and {Raue}, M. and {Rayner}, S.~M. and {Reimer}, A. and {Reimer}, O. and {Ripken}, J. and {Rob}, L. and {Rolland}, L. and {Rowell}, G. and {Sahakian}, V. and {Saug{\'e}}, L. and {Schlenker}, S. and {Schlickeiser}, R. and {Schuster}, C. and {Schwanke}, U. and {Siewert}, M. and {Sol}, H. and {Spangler}, D. and {Steenkamp}, R. and {Stegmann}, C. and {Superina}, G. and {Tavernet}, J.-P. and {Terrier}, R. and {Th{\'e}oret}, C.~G. and {Tluczykont}, M. and {van Eldik}, C. and {Vasileiadis}, G. and {Venter}, C. and {Vincent}, P. and {V{\"o}lk}, H.~J. and {Wagner}, S.~J.},
        title = "{A detailed spectral and morphological study of the gamma-ray supernova remnant <ASTROBJ>RX J1713.7-3946</ASTROBJ> with HESS}",
      journal = {\aap},
         year = 2006,
        month = apr,
       volume = {449},
       number = {1},
        pages = {223-242},
          doi = {10.1051/0004-6361:20054279},
archivePrefix = {arXiv},
       eprint = {astro-ph/0511678},
 primaryClass = {astro-ph},
       adsurl = {https://ui.adsabs.harvard.edu/abs/2006A&A...449..223A}
}

@ARTICLE{2023MNRAS.519..136V,
       author = {{Vieu}, T. and {Reville}, B.},
        title = "{Massive star cluster origin for the galactic cosmic ray population at very-high energies}",
      journal = {\mnras},
         year = 2023,
        month = feb,
       volume = {519},
       number = {1},
        pages = {136-147},
          doi = {10.1093/mnras/stac3469},
archivePrefix = {arXiv},
       eprint = {2211.11625},
 primaryClass = {astro-ph.HE},
       adsurl = {https://ui.adsabs.harvard.edu/abs/2023MNRAS.519..136V}
}

@ARTICLE{1949PhRv...75.1169F,
       author = {{Fermi}, Enrico},
        title = "{On the Origin of the Cosmic Radiation}",
      journal = {Physical Review},
         year = 1949,
        month = apr,
       volume = {75},
       number = {8},
        pages = {1169-1174},
          doi = {10.1103/PhysRev.75.1169},
       adsurl = {https://ui.adsabs.harvard.edu/abs/1949PhRv...75.1169F}
}

@INPROCEEDINGS{2023EPJWC.28304001G,
       author = {{Globus}, No{\'e}mie and {Blandford}, Roger},
        title = "{Ultra High Energy Cosmic Ray Source Models: Successes, Challenges and General Predictions}",
    booktitle = {European Physical Journal Web of Conferences},
         year = 2023,
       series = {European Physical Journal Web of Conferences},
       volume = {283},
        month = oct,
    publisher = {EDP},
          eid = {04001},
        pages = {04001},
          doi = {10.1051/epjconf/202328304001},
archivePrefix = {arXiv},
       eprint = {2302.06791},
 primaryClass = {astro-ph.HE},
       adsurl = {https://ui.adsabs.harvard.edu/abs/2023EPJWC.28304001G}
}

@ARTICLE{2020LRCA....6....1M,
       author = {{Marcowith}, Alexandre and {Ferrand}, Gilles and {Grech}, Mickael and {Meliani}, Zakaria and {Plotnikov}, Illya and {Walder}, Rolf},
        title = "{Multi-scale simulations of particle acceleration in astrophysical systems}",
      journal = {Living Reviews in Computational Astrophysics},
         year = 2020,
        month = mar,
       volume = {6},
       number = {1},
          eid = {1},
        pages = {1},
          doi = {10.1007/s41115-020-0007-6},
archivePrefix = {arXiv},
       eprint = {2002.09411},
 primaryClass = {astro-ph.HE},
       adsurl = {https://ui.adsabs.harvard.edu/abs/2020LRCA....6....1M}
}

@ARTICLE{2021A&A...650A..62C,
       author = {{Cristofari}, P. and {Blasi}, P. and {Caprioli}, D.},
        title = "{Cosmic ray protons and electrons from supernova remnants}",
      journal = {\aap},
         year = 2021,
        month = jun,
       volume = {650},
          eid = {A62},
        pages = {A62},
          doi = {10.1051/0004-6361/202140448},
archivePrefix = {arXiv},
       eprint = {2103.02375},
 primaryClass = {astro-ph.HE},
       adsurl = {https://ui.adsabs.harvard.edu/abs/2021A&A...650A..62C}
}

@ARTICLE{2023PPCF...65a4002B,
       author = {{Bohdan}, Artem},
        title = "{Electron acceleration in supernova remnants}",
      journal = {Plasma Physics and Controlled Fusion},
         year = 2023,
        month = jan,
       volume = {65},
       number = {1},
          eid = {014002},
        pages = {014002},
          doi = {10.1088/1361-6587/aca5b2},
archivePrefix = {arXiv},
       eprint = {2211.13992},
 primaryClass = {astro-ph.HE},
       adsurl = {https://ui.adsabs.harvard.edu/abs/2023PPCF...65a4002B}
}

@ARTICLE{2001AstL...27..625B,
       author = {{Bykov}, A.~M. and {Toptygin}, I.~N.},
        title = "{A Model of Particle Acceleration to High Energies by Multiple Supernova Explosions in OB Associations}",
      journal = {Astronomy Letters},
         year = 2001,
        month = oct,
       volume = {27},
       number = {10},
        pages = {625-633},
          doi = {10.1134/1.1404456},
       adsurl = {https://ui.adsabs.harvard.edu/abs/2001AstL...27..625B}
}

@ARTICLE{2004A&A...424..747P,
       author = {{Parizot}, E. and {Marcowith}, A. and {van der Swaluw}, E. and {Bykov}, A.~M. and {Tatischeff}, V.},
        title = "{Superbubbles and energetic particles in the Galaxy. I. Collective effects of particle acceleration}",
      journal = {\aap},
         year = 2004,
        month = sep,
       volume = {424},
        pages = {747-760},
          doi = {10.1051/0004-6361:20041269},
archivePrefix = {arXiv},
       eprint = {astro-ph/0405531},
 primaryClass = {astro-ph},
       adsurl = {https://ui.adsabs.harvard.edu/abs/2004A&A...424..747P}
}

@ARTICLE{2010A&A...510A.101F,
       author = {{Ferrand}, G. and {Marcowith}, A.},
        title = "{On the shape of the spectrum of cosmic rays accelerated inside superbubbles}",
      journal = {\aap},
         year = 2010,
        month = feb,
       volume = {510},
          eid = {A101},
        pages = {A101},
          doi = {10.1051/0004-6361/200913520},
archivePrefix = {arXiv},
       eprint = {0911.4457},
 primaryClass = {astro-ph.HE},
       adsurl = {https://ui.adsabs.harvard.edu/abs/2010A&A...510A.101F}
}

@ARTICLE{2010ARA&A..48..431P,
       author = {{Portegies Zwart}, Simon F. and {McMillan}, Stephen L.~W. and {Gieles}, Mark},
        title = "{Young Massive Star Clusters}",
      journal = {\araa},
         year = 2010,
        month = sep,
       volume = {48},
        pages = {431-493},
          doi = {10.1146/annurev-astro-081309-130834},
archivePrefix = {arXiv},
       eprint = {1002.1961},
 primaryClass = {astro-ph.GA},
       adsurl = {https://ui.adsabs.harvard.edu/abs/2010ARA&A..48..431P}
}

@ARTICLE{2018NPPP..297..183B,
       author = {{Bykov}, A.~M. and {Ellison}, D.~C. and {Gladilin}, P.~E. and {Osipov}, S.~M.},
        title = "{Supernovae in clusters of massive stars as cosmic ray pevatrons}",
      journal = {Nuclear and Particle Physics Proceedings},
         year = 2018,
        month = apr,
       volume = {297-299},
        pages = {183-193},
          doi = {10.1016/j.nuclphysbps.2018.07.028},
       adsurl = {https://ui.adsabs.harvard.edu/abs/2018NPPP..297..183B}
}

@ARTICLE{2019arXiv190502773C,
       author = {{Cao}, Zhen and {della Volpe}, D. and {Liu}, Siming and {Editors} and {:} and {Bi}, Xiaojun and {Chen}, Yang and {D'Ettorre Piazzoli}, B. and {Feng}, Li and {Jia}, Huanyu and {Li}, Zhuo and {Ma}, Xinhua and {Wang}, Xiangyu and {Zhang}, Xiao and {Referees}, External and {:} and {Qie}, Xiushu and {Hu}, Hongbo and {Referees}, Internal and {:} and {S{\'a}iz}, Alejandro and {Yang}, Ruizhi and {Contributors} and {:} and {Addazi}, Andrea and {Belotsky}, Konstantin and {Beylin}, Vitaly and {Bi}, Yu-Jiang and {Che}, Ming-Jun and {Chen}, Song-Zhan and {Cheng}, Yao-Dong and {Chiavassa}, Andrea and {Cirelli}, Marco and {Di Sciascio}, Giuseppe and {Esmaili}, Arman and {Fang}, Kun and {Fornengo}, Nicolao and {Gou}, Quanbu and {Guo}, Yi-Qing and {Gan}, Qingyu and {Gong}, Guang-Hua and {Gu}, Min-Hao and {He}, Haoning and {He}, Hui-Hai and {Hou}, Chao and {Huang}, Xing-Tao and {Huang}, Wen-Hao and {Kachekriess}, Michael and {Khlopov}, Maxim and {Korchagin}, Vladimir and {Korochkin}, Alexander and {Kuksa}, Vladimir and {Ksenofontov}, Leonid T. and {Liu}, Ye and {Liu}, Ruo-Yu and {Liu}, Cheng and {Marciano}, Antonino and {Martineau-Huynh}, Olivier and {Martraire}, Diane and {Ma}, Lingling and {Neronov}, Andrii and {Panci}, Paolo and {Pasechnick}, Roman and {Ruffolo}, David and {Sakharov}, Alexander and {Sala}, Filippo and {Semikoz}, Dimiri and {Shchegolev}, Oleg and {Serpico}, Pasquale Dario and {Sheng}, Xiang-Dong and {Stenkin}, Yuri V. and {Tam}, P.~H. Thomas and {Vernetto}, Silvia and {Vallania}, Piero and {Volchanskiy}, Nikolay and {Wang}, Zhongxiang and {Wang}, Kai and {Wang}, Xiang-Yu and {Wu}, Han-Rong and {Wu}, Chao-Yong and {Wu}, Sha and {Xiao}, Gang and {Yang}, Rui-zhi and {Yan}, Dahai and {Yao}, Zhi-Guo and {Yin}, Pengfei and {Yuan}, Qiang and {Zhang}, Xiao and {Zeng}, Houdun and {Zhang}, Shou-Shan and {Zhang}, Yi and {Zhou}, Xunxiu and {Zhu}, Hui and {Zuo}, Xiong},
        title = "{The Large High Altitude Air Shower Observatory (LHAASO) Science Book (2021 Edition)}",
      journal = {arXiv e-prints},
         year = 2019,
        month = may,
          eid = {arXiv:1905.02773},
        pages = {arXiv:1905.02773},
          doi = {10.48550/arXiv.1905.02773},
archivePrefix = {arXiv},
       eprint = {1905.02773},
 primaryClass = {astro-ph.HE},
       adsurl = {https://ui.adsabs.harvard.edu/abs/2019arXiv190502773C}
}

@ARTICLE{2019ChPhC..43g5001Y,
       author = {{Yin}, L.~Q. and {Zhang}, S.~S. and {Cao}, Z. and {Bi}, B.~Y. and {Wang}, C. and {Liu}, J.~L. and {Ma}, L.~L. and {Yang}, M.~J. and {Suomij{\"a}rvi}, Tiina and {Zhang}, Y. and {You}, Z.~Y. and {Zong}, Z.~Z. and {the LHAASO Collaboration}},
        title = "{Expected energy spectrum of cosmic ray protons and helium below 4 PeV measured by LHAASO}",
      journal = {Chinese Physics C},
         year = 2019,
        month = jul,
       volume = {43},
       number = {7},
          eid = {075001},
        pages = {075001},
          doi = {10.1088/1674-1137/43/7/075001},
       adsurl = {https://ui.adsabs.harvard.edu/abs/2019ChPhC..43g5001Y}
}

@ARTICLE{2020ChPhC..44f5002J,
       author = {{Jin}, Chao and {Chen}, Song-Zhan and {He}, Hui-Hai and {Collaboration)}, (Lhaaso},
        title = "{Classifying cosmic-ray proton and light groups in LHAASO-KM2A experiment with graph neural network}",
      journal = {Chinese Physics C},
         year = 2020,
        month = jun,
       volume = {44},
       number = {6},
          eid = {065002},
        pages = {065002},
          doi = {10.1088/1674-1137/44/6/065002},
       adsurl = {https://ui.adsabs.harvard.edu/abs/2020ChPhC..44f5002J}
}

@ARTICLE{2024PhRvL.132m1002C,
       author = {{Cao}, Zhen and {Aharonian}, F. and {Axikegu} and {Bai}, Y.~X. and {Bao}, Y.~W. and {Bastieri}, D. and {Bi}, X.~J. and {Bi}, Y.~J. and {Bian}, W. and {Bukevich}, A.~V. and {Cao}, Q. and {Cao}, W.~Y. and {Cao}, Zhe and {Chang}, J. and {Chang}, J.~F. and {Chen}, A.~M. and {Chen}, E.~S. and {Chen}, H.~X. and {Chen}, Liang and {Chen}, Lin and {Chen}, Long and {Chen}, M.~J. and {Chen}, M.~L. and {Chen}, Q.~H. and {Chen}, S. and {Chen}, S.~H. and {Chen}, S.~Z. and {Chen}, T.~L. and {Chen}, Y. and {Cheng}, N. and {Cheng}, Y.~D. and {Cui}, M.~Y. and {Cui}, S.~W. and {Cui}, X.~H. and {Cui}, Y.~D. and {Dai}, B.~Z. and {Dai}, H.~L. and {Dai}, Z.~G. and {Danzengluobu} and {Dong}, X.~Q. and {Duan}, K.~K. and {Fan}, J.~H. and {Fan}, Y.~Z. and {Fang}, J. and {Fang}, J.~H. and {Fang}, K. and {Feng}, C.~F. and {Feng}, H. and {Feng}, L. and {Feng}, S.~H. and {Feng}, X.~T. and {Feng}, Y. and {Feng}, Y.~L. and {Gabici}, S. and {Gao}, B. and {Gao}, C.~D. and {Gao}, Q. and {Gao}, W. and {Gao}, W.~K. and {Ge}, M.~M. and {Geng}, L.~S. and {Giacinti}, G. and {Gong}, G.~H. and {Gou}, Q.~B. and {Gu}, M.~H. and {Guo}, F.~L. and {Guo}, X.~L. and {Guo}, Y.~Q. and {Guo}, Y.~Y. and {Han}, Y.~A. and {Hasan}, M. and {He}, H.~H. and {He}, H.~N. and {He}, J.~Y. and {He}, Y. and {Hor}, Y.~K. and {Hou}, B.~W. and {Hou}, C. and {Hou}, X. and {Hu}, H.~B. and {Hu}, Q. and {Hu}, S.~C. and {Huang}, D.~H. and {Huang}, T.~Q. and {Huang}, W.~J. and {Huang}, X.~T. and {Huang}, X.~Y. and {Huang}, Y. and {Ji}, X.~L. and {Jia}, H.~Y. and {Jia}, K. and {Jiang}, K. and {Jiang}, X.~W. and {Jiang}, Z.~J. and {Jin}, M. and {Kang}, M.~M. and {Karpikov}, I. and {Kuleshov}, D. and {Kurinov}, K. and {Li}, B.~B. and {Li}, C.~M. and {Li}, Cheng and {Li}, Cong and {Li}, D. and {Li}, F. and {Li}, H.~B. and {Li}, H.~C. and {Li}, Jian and {Li}, Jie and {Li}, K. and {Li}, S.~D. and {Li}, W.~L. and {Li}, W.~L. and {Li}, X.~R. and {Li}, Xin and {Li}, Y.~Z. and {Li}, Zhe and {Li}, Zhuo and {Liang}, E.~W. and {Liang}, Y.~F. and {Lin}, S.~J. and {Liu}, B. and {Liu}, C. and {Liu}, D. and {Liu}, D.~B. and {Liu}, H. and {Liu}, H.~D. and {Liu}, J. and {Liu}, J.~L. and {Liu}, M.~Y. and {Liu}, R.~Y. and {Liu}, S.~M. and {Liu}, W. and {Liu}, Y. and {Liu}, Y.~N. and {Luo}, Q. and {Luo}, Y. and {Lv}, H.~K. and {Ma}, B.~Q. and {Ma}, L.~L. and {Ma}, X.~H. and {Mao}, J.~R. and {Min}, Z. and {Mitthumsiri}, W. and {Mu}, H.~J. and {Nan}, Y.~C. and {Neronov}, A. and {Ou}, L.~J. and {Pattarakijwanich}, P. and {Pei}, Z.~Y. and {Qi}, J.~C. and {Qi}, M.~Y. and {Qiao}, B.~Q. and {Qin}, J.~J. and {Raza}, A. and {Ruffolo}, D. and {S{\'a}iz}, A. and {Saeed}, M. and {Semikoz}, D. and {Shao}, L. and {Shchegolev}, O. and {Sheng}, X.~D. and {Shu}, F.~W. and {Song}, H.~C. and {Stenkin}, Yu. V. and {Stepanov}, V. and {Su}, Y. and {Sun}, D.~X. and {Sun}, Q.~N. and {Sun}, X.~N. and {Sun}, Z.~B. and {Takata}, J. and {Tam}, P.~H.~T. and {Tang}, Q.~W. and {Tang}, R. and {Tang}, Z.~B. and {Tian}, W.~W. and {Wang}, C. and {Wang}, C.~B. and {Wang}, G.~W. and {Wang}, H.~G. and {Wang}, H.~H. and {Wang}, J.~C. and {Wang}, Kai and {Wang}, Kai and {Wang}, L.~P. and {Wang}, L.~Y. and {Wang}, P.~H. and {Wang}, R. and {Wang}, W. and {Wang}, X.~G. and {Wang}, X.~Y. and {Wang}, Y. and {Wang}, Y.~D. and {Wang}, Y.~J. and {Wang}, Z.~H. and {Wang}, Z.~X. and {Wang}, Zhen and {Wang}, Zheng and {Wei}, D.~M.},
        title = "{Measurements of All-Particle Energy Spectrum and Mean Logarithmic Mass of Cosmic Rays from 0.3 to 30 PeV with LHAASO-KM2A}",
      journal = {\prl},
         year = 2024,
        month = mar,
       volume = {132},
       number = {13},
          eid = {131002},
        pages = {131002},
          doi = {10.1103/PhysRevLett.132.131002},
archivePrefix = {arXiv},
       eprint = {2403.10010},
 primaryClass = {astro-ph.HE},
       adsurl = {https://ui.adsabs.harvard.edu/abs/2024PhRvL.132m1002C}
}

@ARTICLE{2024PhRvD.110d3030T,
       author = {{Tian}, Xishui and {Li}, Zhuo and {Gou}, Quanbu and {Zhang}, Hengying and {He}, Huihai and {Feng}, Cunfeng and {Di Sciascio}, Giuseppe},
        title = "{Approach for composition measurement of cosmic rays using the muon-to-electron ratio observed by LHAASO-KM2A}",
      journal = {\prd},
         year = 2024,
        month = aug,
       volume = {110},
       number = {4},
          eid = {043030},
        pages = {043030},
          doi = {10.1103/PhysRevD.110.043030},
archivePrefix = {arXiv},
       eprint = {2407.13298},
 primaryClass = {astro-ph.HE},
       adsurl = {https://ui.adsabs.harvard.edu/abs/2024PhRvD.110d3030T}
}

@ARTICLE{2022PhRvD.106l3028Z,
       author = {{Zhang}, Hengying and {He}, Huihai and {Feng}, Cunfeng},
        title = "{Approaches to composition independent energy reconstruction of cosmic rays based on the LHAASO-KM2A detector}",
      journal = {\prd},
         year = 2022,
        month = dec,
       volume = {106},
       number = {12},
          eid = {123028},
        pages = {123028},
          doi = {10.1103/PhysRevD.106.123028},
       adsurl = {https://ui.adsabs.harvard.edu/abs/2022PhRvD.106l3028Z}
}

@ARTICLE{1987ApJ...313..842J,
       author = {{Jokipii}, J.~R.},
        title = "{Rate of Energy Gain and Maximum Energy in Diffusive Shock Acceleration}",
      journal = {\apj},
         year = 1987,
        month = feb,
       volume = {313},
        pages = {842},
          doi = {10.1086/165022},
       adsurl = {https://ui.adsabs.harvard.edu/abs/1987ApJ...313..842J}
}

@ARTICLE{2006A&A...454..687M,
       author = {{Meli}, A. and {Biermann}, P.~L.},
        title = "{Cosmic rays X. The cosmic ray knee and beyond: diffusive acceleration at oblique shocks}",
      journal = {\aap},
         year = 2006,
        month = aug,
       volume = {454},
       number = {3},
        pages = {687-694},
          doi = {10.1051/0004-6361:20064964},
archivePrefix = {arXiv},
       eprint = {astro-ph/0602308},
 primaryClass = {astro-ph},
       adsurl = {https://ui.adsabs.harvard.edu/abs/2006A&A...454..687M}
}

@ARTICLE{1995ARA&A..33..381F,
       author = {{Friel}, E.~D.},
        title = "{The Old Open Clusters Of The Milky Way}",
      journal = {\araa},
         year = 1995,
        month = jan,
       volume = {33},
        pages = {381-414},
          doi = {10.1146/annurev.aa.33.090195.002121},
       adsurl = {https://ui.adsabs.harvard.edu/abs/1995ARA&A..33..381F}
}

@ARTICLE{2013A&A...558A..53K,
       author = {{Kharchenko}, N.~V. and {Piskunov}, A.~E. and {Schilbach}, E. and {R{\"o}ser}, S. and {Scholz}, R.-D.},
        title = "{Global survey of star clusters in the Milky Way. II. The catalogue of basic parameters}",
      journal = {\aap},
         year = 2013,
        month = oct,
       volume = {558},
          eid = {A53},
        pages = {A53},
          doi = {10.1051/0004-6361/201322302},
archivePrefix = {arXiv},
       eprint = {1308.5822},
 primaryClass = {astro-ph.GA},
       adsurl = {https://ui.adsabs.harvard.edu/abs/2013A&A...558A..53K}
}

@ARTICLE{2018A&A...618A..93C,
       author = {{Cantat-Gaudin}, T. and {Jordi}, C. and {Vallenari}, A. and {Bragaglia}, A. and {Balaguer-N{\'u}{\~n}ez}, L. and {Soubiran}, C. and {Bossini}, D. and {Moitinho}, A. and {Castro-Ginard}, A. and {Krone-Martins}, A. and {Casamiquela}, L. and {Sordo}, R. and {Carrera}, R.},
        title = "{A Gaia DR2 view of the open cluster population in the Milky Way}",
      journal = {\aap},
         year = 2018,
        month = oct,
       volume = {618},
          eid = {A93},
        pages = {A93},
          doi = {10.1051/0004-6361/201833476},
archivePrefix = {arXiv},
       eprint = {1805.08726},
 primaryClass = {astro-ph.GA},
       adsurl = {https://ui.adsabs.harvard.edu/abs/2018A&A...618A..93C}
}

@ARTICLE{2020A&A...640A...1C,
       author = {{Cantat-Gaudin}, T. and {Anders}, F. and {Castro-Ginard}, A. and {Jordi}, C. and {Romero-G{\'o}mez}, M. and {Soubiran}, C. and {Casamiquela}, L. and {Tarricq}, Y. and {Moitinho}, A. and {Vallenari}, A. and {Bragaglia}, A. and {Krone-Martins}, A. and {Kounkel}, M.},
        title = "{Painting a portrait of the Galactic disc with its stellar clusters}",
      journal = {\aap},
         year = 2020,
        month = aug,
       volume = {640},
          eid = {A1},
        pages = {A1},
          doi = {10.1051/0004-6361/202038192},
archivePrefix = {arXiv},
       eprint = {2004.07274},
 primaryClass = {astro-ph.GA},
       adsurl = {https://ui.adsabs.harvard.edu/abs/2020A&A...640A...1C}
}

@ARTICLE{2020A&A...633A..99C,
       author = {{Cantat-Gaudin}, T. and {Anders}, F.},
        title = "{Clusters and mirages: cataloguing stellar aggregates in the Milky Way}",
      journal = {\aap},
         year = 2020,
        month = jan,
       volume = {633},
          eid = {A99},
        pages = {A99},
          doi = {10.1051/0004-6361/201936691},
archivePrefix = {arXiv},
       eprint = {1911.07075},
 primaryClass = {astro-ph.SR},
       adsurl = {https://ui.adsabs.harvard.edu/abs/2020A&A...633A..99C}
}

@ARTICLE{2003A&A...404..223B,
       author = {{Bica}, E. and {Dutra}, C.~M. and {Soares}, J. and {Barbuy}, B.},
        title = "{New infrared star clusters in the Northern and Equatorial Milky Way with 2MASS}",
      journal = {\aap},
         year = 2003,
        month = jun,
       volume = {404},
        pages = {223-232},
          doi = {10.1051/0004-6361:20030486},
archivePrefix = {arXiv},
       eprint = {astro-ph/0304379},
 primaryClass = {astro-ph},
       adsurl = {https://ui.adsabs.harvard.edu/abs/2003A&A...404..223B}
}

@ARTICLE{2003A&A...397..177B,
       author = {{Bica}, E. and {Dutra}, C.~M. and {Barbuy}, B.},
        title = "{A Catalogue of infrared star clusters and stellar groups}",
      journal = {\aap},
         year = 2003,
        month = jan,
       volume = {397},
        pages = {177-180},
          doi = {10.1051/0004-6361:20021479},
archivePrefix = {arXiv},
       eprint = {astro-ph/0210302},
 primaryClass = {astro-ph},
       adsurl = {https://ui.adsabs.harvard.edu/abs/2003A&A...397..177B}
}

@ARTICLE{2003A&A...400..533D,
       author = {{Dutra}, C.~M. and {Bica}, E. and {Soares}, J. and {Barbuy}, B.},
        title = "{New infrared star clusters in the southern Milky Way with 2MASS}",
      journal = {\aap},
         year = 2003,
        month = mar,
       volume = {400},
        pages = {533-539},
          doi = {10.1051/0004-6361:20030005},
archivePrefix = {arXiv},
       eprint = {astro-ph/0301221},
 primaryClass = {astro-ph},
       adsurl = {https://ui.adsabs.harvard.edu/abs/2003A&A...400..533D}
}

@ARTICLE{2007MNRAS.374..399F,
       author = {{Froebrich}, D. and {Scholz}, A. and {Raftery}, C.~L.},
        title = "{A systematic survey for infrared star clusters with |b| <20{\textdegree} using 2MASS}",
      journal = {\mnras},
         year = 2007,
        month = jan,
       volume = {374},
       number = {2},
        pages = {399-408},
          doi = {10.1111/j.1365-2966.2006.11148.x},
archivePrefix = {arXiv},
       eprint = {astro-ph/0610146},
 primaryClass = {astro-ph},
       adsurl = {https://ui.adsabs.harvard.edu/abs/2007MNRAS.374..399F}
}

@ARTICLE{2010MNRAS.409.1281F,
       author = {{Froebrich}, D. and {Schmeja}, S. and {Samuel}, D. and {Lucas}, P.~W.},
        title = "{Old star clusters in the FSR catalogue}",
      journal = {\mnras},
         year = 2010,
        month = dec,
       volume = {409},
       number = {3},
        pages = {1281-1288},
          doi = {10.1111/j.1365-2966.2010.17390.x},
archivePrefix = {arXiv},
       eprint = {1007.3410},
 primaryClass = {astro-ph.GA},
       adsurl = {https://ui.adsabs.harvard.edu/abs/2010MNRAS.409.1281F}
}

@ARTICLE{2009MNRAS.400..518M,
       author = {{Mel'Nik}, A.~M. and {Dambis}, A.~K.},
        title = "{Kinematics of OB-associations and the new reduction of the Hipparcos data}",
      journal = {\mnras},
         year = 2009,
        month = nov,
       volume = {400},
       number = {1},
        pages = {518-523},
          doi = {10.1111/j.1365-2966.2009.15484.x},
archivePrefix = {arXiv},
       eprint = {0909.0618},
 primaryClass = {astro-ph.GA},
       adsurl = {https://ui.adsabs.harvard.edu/abs/2009MNRAS.400..518M}
}

@ARTICLE{2011AcA....61..231B,
       author = {{Bukowiecki}, {\L}. and {Maciejewski}, G. and {Konorski}, P. and {Strobel}, A.},
        title = "{Open Clusters in 2MASS Photometry. I. Structural and Basic Astrophysical Parameters}",
      journal = {\actaa},
         year = 2011,
        month = sep,
       volume = {61},
       number = {3},
        pages = {231-246},
          doi = {10.48550/arXiv.1107.5119},
archivePrefix = {arXiv},
       eprint = {1107.5119},
 primaryClass = {astro-ph.SR},
       adsurl = {https://ui.adsabs.harvard.edu/abs/2011AcA....61..231B}
}

@ARTICLE{2009PrPNP..63..293B,
       author = {{Bl{\"u}mer}, Johannes and {Engel}, Ralph and {H{\"o}randel}, J{\"o}rg R.},
        title = "{Cosmic rays from the knee to the highest energies}",
      journal = {Progress in Particle and Nuclear Physics},
         year = 2009,
        month = oct,
       volume = {63},
       number = {2},
        pages = {293-338},
          doi = {10.1016/j.ppnp.2009.05.002},
archivePrefix = {arXiv},
       eprint = {0904.0725},
 primaryClass = {astro-ph.HE},
       adsurl = {https://ui.adsabs.harvard.edu/abs/2009PrPNP..63..293B}
}

@ARTICLE{1957ApJ...125..451V,
       author = {{von Hoerner}, S.},
        title = "{Internal structure of globular clusters}",
      journal = {\apj},
         year = 1957,
        month = mar,
       volume = {125},
        pages = {451},
          doi = {10.1086/146321},
       adsurl = {https://ui.adsabs.harvard.edu/abs/1957ApJ...125..451V}
}

@ARTICLE{2013ApJ...764..124W,
       author = {{Webb}, Jeremy J. and {Harris}, William E. and {Sills}, Alison and {Hurley}, Jarrod R.},
        title = "{The Influence of Orbital Eccentricity on Tidal Radii of Star Clusters}",
      journal = {\apj},
         year = 2013,
        month = feb,
       volume = {764},
       number = {2},
          eid = {124},
        pages = {124},
          doi = {10.1088/0004-637X/764/2/124},
archivePrefix = {arXiv},
       eprint = {1301.0626},
 primaryClass = {astro-ph.GA},
       adsurl = {https://ui.adsabs.harvard.edu/abs/2013ApJ...764..124W}
}

@ARTICLE{2025MNRAS.luana,
       author = {{Padilha}, Luana N. and {Anjos}, Rita C.},
        title = "{Massive Star Clusters as sources of high-energy gamma radiation}",
      journal = {\mnras},
         year = 2025,
        month = dec,
          doi = {10.1093/mnras/staf2192},
archivePrefix = {arXiv},
       eprint = {2512.09743},
 primaryClass = {astro-ph.HE},
       adsurl = {https://ui.adsabs.harvard.edu/abs/2025MNRAS.tmp.2066P}
}

@ARTICLE{1998MNRAS.294..429D,
       author = {{Dehnen}, Walter and {Binney}, James},
        title = "{Mass models of the Milky Way}",
      journal = {\mnras},
         year = 1998,
        month = mar,
       volume = {294},
       number = {3},
        pages = {429-438},
          doi = {10.1046/j.1365-8711.1998.01282.x10.1111/j.1365-8711.1998.01282.x},
archivePrefix = {arXiv},
       eprint = {astro-ph/9612059},
 primaryClass = {astro-ph},
       adsurl = {https://ui.adsabs.harvard.edu/abs/1998MNRAS.294..429D}
}

@ARTICLE{2008ApJ...675..614P,
       author = {{Poelarends}, A.~J.~T. and {Herwig}, F. and {Langer}, N. and {Heger}, A.},
        title = "{The Supernova Channel of Super-AGB Stars}",
      journal = {\apj},
         year = 2008,
        month = mar,
       volume = {675},
       number = {1},
        pages = {614-625},
          doi = {10.1086/520872},
archivePrefix = {arXiv},
       eprint = {0705.4643},
 primaryClass = {astro-ph},
       adsurl = {https://ui.adsabs.harvard.edu/abs/2008ApJ...675..614P}
}

@ARTICLE{1962AJ.....67..471K,
       author = {{King}, Ivan},
        title = "{The structure of star clusters. I. an empirical density law}",
      journal = {\aj},
         year = 1962,
        month = oct,
       volume = {67},
        pages = {471},
          doi = {10.1086/108756},
       adsurl = {https://ui.adsabs.harvard.edu/abs/1962AJ.....67..471K}
}

@ARTICLE{2025A&A...695A.175M,
       author = {{Menchiari}, S. and {Morlino}, G. and {Amato}, E. and {Bucciantini}, N. and {Peron}, G. and {Sacco}, G.},
        title = "{Contribution of young massive stellar clusters to the Galactic diffuse {\ensuremath{\gamma}}-ray emission}",
      journal = {\aap},
         year = 2025,
        month = mar,
       volume = {695},
          eid = {A175},
        pages = {A175},
          doi = {10.1051/0004-6361/202450621},
archivePrefix = {arXiv},
       eprint = {2406.04087},
 primaryClass = {astro-ph.HE},
       adsurl = {https://ui.adsabs.harvard.edu/abs/2025A&A...695A.175M}
}

@ARTICLE{2024A&A...686A.118C,
       author = {{Celli}, S. and {Specovius}, A. and {Menchiari}, S. and {Mitchell}, A. and {Morlino}, G.},
        title = "{Mass and wind luminosity of young Galactic open clusters in Gaia DR2}",
      journal = {\aap},
         year = 2024,
        month = jun,
       volume = {686},
          eid = {A118},
        pages = {A118},
          doi = {10.1051/0004-6361/202348541},
archivePrefix = {arXiv},
       eprint = {2311.09089},
 primaryClass = {astro-ph.GA},
       adsurl = {https://ui.adsabs.harvard.edu/abs/2024A&A...686A.118C}
}

@ARTICLE{2024ApJ...972L..22P,
       author = {{Peron}, Giada and {Morlino}, Giovanni and {Gabici}, Stefano and {Amato}, Elena and {Purushothaman}, Archana and {Brusa}, Marcella},
        title = "{On the Correlation between Young Massive Star Clusters and Gamma-Ray Unassociated Sources}",
      journal = {\apjl},
         year = 2024,
        month = sep,
       volume = {972},
       number = {2},
          eid = {L22},
        pages = {L22},
          doi = {10.3847/2041-8213/ad7024},
archivePrefix = {arXiv},
       eprint = {2408.04973},
 primaryClass = {astro-ph.HE},
       adsurl = {https://ui.adsabs.harvard.edu/abs/2024ApJ...972L..22P}
}

@ARTICLE{2006A&A...445..545P,
       author = {{Piskunov}, A.~E. and {Kharchenko}, N.~V. and {R{\"o}ser}, S. and {Schilbach}, E. and {Scholz}, R.-D.},
        title = "{Revisiting the population of Galactic open clusters}",
      journal = {\aap},
         year = 2006,
        month = jan,
       volume = {445},
       number = {2},
        pages = {545-565},
          doi = {10.1051/0004-6361:20053764},
archivePrefix = {arXiv},
       eprint = {astro-ph/0508575},
 primaryClass = {astro-ph},
       adsurl = {https://ui.adsabs.harvard.edu/abs/2006A&A...445..545P}
}

@ARTICLE{1991MNRAS.249...76B,
       author = {{Battinelli}, Paolo and {Capuzzo-Dolcetta}, Roberto},
        title = "{Formation and evolutionary properties of the galactic open cluster system.}",
      journal = {\mnras},
         year = 1991,
        month = mar,
       volume = {249},
        pages = {76},
          doi = {10.1093/mnras/249.1.76},
       adsurl = {https://ui.adsabs.harvard.edu/abs/1991MNRAS.249...76B}
}

@ARTICLE{2003APh....19..649M,
       author = {{Meli}, A. and {Quenby}, J.~J.},
        title = "{Particle acceleration in ultra-relativistic oblique shock waves}",
      journal = {Astroparticle Physics},
         year = 2003,
        month = aug,
       volume = {19},
       number = {5},
        pages = {649-666},
          doi = {10.1016/S0927-6505(02)00257-8},
archivePrefix = {arXiv},
       eprint = {astro-ph/0212329},
 primaryClass = {astro-ph},
       adsurl = {https://ui.adsabs.harvard.edu/abs/2003APh....19..649M}
}

@ARTICLE{2011MNRAS.418.1208B,
       author = {{Bell}, A.~R. and {Schure}, K.~M. and {Reville}, B.},
        title = "{Cosmic ray acceleration at oblique shocks}",
      journal = {\mnras},
         year = 2011,
        month = dec,
       volume = {418},
       number = {2},
        pages = {1208-1216},
          doi = {10.1111/j.1365-2966.2011.19571.x},
archivePrefix = {arXiv},
       eprint = {1108.0582},
 primaryClass = {astro-ph.HE},
       adsurl = {https://ui.adsabs.harvard.edu/abs/2011MNRAS.418.1208B}
}

@ARTICLE{2022RvMPP...6...29A,
       author = {{Amano}, Takanobu and {Matsumoto}, Yosuke and {Bohdan}, Artem and {Kobzar}, Oleh and {Matsukiyo}, Shuichi and {Oka}, Mitsuo and {Niemiec}, Jacek and {Pohl}, Martin and {Hoshino}, Masahiro},
        title = "{Nonthermal electron acceleration at collisionless quasi-perpendicular shocks}",
      journal = {Reviews of Modern Plasma Physics},
         year = 2022,
        month = dec,
       volume = {6},
       number = {1},
          eid = {29},
        pages = {29},
          doi = {10.1007/s41614-022-00093-1},
archivePrefix = {arXiv},
       eprint = {2209.03521},
 primaryClass = {astro-ph.HE},
       adsurl = {https://ui.adsabs.harvard.edu/abs/2022RvMPP...6...29A}
}

@INPROCEEDINGS{2024hegr.confE..16G,
       author = {{Gabici}, S.},
        title = "{Star clusters as cosmic ray accelerators}",
    booktitle = {7th Heidelberg International Symposium on High-Energy Gamma-Ray Astronomy},
         year = 2024,
        month = dec,
        pages = {16},
          doi = {10.48550/arXiv.2301.06505},
archivePrefix = {arXiv},
       eprint = {2301.06505},
 primaryClass = {astro-ph.HE},
       adsurl = {https://ui.adsabs.harvard.edu/abs/2024hegr.confE..16G}
}

@ARTICLE{2025JCoPh.52313690S,
       author = {{Schween}, Nils W. and {Schulze}, Florian and {Reville}, Brian},
        title = "{Sapphire++: A particle transport code combining a spherical harmonic expansion and the discontinuous Galerkin method}",
      journal = {Journal of Computational Physics},
         year = 2025,
        month = feb,
       volume = {523},
          eid = {113690},
        pages = {113690},
          doi = {10.1016/j.jcp.2024.113690},
archivePrefix = {arXiv},
       eprint = {2501.05110},
 primaryClass = {astro-ph.HE},
       adsurl = {https://ui.adsabs.harvard.edu/abs/2025JCoPh.52313690S}
}

@ARTICLE{2022MNRAS.512.1275V,
       author = {{Vieu}, T. and {Gabici}, S. and {Tatischeff}, V. and {Ravikularaman}, S.},
        title = "{Cosmic ray production in superbubbles}",
      journal = {\mnras},
         year = 2022,
        month = may,
       volume = {512},
       number = {1},
        pages = {1275-1293},
          doi = {10.1093/mnras/stac543},
archivePrefix = {arXiv},
       eprint = {2201.07488},
 primaryClass = {astro-ph.HE},
       adsurl = {https://ui.adsabs.harvard.edu/abs/2022MNRAS.512.1275V}
}

@ARTICLE{1983A&A...125..249L,
       author = {{Lagage}, P.~O. and {Cesarsky}, C.~J.},
        title = "{The maximum energy of cosmic rays accelerated by supernova shocks.}",
      journal = {\aap},
         year = 1983,
        month = sep,
       volume = {125},
        pages = {249-257},
       adsurl = {https://ui.adsabs.harvard.edu/abs/1983A&A...125..249L}
}

@ARTICLE{2005JPhG...31R..95H,
       author = {{Hillas}, A.~M.},
        title = "{TOPICAL REVIEW:  Can diffusive shock acceleration in supernova remnants account for high-energy galactic cosmic rays?}",
      journal = {Journal of Physics G Nuclear Physics},
         year = 2005,
        month = may,
       volume = {31},
       number = {5},
        pages = {R95-R131},
          doi = {10.1088/0954-3899/31/5/R02},
       adsurl = {https://ui.adsabs.harvard.edu/abs/2005JPhG...31R..95H}
}

@ARTICLE{2015A&A...573A..95M,
       author = {{Massi}, F. and {Giannetti}, A. and {Di Carlo}, E. and {Brand}, J. and {Beltr{\'a}n}, M.~T. and {Marconi}, G.},
        title = "{Young open clusters in the Galactic star forming region NGC 6357}",
      journal = {\aap},
         year = 2015,
        month = jan,
       volume = {573},
          eid = {A95},
        pages = {A95},
          doi = {10.1051/0004-6361/201424388},
archivePrefix = {arXiv},
       eprint = {1410.4340},
 primaryClass = {astro-ph.SR},
       adsurl = {https://ui.adsabs.harvard.edu/abs/2015A&A...573A..95M}
}

@INPROCEEDINGS{2015EAS....75...43L,
       author = {{Longmore}, S. and {Barnes}, A. and {Battersby}, C. and {Bally}, J. and {Kruijssen}, J.~M. Diederik and {Dale}, J. and {Henshaw}, J. and {Walker}, D. and {Rathborne}, J. and {Testi}, L. and {Ott}, J. and {Ginsburg}, A.},
        title = "{Using young massive star clusters to understand star formation and feedback in high-redshift-like environments}",
    booktitle = {EAS Publications Series},
         year = 2015,
       series = {EAS Publications Series},
       volume = {75-76},
        month = may,
    publisher = {EDP},
        pages = {43-48},
          doi = {10.1051/eas/1575006},
archivePrefix = {arXiv},
       eprint = {1601.02654},
 primaryClass = {astro-ph.GA},
       adsurl = {https://ui.adsabs.harvard.edu/abs/2015EAS....75...43L}
}

@ARTICLE{2024JApA...45...17S,
       author = {{Soam}, Archana and {Eswaraiah}, Chakali and {Seta}, Amit and {Dewangan}, Lokesh and {Maheswar}, G.},
        title = "{Turbulence and magnetic fields in star formation}",
      journal = {Journal of Astrophysics and Astronomy},
         year = 2024,
        month = jun,
       volume = {45},
       number = {1},
          eid = {17},
        pages = {17},
          doi = {10.1007/s12036-024-10005-z},
archivePrefix = {arXiv},
       eprint = {2402.18840},
 primaryClass = {astro-ph.GA},
       adsurl = {https://ui.adsabs.harvard.edu/abs/2024JApA...45...17S}
}

@ARTICLE{2016A&A...591A..71M,
       author = {{Maurin}, G. and {Marcowith}, A. and {Komin}, N. and {Krayzel}, F. and {Lamanna}, G.},
        title = "{Embedded star clusters as sources of high-energy cosmic rays . Modelling and constraints}",
      journal = {\aap},
         year = 2016,
        month = jun,
       volume = {591},
          eid = {A71},
        pages = {A71},
          doi = {10.1051/0004-6361/201628465},
archivePrefix = {arXiv},
       eprint = {1605.04202},
 primaryClass = {astro-ph.HE},
       adsurl = {https://ui.adsabs.harvard.edu/abs/2016A&A...591A..71M}
}

@ARTICLE{2003APh....19..193H,
       author = {{H{\"o}randel}, J{\"o}rg R.},
        title = "{On the knee in the energy spectrum of cosmic rays}",
      journal = {Astroparticle Physics},
         year = 2003,
        month = may,
       volume = {19},
       number = {2},
        pages = {193-220},
          doi = {10.1016/S0927-6505(02)00198-6},
archivePrefix = {arXiv},
       eprint = {astro-ph/0210453},
 primaryClass = {astro-ph},
       adsurl = {https://ui.adsabs.harvard.edu/abs/2003APh....19..193H}
}

@ARTICLE{2003PASP..115..763C,
       author = {{Chabrier}, Gilles},
        title = "{Galactic Stellar and Substellar Initial Mass Function}",
      journal = {\pasp},
         year = 2003,
        month = jul,
       volume = {115},
       number = {809},
        pages = {763-795},
          doi = {10.1086/376392},
archivePrefix = {arXiv},
       eprint = {astro-ph/0304382},
 primaryClass = {astro-ph},
       adsurl = {https://ui.adsabs.harvard.edu/abs/2003PASP..115..763C}
}

@ARTICLE{1955ApJ...121..161S,
       author = {{Salpeter}, Edwin E.},
        title = "{The Luminosity Function and Stellar Evolution.}",
      journal = {\apj},
         year = 1955,
        month = jan,
       volume = {121},
        pages = {161},
          doi = {10.1086/145971},
       adsurl = {https://ui.adsabs.harvard.edu/abs/1955ApJ...121..161S}
}

@ARTICLE{2009MNRAS.395.1409S,
       author = {{Smartt}, S.~J. and {Eldridge}, J.~J. and {Crockett}, R.~M. and {Maund}, J.~R.},
        title = "{The death of massive stars - I. Observational constraints on the progenitors of Type II-P supernovae}",
      journal = {\mnras},
         year = 2009,
        month = may,
       volume = {395},
       number = {3},
        pages = {1409-1437},
          doi = {10.1111/j.1365-2966.2009.14506.x},
archivePrefix = {arXiv},
       eprint = {0809.0403},
 primaryClass = {astro-ph},
       adsurl = {https://ui.adsabs.harvard.edu/abs/2009MNRAS.395.1409S}
}

@ARTICLE{2021MNRAS.506.4131W,
       author = {{Wirth}, Henriette and {Jerabkova}, Tereza and {Yan}, Zhiqiang and {Kroupa}, Pavel and {Haas}, Jaroslav and {{\v{S}}ubr}, Ladislav},
        title = "{How many explosions does one need? Quantifying supernovae in globular clusters from iron abundance spreads}",
      journal = {\mnras},
         year = 2021,
        month = sep,
       volume = {506},
       number = {3},
        pages = {4131-4138},
          doi = {10.1093/mnras/stab2011},
archivePrefix = {arXiv},
       eprint = {2107.06240},
 primaryClass = {astro-ph.GA},
       adsurl = {https://ui.adsabs.harvard.edu/abs/2021MNRAS.506.4131W}
}

@ARTICLE{2021EPJC...81..966A,
       author = {{Abreu}, P. and {Aglietta}, M. and {Albury}, J.~M. and {Allekotte}, I. and {Almela}, A. and {Alvarez-Mu{\~n}iz}, J. and {Alves Batista}, R. and {Anastasi}, G.~A. and {Anchordoqui}, L. and {Andrada}, B. and {Andringa}, S. and {Aramo}, C. and {Ara{\'u}jo Ferreira}, P.~R. and {Arteaga Vel{\'a}zquez}, J.~C. and {Asorey}, H. and {Assis}, P. and {Avila}, G. and {Badescu}, A.~M. and {Bakalova}, A. and {Balaceanu}, A. and {Barbato}, F. and {Barreira Luz}, R.~J. and {Becker}, K.~H. and {Bellido}, J.~A. and {Berat}, C. and {Bertaina}, M.~E. and {Bertou}, X. and {Biermann}, P.~L. and {Billoir}, P. and {Binet}, V. and {Bismark}, K. and {Bister}, T. and {Biteau}, J. and {Blazek}, J. and {Bleve}, C. and {Boh{\'a}{\v{c}}ov{\'a}}, M. and {Boncioli}, D. and {Bonifazi}, C. and {Bonneau Arbeletche}, L. and {Borodai}, N. and {Botti}, A.~M. and {Brack}, J. and {Bretz}, T. and {Brichetto Orchera}, P.~G. and {Briechle}, F.~L. and {Buchholz}, P. and {Bueno}, A. and {Buitink}, S. and {Buscemi}, M. and {B{\"u}sken}, M. and {Caballero-Mora}, K.~S. and {Caccianiga}, L. and {Canfora}, F. and {Caracas}, I. and {Carceller}, J.~M. and {Caruso}, R. and {Castellina}, A. and {Catalani}, F. and {Cataldi}, G. and {Cazon}, L. and {Cerda}, M. and {Chinellato}, J.~A. and {Chudoba}, J. and {Chytka}, L. and {Clay}, R.~W. and {Cobos Cerutti}, A.~C. and {Colalillo}, R. and {Coleman}, A. and {Coluccia}, M.~R. and {Concei{\c{c}}{\~a}o}, R. and {Condorelli}, A. and {Consolati}, G. and {Contreras}, F. and {Convenga}, F. and {Correia dos Santos}, D. and {Covault}, C.~E. and {Dasso}, S. and {Daumiller}, K. and {Dawson}, B.~R. and {Day}, J.~A. and {de Almeida}, R.~M. and {de Jes{\'u}s}, J. and {de Jong}, S.~J. and {De Mauro}, G. and {de Mello Neto}, J.~R.~T. and {De Mitri}, I. and {de Oliveira}, J. and {de Oliveira Franco}, D. and {de Palma}, F. and {de Souza}, V. and {De Vito}, E. and {del R{\'\i}o}, M. and {Deligny}, O. and {Di Matteo}, A. and {Dobrigkeit}, C. and {D'Olivo}, J.~C. and {Domingues Mendes}, L.~M. and {dos Anjos}, R.~C. and {dos Santos}, D. and {Dova}, M.~T. and {Ebr}, J. and {Engel}, R. and {Epicoco}, I. and {Erdmann}, M. and {Escobar}, C.~O. and {Etchegoyen}, A. and {Falcke}, H. and {Farmer}, J. and {Farrar}, G. and {Fauth}, A.~C. and {Fazzini}, N. and {Feldbusch}, F. and {Fenu}, F. and {Fick}, B. and {Figueira}, J.~M. and {Filip{\v{c}}i{\v{c}}}, A. and {Fitoussi}, T. and {Fodran}, T. and {Freire}, M.~M. and {Fujii}, T. and {Fuster}, A. and {Galea}, C. and {Galelli}, C. and {Garc{\'\i}a}, B. and {Garcia Vegas}, A.~L. and {Gemmeke}, H. and {Gesualdi}, F. and {Gherghel-Lascu}, A. and {Ghia}, P.~L. and {Giaccari}, U. and {Giammarchi}, M. and {Glombitza}, J. and {Gobbi}, F. and {Gollan}, F. and {Golup}, G. and {G{\'o}mez Berisso}, M. and {G{\'o}mez Vitale}, P.~F. and {Gongora}, J.~P. and {Gonz{\'a}lez}, J.~M. and {Gonz{\'a}lez}, N. and {Goos}, I. and {G{\'o}ra}, D. and {Gorgi}, A. and {Gottowik}, M. and {Grubb}, T.~D. and {Guarino}, F. and {Guedes}, G.~P. and {Guido}, E. and {Hahn}, S. and {Hamal}, P. and {Hampel}, M.~R. and {Hansen}, P. and {Harari}, D. and {Harvey}, V.~M. and {Haungs}, A. and {Hebbeker}, T. and {Heck}, D. and {Hill}, G.~C. and {Hojvat}, C. and {H{\"o}randel}, J.~R. and {Horvath}, P. and {Hrabovsk{\'y}}, M. and {Huege}, T. and {Insolia}, A. and {Isar}, P.~G. and {Janecek}, P. and {Johnsen}, J.~A. and {Jurysek}, J. and {K{\"a}{\"a}p{\"a}}, A. and {Kampert}, K.~H. and {Karastathis}, N. and {Keilhauer}, B. and {Kemp}, J. and {Khakurdikar}, A. and {Kizakke Covilakam}, V.~V. and {Klages}, H.~O. and {Kleifges}, M. and {Kleinfeller}, J. and {K{\"o}pke}, M. and {Kunka}, N. and {Lago}, B.~L. and {Lang}, R.~G. and {Langner}, N. and {Leigui de Oliveira}, M.~A. and {Lenok}, V. and {Letessier-Selvon}, A. and {Lhenry-Yvon}, I. and {Lo Presti}, D. and {Lopes}, L. and {L{\'o}pez}, R. and {Lu}, L. and {Luce}, Q. and {Lundquist}, J.~P. and {Machado Payeras}, A. and {Mancarella}, G. and {Mandat}, D. and {Manning}, B.~C. and {Manshanden}, J. and {Mantsch}, P. and {Marafico}, S.},
        title = "{The energy spectrum of cosmic rays beyond the turn-down around {}10$^{17}$ eV as measured with the surface detector of the Pierre Auger Observatory}",
      journal = {European Physical Journal C},
         year = 2021,
        month = nov,
       volume = {81},
       number = {11},
          eid = {966},
        pages = {966},
          doi = {10.1140/epjc/s10052-021-09700-w},
archivePrefix = {arXiv},
       eprint = {2109.13400},
 primaryClass = {astro-ph.HE},
       adsurl = {https://ui.adsabs.harvard.edu/abs/2021EPJC...81..966A}
}

@ARTICLE{2007ARNPS..57..285S,
       author = {{Strong}, Andrew W. and {Moskalenko}, Igor V. and {Ptuskin}, Vladimir S.},
        title = "{Cosmic-Ray Propagation and Interactions in the Galaxy}",
      journal = {Annual Review of Nuclear and Particle Science},
         year = 2007,
        month = nov,
       volume = {57},
       number = {1},
        pages = {285-327},
          doi = {10.1146/annurev.nucl.57.090506.123011},
archivePrefix = {arXiv},
       eprint = {astro-ph/0701517},
 primaryClass = {astro-ph},
       adsurl = {https://ui.adsabs.harvard.edu/abs/2007ARNPS..57..285S}
}

@ARTICLE{2011ApJ...729..106T,
       author = {{Trotta}, R. and {J{\'o}hannesson}, G. and {Moskalenko}, I.~V. and {Porter}, T.~A. and {Ruiz de Austri}, R. and {Strong}, A.~W.},
        title = "{Constraints on Cosmic-ray Propagation Models from A Global Bayesian Analysis}",
      journal = {\apj},
         year = 2011,
        month = mar,
       volume = {729},
       number = {2},
          eid = {106},
        pages = {106},
          doi = {10.1088/0004-637X/729/2/106},
archivePrefix = {arXiv},
       eprint = {1011.0037},
 primaryClass = {astro-ph.HE},
       adsurl = {https://ui.adsabs.harvard.edu/abs/2011ApJ...729..106T}
}

@ARTICLE{2016ApJ...824...16J,
       author = {{J{\'o}hannesson}, G. and {Ruiz de Austri}, R. and {Vincent}, A.~C. and {Moskalenko}, I.~V. and {Orlando}, E. and {Porter}, T.~A. and {Strong}, A.~W. and {Trotta}, R. and {Feroz}, F. and {Graff}, P. and {Hobson}, M.~P.},
        title = "{Bayesian Analysis of Cosmic Ray Propagation: Evidence against Homogeneous Diffusion}",
      journal = {\apj},
         year = 2016,
        month = jun,
       volume = {824},
       number = {1},
          eid = {16},
        pages = {16},
          doi = {10.3847/0004-637X/824/1/16},
archivePrefix = {arXiv},
       eprint = {1602.02243},
 primaryClass = {astro-ph.HE},
       adsurl = {https://ui.adsabs.harvard.edu/abs/2016ApJ...824...16J}
}

@ARTICLE{2024A&A...692A..20R,
       author = {{Recchia}, S. and {Gabici}, S.},
        title = "{Origin of the spectral features observed in the cosmic-ray spectrum}",
      journal = {\aap},
         year = 2024,
        month = dec,
       volume = {692},
          eid = {A20},
        pages = {A20},
          doi = {10.1051/0004-6361/202349005},
archivePrefix = {arXiv},
       eprint = {2312.11397},
 primaryClass = {astro-ph.HE},
       adsurl = {https://ui.adsabs.harvard.edu/abs/2024A&A...692A..20R}
}

@ARTICLE{2021MNRAS.504.6096M,
       author = {{Morlino}, G. and {Blasi}, P. and {Peretti}, E. and {Cristofari}, P.},
        title = "{Particle acceleration in winds of star clusters}",
      journal = {\mnras},
         year = 2021,
        month = jul,
       volume = {504},
       number = {4},
        pages = {6096-6105},
          doi = {10.1093/mnras/stab690},
archivePrefix = {arXiv},
       eprint = {2102.09217},
 primaryClass = {astro-ph.HE},
       adsurl = {https://ui.adsabs.harvard.edu/abs/2021MNRAS.504.6096M}
}

@ARTICLE{2024arXiv240316650M,
       author = {{Mitchell}, Alison M.~W. and {Morlino}, Giovanni and {Celli}, Silvia and {Menchiari}, Stefano and {Specovius}, Andreas},
        title = "{Probing Stellar Clusters from Gaia DR2 as Galactic PeVatrons: I -- Expected Gamma-ray and Neutrino Emission}",
      journal = {arXiv e-prints},
         year = 2024,
        month = mar,
          eid = {arXiv:2403.16650},
        pages = {arXiv:2403.16650},
          doi = {10.48550/arXiv.2403.16650},
archivePrefix = {arXiv},
       eprint = {2403.16650},
 primaryClass = {astro-ph.HE},
       adsurl = {https://ui.adsabs.harvard.edu/abs/2024arXiv240316650M}
}

@ARTICLE{2024NatAs...8..530P,
       author = {{Peron}, Giada and {Casanova}, Sabrina and {Gabici}, Stefano and {Baghmanyan}, Vardan and {Aharonian}, Felix},
        title = "{The contribution of winds from star clusters to the Galactic cosmic-ray population}",
      journal = {Nature Astronomy},
         year = 2024,
        month = apr,
       volume = {8},
        pages = {530-537},
          doi = {10.1038/s41550-023-02168-6},
archivePrefix = {arXiv},
       eprint = {2407.07509},
 primaryClass = {astro-ph.HE},
       adsurl = {https://ui.adsabs.harvard.edu/abs/2024NatAs...8..530P}
}

@ARTICLE{2009MNRAS.396.2065C,
       author = {{Caprioli}, D. and {Blasi}, P. and {Amato}, E.},
        title = "{On the escape of particles from cosmic ray modified shocks}",
      journal = {\mnras},
         year = 2009,
        month = jul,
       volume = {396},
       number = {4},
        pages = {2065-2073},
          doi = {10.1111/j.1365-2966.2008.14298.x},
archivePrefix = {arXiv},
       eprint = {0807.4259},
 primaryClass = {astro-ph},
       adsurl = {https://ui.adsabs.harvard.edu/abs/2009MNRAS.396.2065C}
}

@ARTICLE{2022MNRAS.515.2256V,
       author = {{Vieu}, T. and {Reville}, B. and {Aharonian}, F.},
        title = "{Can superbubbles accelerate ultrahigh energy protons?}",
      journal = {\mnras},
         year = 2022,
        month = sep,
       volume = {515},
       number = {2},
        pages = {2256-2265},
          doi = {10.1093/mnras/stac1901},
archivePrefix = {arXiv},
       eprint = {2207.01432},
 primaryClass = {astro-ph.HE},
       adsurl = {https://ui.adsabs.harvard.edu/abs/2022MNRAS.515.2256V}
}

@ARTICLE{2022MNRAS.517.2818B,
       author = {{Badmaev}, D.~V. and {Bykov}, A.~M. and {Kalyashova}, M.~E.},
        title = "{Inside the core of a young massive star cluster: 3D MHD simulations}",
      journal = {\mnras},
         year = 2022,
        month = dec,
       volume = {517},
       number = {2},
        pages = {2818-2830},
          doi = {10.1093/mnras/stac2738},
archivePrefix = {arXiv},
       eprint = {2209.11465},
 primaryClass = {astro-ph.HE},
       adsurl = {https://ui.adsabs.harvard.edu/abs/2022MNRAS.517.2818B}
}

@INPROCEEDINGS{2019ICRC...36..450V,
       author = {{Verzi}, V.},
        title = "{Measurement of the energy spectrum of ultra-high energy cosmic rays using the Pierre Auger Observatory}",
    booktitle = {36th International Cosmic Ray Conference (ICRC2019)},
         year = 2019,
       series = {International Cosmic Ray Conference},
       volume = {36},
        month = jul,
          eid = {450},
        pages = {450},
          doi = {10.22323/1.358.0450},
       adsurl = {https://ui.adsabs.harvard.edu/abs/2019ICRC...36..450V}
}

@INPROCEEDINGS{2015ICRC...34..359B,
       author = {{Bertaina}, M. and {Apel}, W.~D. and {Arteaga-Vel{\'a}zquez}, J.~C. and {Bekk}, K. and {Bl{\"u}mer}, J. and {Bozdog}, H. and {Brancus}, I.~M. and {Cantoni}, E. and {Chiavassa}, A. and {Cossavella}, F. and {Daumiller}, K. and {de Souza}, V. and {di Pierro}, F. and {Doll}, P. and {Engel}, R. and {Fuhrmann}, D. and {Gherghel-Lascu}, A. and {Gils}, H.~J. and {Glasstetter}, R. and {Grupen}, C. and {Haungs}, A. and {Heck}, D. and {H{\"o}randel}, J.~R. and {Huber}, D. and {Huege}, T. and {Kampert}, K.-H. and {Kang}, D. and {Klages}, H.~O. and {Link}, K. and {{\L}uczak}, P. and {Mathes}, H.~J. and {Mayer}, H.~J. and {Milke}, J. and {Mitrica}, B. and {Morello}, C. and {Oehlschl{\"a}ger}, J. and {Ostapchenko}, S. and {Palmieri}, N. and {Pierog}, T. and {Rebel}, H. and {Roth}, M. and {Schieler}, H. and {Schoo}, S. and {Schr{\"o}der}, F.~G. and {Sima3}, O. and {Toma}, G. and {Trinchero}, G.~C. and {Ulrich}, H. and {Weindl}, A. and {Wochele}, J. and {Zabierowski}, J. and {KASCADE-Grande Collaboration}},
        title = "{KASCADE-Grande energy spectrum of cosmic rays interpreted with post-LHC hadronic interaction models}",
    booktitle = {34th International Cosmic Ray Conference (ICRC2015)},
         year = 2015,
       series = {International Cosmic Ray Conference},
       volume = {34},
        month = jul,
          eid = {359},
        pages = {359},
          doi = {10.22323/1.236.0359},
       adsurl = {https://ui.adsabs.harvard.edu/abs/2015ICRC...34..359B}
}

@ARTICLE{2019PhRvD.100h2002A,
       author = {{Aartsen}, M.~G. and {Ackermann}, M. and {Adams}, J. and {Aguilar}, J.~A. and {Ahlers}, M. and {Ahrens}, M. and {Alispach}, C. and {Andeen}, K. and {Anderson}, T. and {Ansseau}, I. and {Anton}, G. and {Arg{\"u}elles}, C. and {Auffenberg}, J. and {Axani}, S. and {Backes}, P. and {Bagherpour}, H. and {Bai}, X. and {Barbano}, A. and {Barwick}, S.~W. and {Baum}, V. and {Baur}, S. and {Bay}, R. and {Beatty}, J.~J. and {Becker}, K.-H. and {Becker Tjus}, J. and {BenZvi}, S. and {Berley}, D. and {Bernardini}, E. and {Besson}, D.~Z. and {Binder}, G. and {Bindig}, D. and {Blaufuss}, E. and {Blot}, S. and {Bohm}, C. and {B{\"o}rner}, M. and {B{\"o}ser}, S. and {Botner}, O. and {B{\"o}ttcher}, J. and {Bourbeau}, E. and {Bourbeau}, J. and {Bradascio}, F. and {Braun}, J. and {Bretz}, H.-P. and {Bron}, S. and {Brostean-Kaiser}, J. and {Burgman}, A. and {Buscher}, J. and {Busse}, R.~S. and {Carver}, T. and {Chen}, C. and {Cheung}, E. and {Chirkin}, D. and {Clark}, K. and {Classen}, L. and {Collin}, G.~H. and {Conrad}, J.~M. and {Coppin}, P. and {Correa}, P. and {Cowen}, D.~F. and {Cross}, R. and {Dave}, P. and {de Andr{\'e}}, J.~P.~A.~M. and {De Clercq}, C. and {DeLaunay}, J.~J. and {Dembinski}, H. and {Deoskar}, K. and {De Ridder}, S. and {Desiati}, P. and {de Vries}, K.~D. and {de Wasseige}, G. and {de With}, M. and {DeYoung}, T. and {Diaz}, A. and {D{\'\i}az-V{\'e}lez}, J.~C. and {Dujmovic}, H. and {Dunkman}, M. and {Dvorak}, E. and {Eberhardt}, B. and {Ehrhardt}, T. and {Eller}, P. and {Evenson}, P.~A. and {Fahey}, S. and {Fazely}, A.~R. and {Felde}, J. and {Feusels}, T. and {Filimonov}, K. and {Finley}, C. and {Franckowiak}, A. and {Friedman}, E. and {Fritz}, A. and {Gaisser}, T.~K. and {Gallagher}, J. and {Ganster}, E. and {Garrappa}, S. and {Gerhardt}, L. and {Ghorbani}, K. and {Glauch}, T. and {Gl{\"u}senkamp}, T. and {Goldschmidt}, A. and {Gonzalez}, J.~G. and {Grant}, D. and {Griffith}, Z. and {G{\"u}nder}, M. and {G{\"u}nd{\"u}z}, M. and {Haack}, C. and {Hallgren}, A. and {Halve}, L. and {Halzen}, F. and {Hanson}, K. and {Hebecker}, D. and {Heereman}, D. and {Heix}, P. and {Helbing}, K. and {Hellauer}, R. and {Henningsen}, F. and {Hickford}, S. and {Hignight}, J. and {Hill}, G.~C. and {Hoffman}, K.~D. and {Hoffmann}, R. and {Hoinka}, T. and {Hokanson-Fasig}, B. and {Hoshina}, K. and {Huang}, F. and {Huber}, M. and {Hultqvist}, K. and {H{\"u}nnefeld}, M. and {Hussain}, R. and {In}, S. and {Iovine}, N. and {Ishihara}, A. and {Jacobi}, E. and {Japaridze}, G.~S. and {Jeong}, M. and {Jero}, K. and {Jones}, B.~J.~P. and {Jonske}, F. and {Joppe}, R. and {Kang}, W. and {Kappes}, A. and {Kappesser}, D. and {Karg}, T. and {Karl}, M. and {Karle}, A. and {Katz}, U. and {Kauer}, M. and {Kelley}, J.~L. and {Kheirandish}, A. and {Kim}, J. and {Kintscher}, T. and {Kiryluk}, J. and {Kittler}, T. and {Klein}, S.~R. and {Koirala}, R. and {Kolanoski}, H. and {K{\"o}pke}, L. and {Kopper}, C. and {Kopper}, S. and {Koskinen}, D.~J. and {Kowalski}, M. and {Krings}, K. and {Kr{\"u}ckl}, G. and {Kulacz}, N. and {Kunwar}, S. and {Kurahashi}, N. and {Kyriacou}, A. and {Labare}, M. and {Lanfranchi}, J.~L. and {Larson}, M.~J. and {Lauber}, F. and {Lazar}, J.~P. and {Leonard}, K. and {Leuermann}, M. and {Liu}, Q.~R. and {Lohfink}, E. and {Lozano Mariscal}, C.~J. and {Lu}, L. and {Lucarelli}, F. and {L{\"u}nemann}, J. and {Luszczak}, W. and {Madsen}, J. and {Maggi}, G. and {Mahn}, K.~B.~M. and {Makino}, Y. and {Mallik}, P. and {Mallot}, K. and {Mancina}, S. and {Mari{\c{s}}}, I.~C. and {Maruyama}, R. and {Mase}, K. and {Maunu}, R. and {Meagher}, K. and {Medici}, M. and {Medina}, A. and {Meier}, M. and {Meighen-Berger}, S. and {Menne}, T. and {Merino}, G. and {Meures}, T. and {Miarecki}, S.},
        title = "{Cosmic ray spectrum and composition from PeV to EeV using 3 years of data from IceTop and IceCube}",
      journal = {\prd},
         year = 2019,
        month = oct,
       volume = {100},
       number = {8},
          eid = {082002},
        pages = {082002},
          doi = {10.1103/PhysRevD.100.082002},
archivePrefix = {arXiv},
       eprint = {1906.04317},
 primaryClass = {astro-ph.HE},
       adsurl = {https://ui.adsabs.harvard.edu/abs/2019PhRvD.100h2002A}
}

@ARTICLE{2020APh...11702406B,
       author = {{Budnev}, N.~M. and {Chiavassa}, A. and {Gress}, O.~A. and {Gress}, T.~I. and {Dyachok}, A.~N. and {Karpov}, N.~I. and {Kalmykov}, N.~N. and {Korosteleva}, E.~E. and {Kozhin}, V.~A. and {Kuzmichev}, L.~A. and {Lubsandorzhiev}, B.~K. and {Lubsandorzhiev}, N.~B. and {Mirgazov}, R.~R. and {Osipova}, E.~A. and {Panasyuk}, M.~I. and {Pankov}, L.~V. and {Popova}, E.~G. and {Prosin}, V.~V. and {Ptuskin}, V.~S. and {Semeney}, Yu. A. and {Silaev}, A.~A. and {Silaev(junior)}, A.~A. and {Skurikhin}, A.~V. and {Spiering}, C. and {Sveshnikova}, L.~G. and {Zagorodnikov}, A.~V.},
        title = "{The primary cosmic-ray energy spectrum measured with the Tunka-133 array}",
      journal = {Astroparticle Physics},
         year = 2020,
        month = jan,
       volume = {117},
          eid = {102406},
        pages = {102406},
          doi = {10.1016/j.astropartphys.2019.102406},
archivePrefix = {arXiv},
       eprint = {2104.03599},
 primaryClass = {astro-ph.HE},
       adsurl = {https://ui.adsabs.harvard.edu/abs/2020APh...11702406B}
}

@ARTICLE{2017PhRvD..96l2001A,
       author = {{Alfaro}, R. and {Alvarez}, C. and {{\'A}lvarez}, J.~D. and {Arceo}, R. and {Arteaga-Vel{\'a}zquez}, J.~C. and {Avila Rojas}, D. and {Ayala Solares}, H.~A. and {Barber}, A.~S. and {Becerril}, A. and {Belmont-Moreno}, E. and {BenZvi}, S.~Y. and {Brisbois}, C. and {Caballero-Mora}, K.~S. and {Capistr{\'a}n}, T. and {Carrami{\~n}ana}, A. and {Casanova}, S. and {Castillo}, M. and {Cotti}, U. and {Cotzomi}, J. and {Couti{\~n}o de Le{\'o}n}, S. and {De Le{\'o}n}, C. and {De la Fuente}, E. and {Diaz Hernandez}, R. and {Dichiara}, S. and {Dingus}, B.~L. and {DuVernois}, M.~A. and {D{\'\i}az-V{\'e}lez}, J.~C. and {Ellsworth}, R.~W. and {Enriquez-Rivera}, O. and {Fiorino}, D.~W. and {Fleischhack}, H. and {Fraija}, N. and {Garc{\'\i}a-Gonz{\'a}lez}, J.~A. and {Gonz{\'a}lez Mu{\~n}oz}, A. and {Gonz{\'a}lez}, M.~M. and {Goodman}, J.~A. and {Hampel-Arias}, Z. and {Harding}, J.~P. and {Hernandez-Almada}, A. and {Hinton}, J. and {Hueyotl-Zahuantitla}, F. and {Hui}, C.~M. and {H{\"u}ntemeyer}, P. and {Iriarte}, A. and {Jardin-Blicq}, A. and {Joshi}, V. and {Kaufmann}, S. and {Lara}, A. and {Lauer}, R.~J. and {Lennarz}, D. and {Le{\'o}n Vargas}, H. and {Linnemann}, J.~T. and {Longinotti}, A.~L. and {Luis Raya}, G. and {Luna-Garc{\'\i}a}, R. and {L{\'o}pez-C{\'a}mara}, D. and {L{\'o}pez-Coto}, R. and {Malone}, K. and {Marinelli}, S.~S. and {Martinez}, O. and {Martinez-Castellanos}, I. and {Mart{\'\i}nez-Castro}, J. and {Mart{\'\i}nez-Huerta}, H. and {Matthews}, J.~A. and {Miranda-Romagnoli}, P. and {Moreno}, E. and {Mostaf{\'a}}, M. and {Nellen}, L. and {Newbold}, M. and {Nisa}, M.~U. and {Noriega-Papaqui}, R. and {Pelayo}, R. and {Pretz}, J. and {P{\'e}rez-P{\'e}rez}, E.~G. and {Ren}, Z. and {Rho}, C.~D. and {Rivi{\`e}re}, C. and {Rosa-Gonz{\'a}lez}, D. and {Rosenberg}, M. and {Ruiz-Velasco}, E. and {Salesa Greus}, F. and {Sandoval}, A. and {Schneider}, M. and {Schoorlemmer}, H. and {Sinnis}, G. and {Smith}, A.~J. and {Springer}, R.~W. and {Surajbali}, P. and {Taboada}, I. and {Tibolla}, O. and {Tollefson}, K. and {Torres}, I. and {Ukwatta}, T.~N. and {Villase{\~n}or}, L. and {Weisgarber}, T. and {Westerhoff}, S. and {Wood}, J. and {Yapici}, T. and {Zepeda}, A. and {Zhou}, H. and {HAWC Collaboration}},
        title = "{All-particle cosmic ray energy spectrum measured by the HAWC experiment from 10 to 500 TeV}",
      journal = {\prd},
         year = 2017,
        month = dec,
       volume = {96},
       number = {12},
          eid = {122001},
        pages = {122001},
          doi = {10.1103/PhysRevD.96.122001},
archivePrefix = {arXiv},
       eprint = {1710.00890},
 primaryClass = {astro-ph.HE},
       adsurl = {https://ui.adsabs.harvard.edu/abs/2017PhRvD..96l2001A}
}

@misc{The_CR_spectrum,
  author       = {Evoli, Carmelo},
  title        = {The Cosmic-Ray Energy Spectrum},
  month        = dec,
  year         = 2020,
  publisher    = {Zenodo},
  doi          = {10.5281/zenodo.4396125},
  url          = {https://doi.org/10.5281/zenodo.4396125}
}

@ARTICLE{2018A&A...611A..77Y,
       author = {{Yang}, Rui-zhi and {de O{\~n}a Wilhelmi}, Emma and {Aharonian}, Felix},
        title = "{Diffuse {\ensuremath{\gamma}}-ray emission in the vicinity of young star cluster Westerlund 2}",
      journal = {\aap},
         year = 2018,
        month = apr,
       volume = {611},
          eid = {A77},
        pages = {A77},
          doi = {10.1051/0004-6361/201732045},
archivePrefix = {arXiv},
       eprint = {1710.02803},
 primaryClass = {astro-ph.HE},
       adsurl = {https://ui.adsabs.harvard.edu/abs/2018A&A...611A..77Y}
}

@ARTICLE{2021NatAs...5..465A,
       author = {{Abeysekara}, A.~U. and {Albert}, A. and {Alfaro}, R. and {Alvarez}, C. and {Camacho}, J.~R. Angeles and {Arteaga-Vel{\'a}zquez}, J.~C. and {Arunbabu}, K.~P. and {Rojas}, D. Avila and {Solares}, H.~A. Ayala and {Baghmanyan}, V. and {Belmont-Moreno}, E. and {BenZvi}, S.~Y. and {Blandford}, R. and {Brisbois}, C. and {Caballero-Mora}, K.~S. and {Capistr{\'a}n}, T. and {Carrami{\~n}ana}, A. and {Casanova}, S. and {Cotti}, U. and {Le{\'o}n}, S. Couti{\~n}o de and {De la Fuente}, E. and {Hernandez}, R. Diaz and {Dingus}, B.~L. and {DuVernois}, M.~A. and {Durocher}, M. and {D{\'\i}az-V{\'e}lez}, J.~C. and {Ellsworth}, R.~W. and {Engel}, K. and {Espinoza}, C. and {Fan}, K.~L. and {Fang}, K. and {Fleischhack}, H. and {Fraija}, N. and {Galv{\'a}n-G{\'a}mez}, A. and {Garcia}, D. and {Garc{\'\i}a-Gonz{\'a}lez}, J.~A. and {Garfias}, F. and {Giacinti}, G. and {Gonz{\'a}lez}, M.~M. and {Goodman}, J.~A. and {Harding}, J.~P. and {Hernandez}, S. and {Hinton}, J. and {Hona}, B. and {Huang}, D. and {Hueyotl-Zahuantitla}, F. and {H{\"u}ntemeyer}, P. and {Iriarte}, A. and {Jardin-Blicq}, A. and {Joshi}, V. and {Kieda}, D. and {Lara}, A. and {Lee}, W.~H. and {Vargas}, H. Le{\'o}n and {Linnemann}, J.~T. and {Longinotti}, A.~L. and {Luis-Raya}, G. and {Lundeen}, J. and {Malone}, K. and {Martinez}, O. and {Martinez-Castellanos}, I. and {Mart{\'\i}nez-Castro}, J. and {Matthews}, J.~A. and {Miranda-Romagnoli}, P. and {Morales-Soto}, J.~A. and {Moreno}, E. and {Mostaf{\'a}}, M. and {Nayerhoda}, A. and {Nellen}, L. and {Newbold}, M. and {Nisa}, M.~U. and {Noriega-Papaqui}, R. and {Olivera-Nieto}, L. and {Omodei}, N. and {Peisker}, A. and {P{\'e}rez Araujo}, Y. and {P{\'e}rez-P{\'e}rez}, E.~G. and {Ren}, Z. and {Rho}, C.~D. and {Rosa-Gonz{\'a}lez}, D. and {Ruiz-Velasco}, E. and {Salazar}, H. and {Greus}, F. Salesa and {Sandoval}, A. and {Schneider}, M. and {Schoorlemmer}, H. and {Serna}, F. and {Smith}, A.~J. and {Springer}, R.~W. and {Surajbali}, P. and {Tollefson}, K. and {Torres}, I. and {Torres-Escobedo}, R. and {Ure{\~n}a-Mena}, F. and {Weisgarber}, T. and {Werner}, F. and {Willox}, E. and {Zepeda}, A. and {Zhou}, H. and {De Le{\'o}n}, C. and {{\'A}lvarez}, J.~D.},
        title = "{HAWC observations of the acceleration of very-high-energy cosmic rays in the Cygnus Cocoon}",
      journal = {Nature Astronomy},
         year = 2021,
        month = may,
       volume = {5},
        pages = {465-471},
          doi = {10.1038/s41550-021-01318-y},
archivePrefix = {arXiv},
       eprint = {2103.06820},
 primaryClass = {astro-ph.HE},
       adsurl = {https://ui.adsabs.harvard.edu/abs/2021NatAs...5..465A}
}

@ARTICLE{2022A&A...666A.124A,
       author = {{Aharonian}, F. and {Ashkar}, H. and {Backes}, M. and {Barbosa Martins}, V. and {Becherini}, Y. and {Berge}, D. and {Bi}, B. and {B{\"o}ttcher}, M. and {de Bony de Lavergne}, M. and {Bradascio}, F. and {Brose}, R. and {Brun}, F. and {Bulik}, T. and {Burger-Scheidlin}, C. and {Cangemi}, F. and {Caroff}, S. and {Casanova}, S. and {Cerruti}, M. and {Chand}, T. and {Chandra}, S. and {Chen}, A. and {Chibueze}, O. and {Cristofari}, P. and {Damascene Mbarubucyeye}, J. and {Djannati-Ata{\"\i}}, A. and {Ernenwein}, J.-P. and {Feijen}, K. and {Fichet de Clairfontaine}, G. and {Fontaine}, G. and {Funk}, S. and {Gabici}, S. and {Gallant}, Y.~A. and {Ghafourizadeh}, S. and {Giavitto}, G. and {Giunti}, L. and {Glawion}, D. and {Glicenstein}, J.~F. and {Goswami}, P. and {Grondin}, M.-H. and {H{\"a}rer}, L.~K. and {Haupt}, M. and {Hinton}, J.~A. and {H{\"o}rbe}, M. and {Hofmann}, W. and {Holch}, T.~L. and {Holler}, M. and {Horns}, D. and {Jamrozy}, M. and {Joshi}, V. and {Jung-Richardt}, I. and {Kasai}, E. and {Katarzy{\'n}ski}, K. and {Katz}, U. and {Kh{\'e}lifi}, B. and {Klu{\'z}niak}, W. and {Komin}, Nu. and {Kosack}, K. and {Kostunin}, D. and {Kukec Mezek}, G. and {Lang}, R.~G. and {Le Stum}, S. and {Lemi{\`e}re}, A. and {Lemoine-Goumard}, M. and {Lenain}, J.-P. and {Leuschner}, F. and {Lohse}, T. and {Luashvili}, A. and {Lypova}, I. and {Mackey}, J. and {Majumdar}, J. and {Malyshev}, D. and {Marandon}, V. and {Marchegiani}, P. and {Marcowith}, A. and {Mart{\'\i}-Devesa}, G. and {Marx}, R. and {Maurin}, G. and {Meyer}, M. and {Mitchell}, A. and {Moderski}, R. and {Mohrmann}, L. and {Montanari}, A. and {Moulin}, E. and {Muller}, J. and {Murach}, T. and {Nakashima}, K. and {de Naurois}, M. and {Nayerhoda}, A. and {Niemiec}, J. and {Ohm}, S. and {Olivera-Nieto}, L. and {de Ona Wilhelmi}, E. and {Ostrowski}, M. and {Panny}, S. and {Panter}, M. and {Parsons}, R.~D. and {Peron}, G. and {Prokhorov}, D.~A. and {P{\"u}hlhofer}, G. and {Punch}, M. and {Quirrenbach}, A. and {Rauth}, R. and {Reichherzer}, P. and {Reimer}, A. and {Reimer}, O. and {Renaud}, M. and {Reville}, B. and {Rieger}, F. and {Rowell}, G. and {Rudak}, B. and {Ruiz-Velasco}, E. and {Sahakian}, V. and {Salzmann}, H. and {Sanchez}, D.~A. and {Santangelo}, A. and {Sasaki}, M. and {Sch{\"u}ssler}, F. and {Schutte}, H.~M. and {Schwanke}, U. and {Shapopi}, J.~N.~S. and {Specovius}, A. and {Spencer}, S. and {Stawarz}, {\L}. and {Steenkamp}, R. and {Steinmassl}, S. and {Steppa}, C. and {Sushch}, I. and {Suzuki}, H. and {Takahashi}, T. and {Tanaka}, T. and {Terrier}, R. and {Thorpe-Morgan}, C. and {Tsirou}, M. and {Tsuji}, N. and {Tuffs}, R. and {Unbehaun}, T. and {van Eldik}, C. and {van Soelen}, B. and {Vecchi}, M. and {Veh}, J. and {Venter}, C. and {Vink}, J. and {Wagner}, S.~J. and {White}, R. and {Wierzcholska}, A. and {Wong}, Y. Wun and {Zacharias}, M. and {Zargaryan}, D. and {Zdziarski}, A.~A. and {Zhu}, S.~J. and {Zouari}, S. and {{\.Z}ywucka}, N. and {Blackwell}, R. and {Braiding}, C. and {Burton}, M. and {Cubuk}, K. and {Filipovi{\'c}}, M. and {Tothill}, N. and {Wong}, G.},
        title = "{A deep spectromorphological study of the {\ensuremath{\gamma}}-ray emission surrounding the young massive stellar cluster Westerlund 1}",
      journal = {\aap},
         year = 2022,
        month = oct,
       volume = {666},
          eid = {A124},
        pages = {A124},
          doi = {10.1051/0004-6361/202244323},
archivePrefix = {arXiv},
       eprint = {2207.10921},
 primaryClass = {astro-ph.HE},
       adsurl = {https://ui.adsabs.harvard.edu/abs/2022A&A...666A.124A}
}

@ARTICLE{2005APh....24....1A,
       author = {{Antoni}, T. and {Apel}, W.~D. and {Badea}, A.~F. and {Bekk}, K. and {Bercuci}, A. and {Bl{\"u}mer}, J. and {Bozdog}, H. and {Brancus}, I.~M. and {Chilingarian}, A. and {Daumiller}, K. and {Doll}, P. and {Engel}, R. and {Engler}, J. and {Fe{\ss}ler}, F. and {Gils}, H.~J. and {Glasstetter}, R. and {Haungs}, A. and {Heck}, D. and {H{\"o}randel}, J.~R. and {Kampert}, K.-H. and {Klages}, H.~O. and {Maier}, G. and {Mathes}, H.~J. and {Mayer}, H.~J. and {Milke}, J. and {M{\"u}ller}, M. and {Obenland}, R. and {Oehlschl{\"a}ger}, J. and {Ostapchenko}, S. and {Petcu}, M. and {Rebel}, H. and {Risse}, A. and {Risse}, M. and {Roth}, M. and {Schatz}, G. and {Schieler}, H. and {Scholz}, J. and {Thouw}, T. and {Ulrich}, H. and {van Buren}, J. and {Vardanyan}, A. and {Weindl}, A. and {Wochele}, J. and {Zabierowski}, J.},
        title = "{KASCADE measurements of energy spectra for elemental groups of cosmic rays: Results and open problems}",
      journal = {Astroparticle Physics},
         year = 2005,
        month = sep,
       volume = {24},
       number = {1-2},
        pages = {1-25},
          doi = {10.1016/j.astropartphys.2005.04.001},
archivePrefix = {arXiv},
       eprint = {astro-ph/0505413},
 primaryClass = {astro-ph},
       adsurl = {https://ui.adsabs.harvard.edu/abs/2005APh....24....1A}
}

@ARTICLE{2016NPPP..279....7M,
       author = {{Montini}, Paolo and {ARGO-YBJ Collaboration}},
        title = "{Cosmic ray physics with ARGO-YBJ}",
      journal = {Nuclear and Particle Physics Proceedings},
         year = 2016,
        month = oct,
       volume = {279-281},
        pages = {7-14},
          doi = {10.1016/j.nuclphysbps.2016.10.003},
archivePrefix = {arXiv},
       eprint = {1608.01251},
 primaryClass = {astro-ph.HE},
       adsurl = {https://ui.adsabs.harvard.edu/abs/2016NPPP..279....7M}
}

@ARTICLE{1999APh....10..291G,
       author = {{Glasmacher}, M.~A.~K. and {Catanese}, M.~A. and {Chantell}, M.~C. and {Covault}, C.~E. and {Cronin}, J.~W. and {Fick}, B.~E. and {Fortson}, L.~F. and {Fowler}, J.~W. and {Green}, K.~D. and {Kieda}, D.~B. and {Matthews}, J. and {Newport}, B.~J. and {Nitz}, D.~F. and {Ong}, R.~A. and {Oser}, S. and {Sinclair}, D. and {van der Velde}, J.~C.},
        title = "{The cosmic ray energy spectrum between 10 $^{14}$ and 10 $^{16}$ eV}",
      journal = {Astroparticle Physics},
         year = 1999,
        month = may,
       volume = {10},
       number = {4},
        pages = {291-302},
          doi = {10.1016/S0927-6505(98)00070-X},
       adsurl = {https://ui.adsabs.harvard.edu/abs/1999APh....10..291G}
}

@ARTICLE{2020PhRvD.102l2001A,
       author = {{Aartsen}, M.~G. and {Abbasi}, R. and {Ackermann}, M. and {Adams}, J. and {Aguilar}, J.~A. and {Ahlers}, M. and {Ahrens}, M. and {Alispach}, C. and {Amin}, N.~M. and {Andeen}, K. and {Anderson}, T. and {Ansseau}, I. and {Anton}, G. and {Arg{\"u}elles}, C. and {Auffenberg}, J. and {Axani}, S. and {Bagherpour}, H. and {Bai}, X. and {Balagopal V.}, A. and {Barbano}, A. and {Barwick}, S.~W. and {Bastian}, B. and {Baum}, V. and {Baur}, S. and {Bay}, R. and {Beatty}, J.~J. and {Becker}, K.-H. and {Becker Tjus}, J. and {BenZvi}, S. and {Berley}, D. and {Bernardini}, E. and {Besson}, D.~Z. and {Binder}, G. and {Bindig}, D. and {Blaufuss}, E. and {Blot}, S. and {Bohm}, C. and {B{\"o}ser}, S. and {Botner}, O. and {B{\"o}ttcher}, J. and {Bourbeau}, E. and {Bourbeau}, J. and {Bradascio}, F. and {Braun}, J. and {Bron}, S. and {Brostean-Kaiser}, J. and {Burgman}, A. and {Buscher}, J. and {Busse}, R.~S. and {Carver}, T. and {Chen}, C. and {Cheung}, E. and {Chirkin}, D. and {Choi}, S. and {Clark}, B.~A. and {Clark}, K. and {Classen}, L. and {Coleman}, A. and {Collin}, G.~H. and {Conrad}, J.~M. and {Coppin}, P. and {Correa}, P. and {Cowen}, D.~F. and {Cross}, R. and {Dave}, P. and {De Clercq}, C. and {DeLaunay}, J.~J. and {Dembinski}, H. and {Deoskar}, K. and {De Ridder}, S. and {Desiati}, P. and {de Vries}, K.~D. and {de Wasseige}, G. and {de With}, M. and {DeYoung}, T. and {Dharani}, S. and {Diaz}, A. and {D{\'\i}az-V{\'e}lez}, J.~C. and {Dujmovic}, H. and {Dvorak}, E. and {Eberhardt}, B. and {Ehrhardt}, T. and {Eller}, P. and {Engel}, R. and {Evenson}, P.~A. and {Fahey}, S. and {Fazely}, A.~R. and {Felde}, J. and {Fienberg}, A.~T. and {Filimonov}, K. and {Finley}, C. and {Fox}, D. and {Franckowiak}, A. and {Friedman}, E. and {Fritz}, A. and {Gaisser}, T.~K. and {Gallagher}, J. and {Ganster}, E. and {Garrappa}, S. and {Gerhardt}, L. and {Ghorbani}, K. and {Glauch}, T. and {Gl{\"u}senkamp}, T. and {Goldschmidt}, A. and {Gonzalez}, J.~G. and {Grant}, D. and {Gr{\'e}goire}, T. and {Griffith}, Z. and {Griswold}, S. and {G{\"u}nder}, M. and {G{\"u}nd{\"u}z}, M. and {Haack}, C. and {Hallgren}, A. and {Halliday}, R. and {Halve}, L. and {Halzen}, F. and {Hanson}, K. and {Haungs}, A. and {Hauser}, S. and {Hebecker}, D. and {Heereman}, D. and {Heix}, P. and {Helbing}, K. and {Hellauer}, R. and {Henningsen}, F. and {Hickford}, S. and {Hignight}, J. and {Hill}, C. and {Hill}, G.~C. and {Hoffman}, K.~D. and {Hoffmann}, R. and {Hoinka}, T. and {Hokanson-Fasig}, B. and {Hoshina}, K. and {Huber}, M. and {Huber}, T. and {Hultqvist}, K. and {H{\"u}nnefeld}, M. and {Hussain}, R. and {In}, S. and {Iovine}, N. and {Ishihara}, A. and {Jansson}, M. and {Japaridze}, G.~S. and {Jeong}, M. and {Jero}, K. and {Jones}, B.~J.~P. and {Jonske}, F. and {Joppe}, R. and {Kang}, D. and {Kang}, W. and {Kappes}, A. and {Kappesser}, D. and {Karg}, T. and {Karl}, M. and {Karle}, A. and {Katz}, U. and {Kauer}, M. and {Kellermann}, M. and {Kelley}, J.~L. and {Kheirandish}, A. and {Kim}, J. and {Kintscher}, T. and {Kiryluk}, J. and {Kittler}, T. and {Klein}, S.~R. and {Koirala}, R. and {Kolanoski}, H. and {K{\"o}pke}, L. and {Kopper}, C. and {Kopper}, S. and {Koskinen}, D.~J. and {Koundal}, P. and {Kowalski}, M. and {Krings}, K. and {Kr{\"u}ckl}, G. and {Kulacz}, N. and {Kurahashi}, N. and {Kyriacou}, A. and {Lanfranchi}, J.~L. and {Larson}, M.~J. and {Lauber}, F. and {Lazar}, J.~P. and {Leonard}, K. and {Leszczy{\'n}ska}, A. and {Li}, Y. and {Liu}, Q.~R. and {Lohfink}, E. and {Lozano Mariscal}, C.~J. and {Lu}, L. and {Lucarelli}, F. and {Ludwig}, A. and {L{\"u}nemann}, J. and {Luszczak}, W. and {Lyu}, Y. and {Ma}, W.~Y. and {Madsen}, J. and {Maggi}, G. and {Mahn}, K.~B.~M. and {Mallik}, P.},
        title = "{Cosmic ray spectrum from 250 TeV to 10 PeV using IceTop}",
      journal = {\prd},
         year = 2020,
        month = dec,
       volume = {102},
       number = {12},
          eid = {122001},
        pages = {122001},
          doi = {10.1103/PhysRevD.102.122001},
archivePrefix = {arXiv},
       eprint = {2006.05215},
 primaryClass = {astro-ph.HE},
       adsurl = {https://ui.adsabs.harvard.edu/abs/2020PhRvD.102l2001A}
}

@ARTICLE{2001AdSpR..27..803E,
       author = {{Erlykin}, A.~D. and {Wolfendale}, A.~W.},
        title = "{Models for the origin of the knee in the cosmic-ray spectrum}",
      journal = {Advances in Space Research},
         year = 2001,
        month = jan,
       volume = {27},
       number = {4},
        pages = {803-812},
          doi = {10.1016/S0273-1177(01)00125-9},
archivePrefix = {arXiv},
       eprint = {astro-ph/0011057},
 primaryClass = {astro-ph},
       adsurl = {https://ui.adsabs.harvard.edu/abs/2001AdSpR..27..803E}
}

@ARTICLE{1961NCim...22..800P,
       author = {{Peters}, B.},
        title = "{Primary cosmic radiation and extensive air showers}",
      journal = {Il Nuovo Cimento},
         year = 1961,
        month = nov,
       volume = {22},
       number = {4},
        pages = {800-819},
          doi = {10.1007/BF02783106},
       adsurl = {https://ui.adsabs.harvard.edu/abs/1961NCim...22..800P}
}

@ARTICLE{2004APh....21..241H,
       author = {{H{\"o}randel}, J{\"o}rg R.},
        title = "{Models of the knee in the energy spectrum of cosmic rays}",
      journal = {Astroparticle Physics},
         year = 2004,
        month = jun,
       volume = {21},
       number = {3},
        pages = {241-265},
          doi = {10.1016/j.astropartphys.2004.01.004},
archivePrefix = {arXiv},
       eprint = {astro-ph/0402356},
 primaryClass = {astro-ph},
       adsurl = {https://ui.adsabs.harvard.edu/abs/2004APh....21..241H}
}

@ARTICLE{2024arXiv241113793H,
       author = {{He}, Huihai and {Zhang}, Hengying and {Cheng}, Qinyi and {Ma}, Lingling and {Feng}, Cunfeng},
        title = "{Unveiling the nature of the knee in the cosmic ray energy spectrum}",
      journal = {arXiv e-prints},
         year = 2024,
        month = nov,
          eid = {arXiv:2411.13793},
        pages = {arXiv:2411.13793},
          doi = {10.48550/arXiv.2411.13793},
archivePrefix = {arXiv},
       eprint = {2411.13793},
 primaryClass = {astro-ph.HE},
       adsurl = {https://ui.adsabs.harvard.edu/abs/2024arXiv241113793H}
}

@article{PhysRevD.106.123028,
  title = {Approaches to composition independent energy reconstruction of cosmic rays based on the LHAASO-KM2A detector},
  author = {Zhang, Hengying and He, Huihai and Feng, Cunfeng},
  journal = {Phys. Rev. D},
  volume = {106},
  issue = {12},
  pages = {123028},
  numpages = {9},
  year = {2022},
  month = {Dec},
  publisher = {American Physical Society},
  doi = {10.1103/PhysRevD.106.123028},
  url = {https://link.aps.org/doi/10.1103/PhysRevD.106.123028}
}

@ARTICLE{1998ApJ...492..219G,
       author = {{Gaisser}, T.~K. and {Protheroe}, R.~J. and {Stanev}, Todor},
        title = "{Gamma-Ray Production in Supernova Remnants}",
      journal = {\apj},
         year = 1998,
        month = jan,
       volume = {492},
       number = {1},
        pages = {219-227},
          doi = {10.1086/305011},
archivePrefix = {arXiv},
       eprint = {astro-ph/9609044},
 primaryClass = {astro-ph},
       adsurl = {https://ui.adsabs.harvard.edu/abs/1998ApJ...492..219G}
}

@ARTICLE{2014PhRvD..90l3014K,
       author = {{Kafexhiu}, Ervin and {Aharonian}, Felix and {Taylor}, Andrew M. and {Vila}, Gabriela S.},
        title = "{Parametrization of gamma-ray production cross sections for p p interactions in a broad proton energy range from the kinematic threshold to PeV energies}",
      journal = {\prd},
         year = 2014,
        month = dec,
       volume = {90},
       number = {12},
          eid = {123014},
        pages = {123014},
          doi = {10.1103/PhysRevD.90.123014},
archivePrefix = {arXiv},
       eprint = {1406.7369},
 primaryClass = {astro-ph.HE},
       adsurl = {https://ui.adsabs.harvard.edu/abs/2014PhRvD..90l3014K}
}

@ARTICLE{1992PhRvD..46.5013E,
       author = {{Engel}, J. and {Gaisser}, T.~K. and {Lipari}, Paolo and {Stanev}, Todor},
        title = "{Nucleus-nucleus collisions and interpretation of cosmic-ray cascades}",
      journal = {\prd},
         year = 1992,
        month = dec,
       volume = {46},
       number = {11},
        pages = {5013-5025},
          doi = {10.1103/PhysRevD.46.5013},
       adsurl = {https://ui.adsabs.harvard.edu/abs/1992PhRvD..46.5013E}
}

@ARTICLE{2009PhRvD..80i4003A,
       author = {{Ahn}, Eun-Joo and {Engel}, Ralph and {Gaisser}, Thomas K. and {Lipari}, Paolo and {Stanev}, Todor},
        title = "{Cosmic ray interaction event generator SIBYLL 2.1}",
      journal = {\prd},
         year = 2009,
        month = nov,
       volume = {80},
       number = {9},
          eid = {094003},
        pages = {094003},
          doi = {10.1103/PhysRevD.80.094003},
archivePrefix = {arXiv},
       eprint = {0906.4113},
 primaryClass = {hep-ph},
       adsurl = {https://ui.adsabs.harvard.edu/abs/2009PhRvD..80i4003A}
}

@ARTICLE{2016SAAS...43...85K,
       author = {{Klessen}, Ralf S. and {Glover}, Simon C.~O.},
        title = "{Physical Processes in the Interstellar Medium}",
      journal = {Saas-Fee Advanced Course},
         year = 2016,
        month = jan,
       volume = {43},
        pages = {85},
          doi = {10.1007/978-3-662-47890-5_2},
archivePrefix = {arXiv},
       eprint = {1412.5182},
 primaryClass = {astro-ph.GA},
       adsurl = {https://ui.adsabs.harvard.edu/abs/2016SAAS...43...85K}
}

@ARTICLE{2025A&A...694A.244B,
       author = {{Blasi}, Pasquale},
        title = "{Gamma rays from star clusters and implications for the origin of Galactic cosmic rays}",
      journal = {\aap},
         year = 2025,
        month = feb,
       volume = {694},
          eid = {A244},
        pages = {A244},
          doi = {10.1051/0004-6361/202453017},
archivePrefix = {arXiv},
       eprint = {2501.16097},
 primaryClass = {astro-ph.HE},
       adsurl = {https://ui.adsabs.harvard.edu/abs/2025A&A...694A.244B}
}

@ARTICLE{2024MNRAS.533..561B,
       author = {{Blasi}, Pasquale and {Morlino}, Giovanni},
        title = "{Different spectra of cosmic ray H, He, and heavier nuclei escaping compact star clusters}",
      journal = {\mnras},
         year = 2024,
        month = sep,
       volume = {533},
       number = {1},
        pages = {561-571},
          doi = {10.1093/mnras/stae1782},
archivePrefix = {arXiv},
       eprint = {2307.11663},
 primaryClass = {astro-ph.HE},
       adsurl = {https://ui.adsabs.harvard.edu/abs/2024MNRAS.533..561B}
}

@ARTICLE{2022A&A...667A..69S,
       author = {{Suin}, Paolo and {Shore}, Steven N. and {Pavl{\'\i}k}, V{\'a}clav},
        title = "{Environmental effects on the dynamical evolution of star clusters in turbulent molecular clouds}",
      journal = {\aap},
         year = 2022,
        month = nov,
       volume = {667},
          eid = {A69},
        pages = {A69},
          doi = {10.1051/0004-6361/202243579},
archivePrefix = {arXiv},
       eprint = {2207.01634},
 primaryClass = {astro-ph.GA},
       adsurl = {https://ui.adsabs.harvard.edu/abs/2022A&A...667A..69S}
}

@ARTICLE{2014PhRvD..90b3010A,
       author = {{Ahlers}, Markus and {Murase}, Kohta},
        title = "{Probing the Galactic origin of the IceCube excess with gamma rays}",
      journal = {\prd},
         year = 2014,
        month = jul,
       volume = {90},
       number = {2},
          eid = {023010},
        pages = {023010},
          doi = {10.1103/PhysRevD.90.023010},
archivePrefix = {arXiv},
       eprint = {1309.4077},
 primaryClass = {astro-ph.HE},
       adsurl = {https://ui.adsabs.harvard.edu/abs/2014PhRvD..90b3010A}
}

@ARTICLE{2023MNRAS.521.1144S,
       author = {{Sarmah}, Prantik and {Chakraborty}, Sovan and {Joshi}, Jagdish C.},
        title = "{Probing LHAASO galactic PeVatrons through gamma-ray and neutrino correspondence}",
      journal = {\mnras},
         year = 2023,
        month = may,
       volume = {521},
       number = {1},
        pages = {1144-1151},
          doi = {10.1093/mnras/stad609},
archivePrefix = {arXiv},
       eprint = {2301.04161},
 primaryClass = {astro-ph.HE},
       adsurl = {https://ui.adsabs.harvard.edu/abs/2023MNRAS.521.1144S}
}

@ARTICLE{2006ApJ...640L.155M,
       author = {{Moskalenko}, Igor V. and {Porter}, Troy A. and {Strong}, Andrew W.},
        title = "{Attenuation of Very High Energy Gamma Rays by the Milky Way Interstellar Radiation Field}",
      journal = {\apjl},
         year = 2006,
        month = apr,
       volume = {640},
       number = {2},
        pages = {L155-L158},
          doi = {10.1086/503524},
archivePrefix = {arXiv},
       eprint = {astro-ph/0511149},
 primaryClass = {astro-ph},
       adsurl = {https://ui.adsabs.harvard.edu/abs/2006ApJ...640L.155M}
}

@ARTICLE{2017NPPP..291....9G,
       author = {{Grasso}, D. and {Gaggero}, D. and {Marinelli}, A. and {Urbano}, A. and {Valli}, M.},
        title = "{Gamma-ray and Neutrino Diffuse Emissions of the Galaxy at very High Energy}",
      journal = {Nuclear and Particle Physics Proceedings},
         year = 2017,
        month = oct,
       volume = {291-293},
        pages = {9-14},
          doi = {10.1016/j.nuclphysbps.2017.06.003},
       adsurl = {https://ui.adsabs.harvard.edu/abs/2017NPPP..291....9G}
}
\bibliographystyle{aasjournalv7}

\appendix
\section{Corner Plots}

\begin{figure}[!htb]
   \centering
   \subfloat[$\chi^{2} = 2.631$]{
   \includegraphics[width=0.445\textwidth]{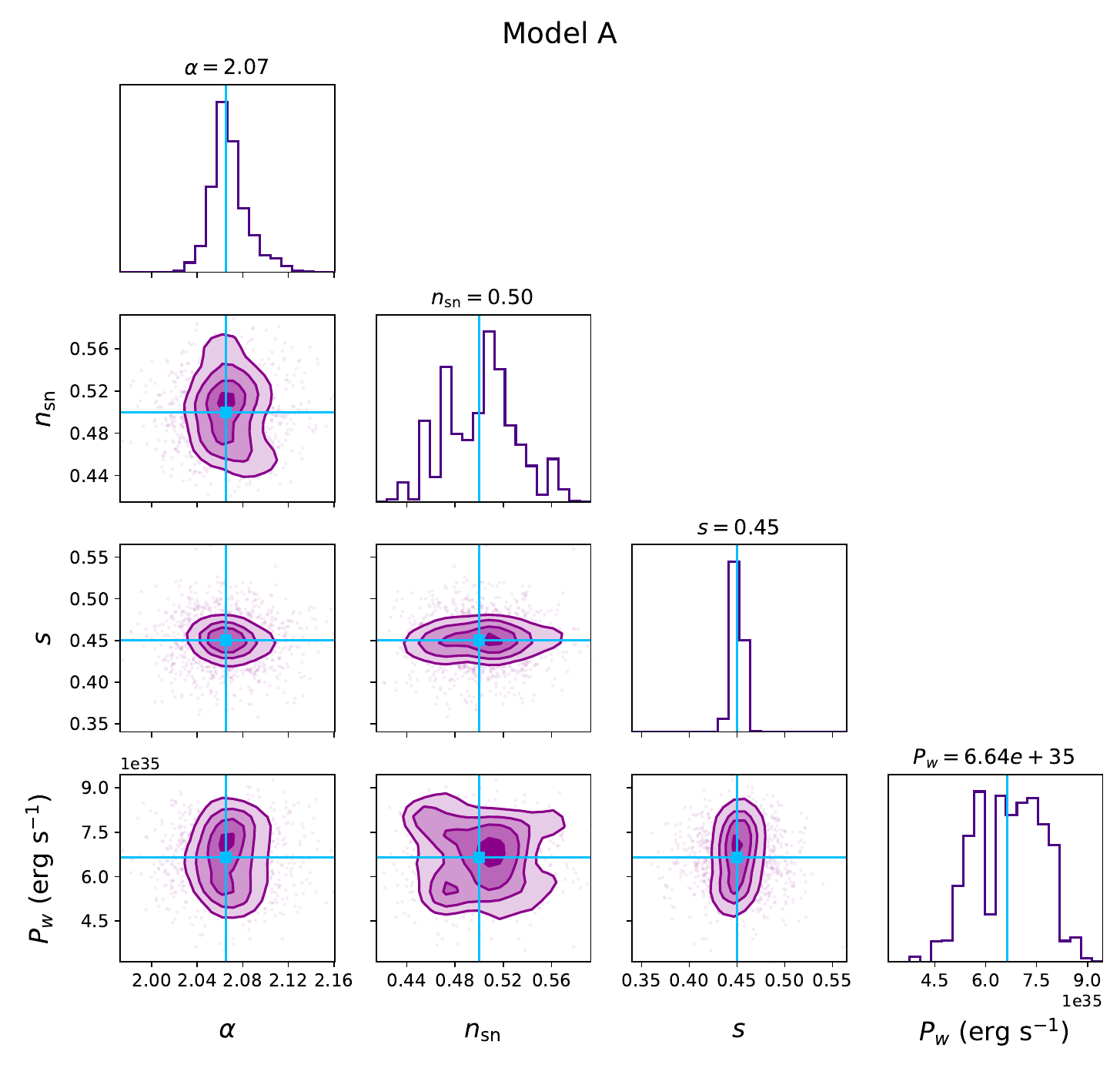}}
   \subfloat[$\chi^{2} = 1.608$]{\includegraphics[width=0.445\textwidth]  {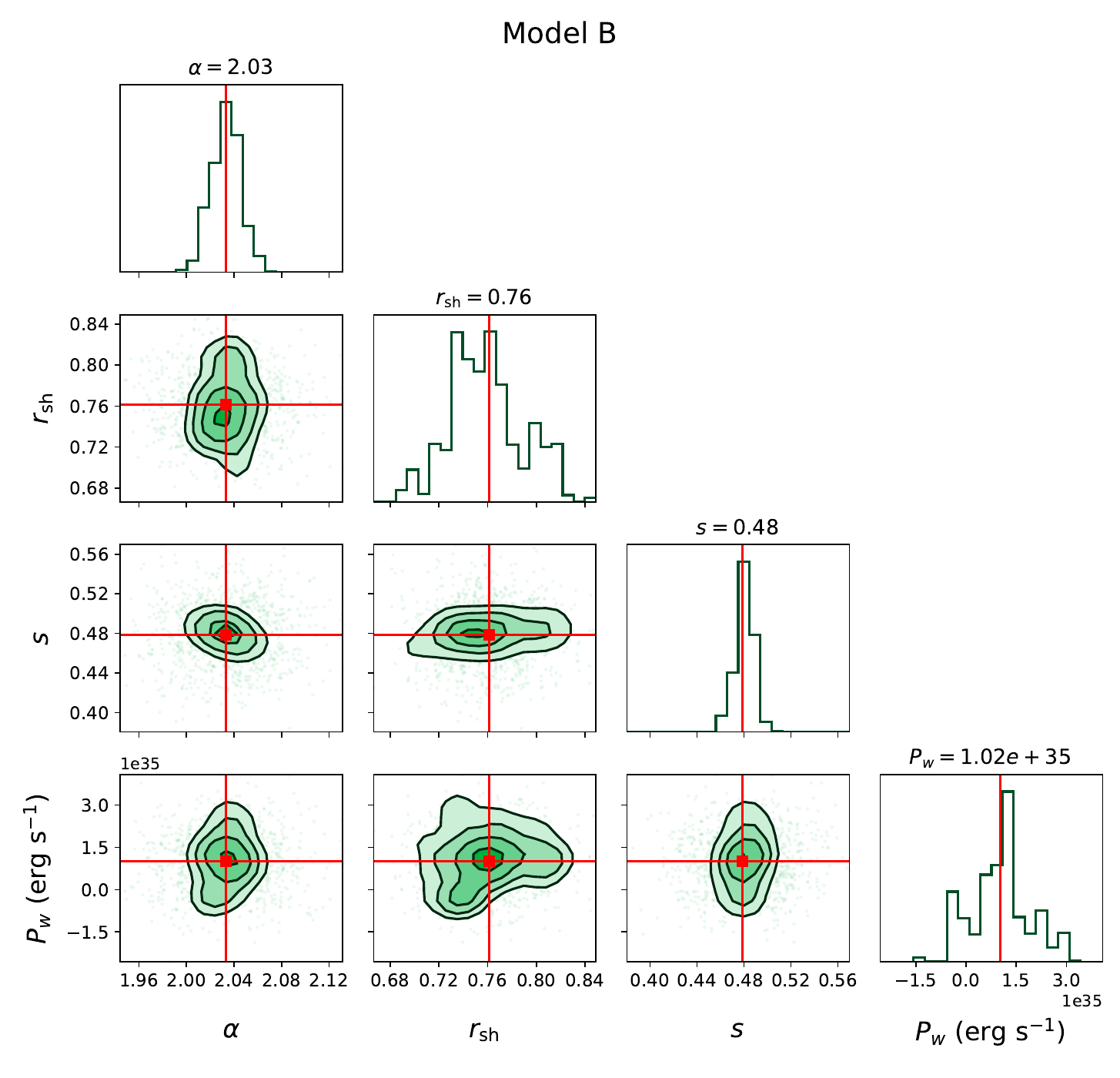}}\hfill
    \subfloat[$\chi^{2} = 1.938$]{\includegraphics[width=0.445\textwidth]{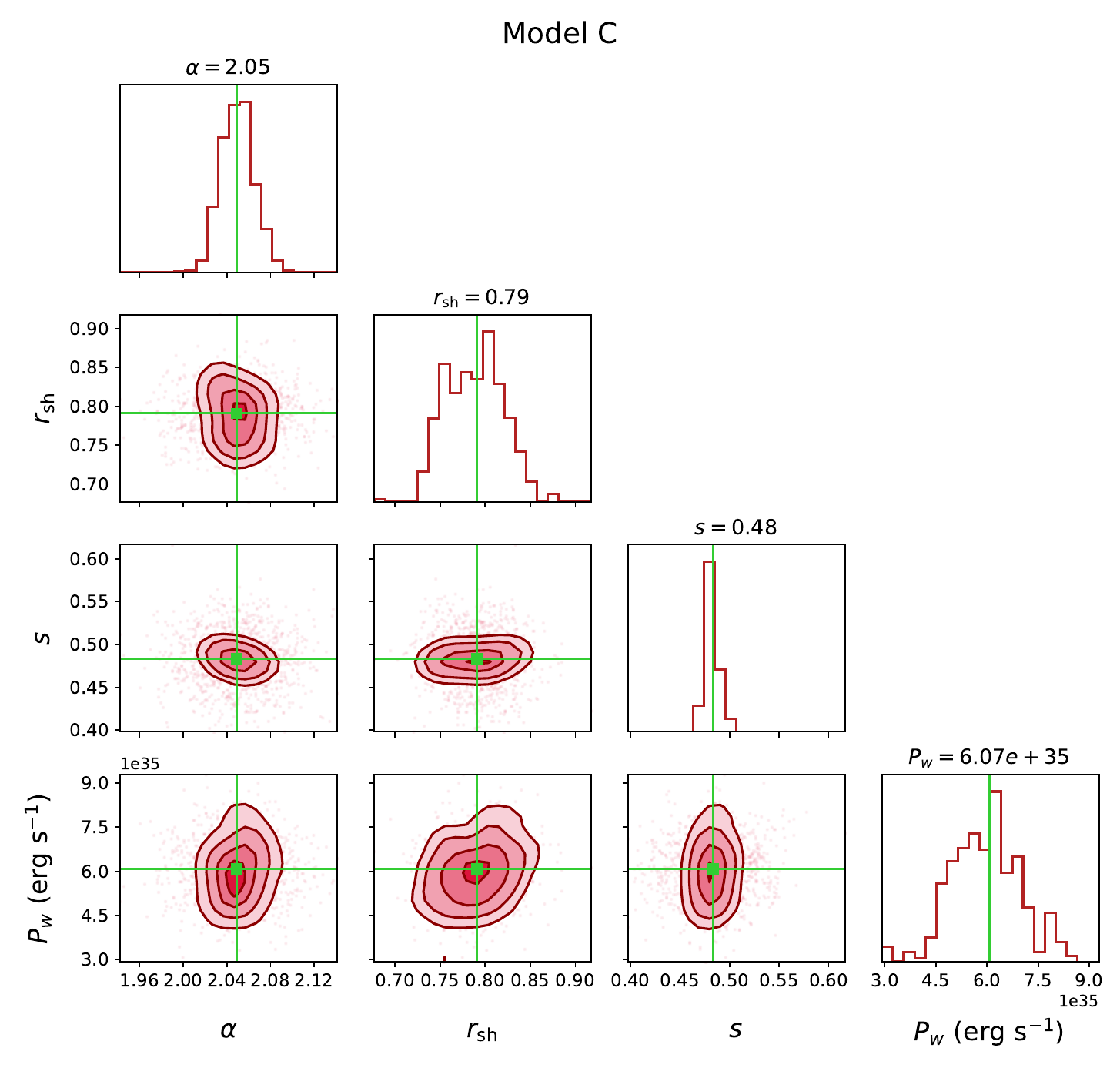}} 
    \subfloat[$\chi^{2} = 2.691$]{\includegraphics[width=0.445\textwidth]{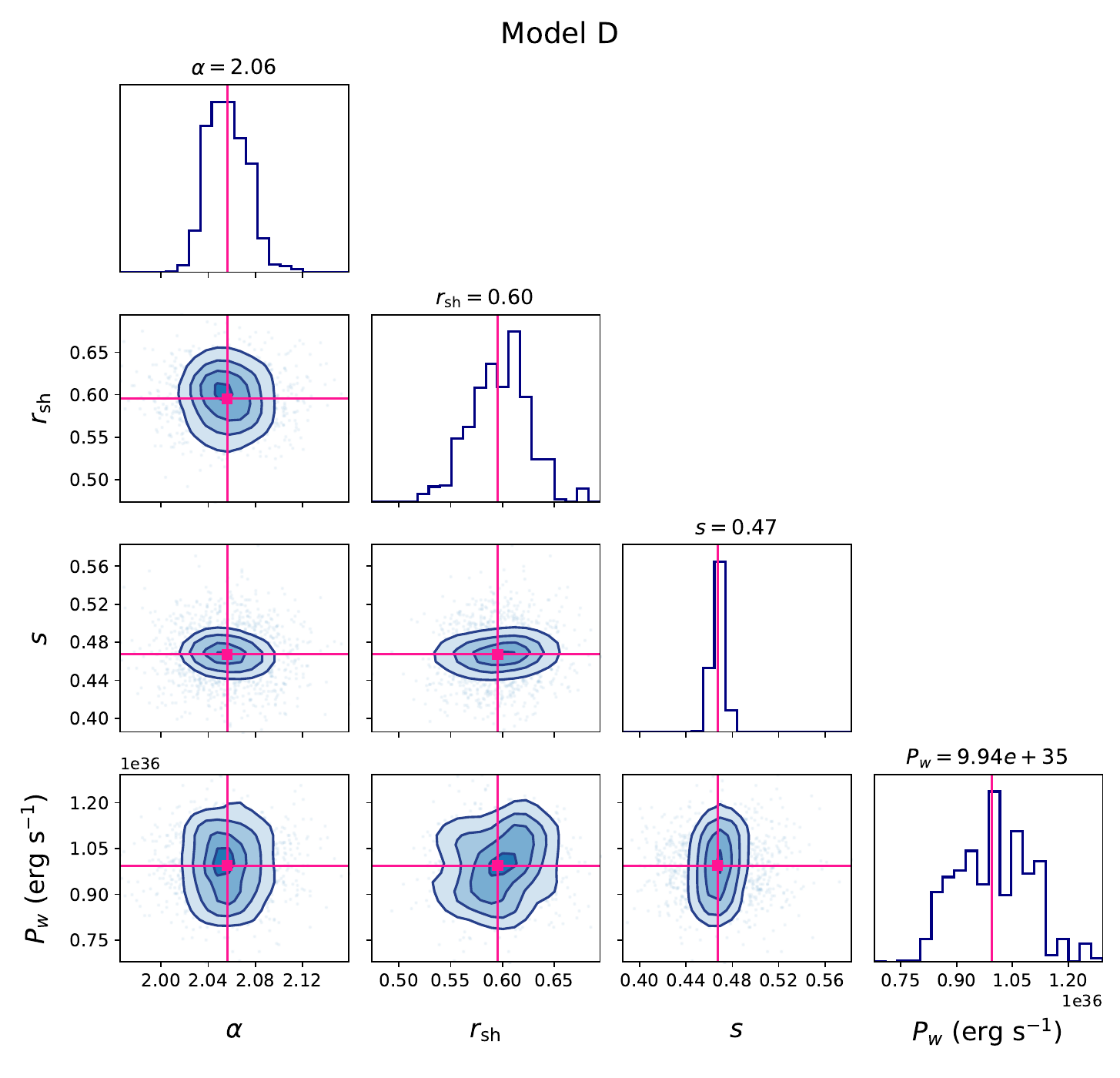}}\hfill
    \caption{\small \justifying  Corner plots for Models A to D showing the local exploration of parameter space around the best-fit values, weighted by the \(\chi^2\) difference. The diagonal panels display one dimensional posteriors; vertical lines mark the median and the central 68\% confidence interval. The off diagonal panels show joint posteriors with contours at 68\% and 95\% credibility. Parameters include the wind power \(P_{w}\) (in erg\,s\(^{-1}\)), the transport index \(s\), the characteristic shock radius \(r_{\rm sh}\) (in pc), the source index \(\alpha\), and, for Model A only, the supernova proximity factor \(n_{\rm sn}\). The numerical values printed in each panel correspond to the marginalized medians, and the number \(\chi^{2}_{\nu}\) beneath each set summarizes the fit quality for the corresponding model. Angular weights are not shown in this corner plot for clarity, they are presented in Table~\ref{tab:angle} and discussed in Section~\ref{ssec:inj_prop}}.
    \label{fig:corners}
\end{figure}

\end{document}